\documentclass[structabstract]{aa}  
\usepackage{graphicx}
\usepackage{txfonts}
\begin{document}
   \title{Young open clusters in the Galactic star forming region NGC~6357\thanks{Based on 
         observations made with ESO telescopes at the La Silla Paranal Observatory under 
         programme ID 63.L--0717}\fnmsep\thanks{Tables~1 and~2 are only available in electronic
		form at the CDS via anonymous ftp to cdsarc.u-strasbg.fr (130.79.128.5)
		or via http://cdsweb.u-strasbg.fr/cgi-bin/qcat?J/A+A/}}


   \author{F. Massi 
          \inst{1}
          \and
	  A. Giannetti
	  \inst{2}\fnmsep\inst{5}
	  \and 
	  E. di Carlo
	  \inst{3}
	  \and
	  J. Brand
	  \inst{2}\fnmsep\inst{5}
	\and
          M. T. Beltr\'an
	\inst{1}
	  \and
          G. Marconi
         \inst{4}
          }

   \institute{INAF - Osservatorio Astrofisico di Arcetri, Largo E.\ Fermi 5,
              I-50125 Firenze, Italy\\
              \email{fmassi,mbeltran@arcetri.astro.it}
         \and
    	      INAF - Istituto di Radioastronomia, via Gobetti
 		101, I-40129 Bologna, Italy\\	
             \email{agianne,brand@ira.inaf.it}
	   \and
	      INAF - Osservatorio Astronomico di Collurania-Teramo,
             Via M.\ Maggini, I--64100 Teramo, Italy\\
             \email{dicarlo@oa-teramo.inaf.it}
	   \and
              ESO, Al\'{o}nso de C\'{o}rdova 3107, Vitacura, Santiago de Chile, Chile\\
             \email{gmarconi@eso.org}
	   \and
	      Italian ALMA Regional Centre, via Gobetti
                101, I-40129 Bologna, Italy\\ 
             }

\date{Received ; accepted }

 
  \abstract
   {NGC~6357 is an active star forming region with very young
    massive open clusters. These clusters  contain some of the most massive
    stars in the Galaxy and strongly interact with nearby
   giant molecular clouds.}
   {We study the young stellar populations of the region 
    and of the open cluster Pismis~24, focusing on
    their relationship with the nearby giant molecular clouds.
    We seek evidence of triggered star formation
	``propagating'' from the clusters.}
   {We used new deep $JHK_{s}$ photometry, along with unpublished
   deep IRAC/Spitzer mid-infrared photometry, complemented with optical
   HST/WFPC2 high spatial resolution photometry and X-ray Chandra observations, 
   to constrain age, initial mass function, and star formation modes in progress. We carefully examine
   and discuss all sources of bias (saturation, confusion, different sensitivities, extinction).}
   {NGC~6357 hosts three large young stellar clusters, of which Pismis~24 is the most prominent. 
   We found that Pismis~24 is a very young ($\sim 1-3$ Myr)
    open cluster with a Salpeter-like IMF and a few
	thousand members. A comparison between optical and infrared
        photometry indicates that the fraction of members with a near-infrared
        excess (i. e., with a circumstellar disk) is in the range $0.3-0.6$,
        consistent with its photometrically derived age. 
        We also find that Pismis~24 is likely subdivided into 
	a few different sub-clusters, one of which contains almost all the
	massive members. There are indications of current star formation
	triggered by these massive stars, but clear age trends could not be derived
        (although the fraction of stars with a near-infrared excess does increase towards
        the H\textsc{ii} region associated with the cluster). 
    	The gas out of which Pismis~24 formed
	must have been distributed in dense clumps within a cloud
	of less dense gas $\sim 1$ pc in radius.}
   {Our findings provide some new insight into how young stellar populations
    and massive stars emerge, and evolve in the first few Myr after birth, 
    from a giant molecular cloud complex.}


	\titlerunning{Young open clusters in NGC~6357}
   \authorrunning{Massi et al.}
   \maketitle
%

\section{Introduction}

It is becoming clear that even stellar systems that are smaller than galaxies exhibit
complex patterns of star formation during their evolution. 
The most notable example is that of some
globular clusters, for which recent observations have shown 
that they can no longer be considered as made up of
single stellar populations, i. e., ensembles of stars of the  same age and chemical 
composition. Rather, it has been found that globular clusters host multiple stellar 
populations with different chemical signatures
(for an updated review on the subject, see Gratton et al.\ \cite{gratton}),
with age differences of $\sim 10^{8}$ yrs.  

In the context of the much younger star forming regions and open clusters 
populating the Galactic disk, $\la 10$
Myr in age, the concept of multiple stellar populations with different 
chemical compositions, as envisaged for globular clusters,
cannot be applied (e.g. De Silva et al.\ \cite{deSilva}). 
Nevertheless, observations are accumulating of young clusters and star forming regions 
that underwent several episodes of star formation during their early evolution, i. e., various generations of
stars formed in different bursts (maybe $\sim 1$ Myr or less apart) inside 
different nearby gaseous clumps,
which may or may not have been triggered by a previous nearby episode. 
Examples of star formation activity
occurring in several bursts or lasting for quite a long time can be
found in the literature.  Both an age spread and co-existing younger ($\sim 1$ Myr)
and older ($\sim 10$ Myr) stellar populations
were found in \object{NGC 3603} (Beccari et al.\ \cite{becca10}). Two generations of
pre-main sequence (PMS) stars, one $\sim 1$ Myr and one $\sim 10$ Myr old, were found
towards \object{NGC~346} (De Marchi et al.\ \cite{dema11b}).
De Marchi et al.\ (\cite{dema13}) found a bimodal age distribution in
\object{NGC 602}, with stars younger than 5 Myr and stars older than 30 Myr.
A younger ($\sim 4$ Myr) and an older ($> 12$ Myr)
PMS star population were also found in 30 Doradus, although with
different spatial distributions (De Marchi et al.\ \cite{dema11a}).
Jeffries et al.\ (\cite{jeffries2}) found two kinematically distinct young
stellar populations around the Wolf-Rayet binary system $\gamma^{2}$ Velorum,
with this binary system at least a few Myr younger than most
of the surrounding stars.
The observations therefore would point to multiple and/or lasting star formation episodes
occurring in a single region, on timescales ranging from 1--10 Myr in the smallest systems
(e. g., open clusters) and up to 100 Myr in the largest ones
(i. e., globular clusters).

Focusing on the shortest timescales ($\sim 10$ Myr), however, at least some of
the observational results are still debated. 
How long star formation goes on in a particular region before it runs out 
of dense gas and whether it occurs in a single burst
or in multiple episodes, or rather continuously, are often difficult questions to tackle 
observationally. 
For example, the most commonly used method for age determination
consists of comparing Hertzsprung-Russel or colour-magnitude 
diagrams with theoretical evolutionary tracks (e. g., Hillenbrand \cite{hillen}). 
Nevertheless, current theoretical modelling of PMS stars is not reliable enough in the first few 
Myr of life to pinpoint differences $\la 1$ Myr. 
In this respect, young ``multiple'' populations are therefore much more
difficult to identify than in evolved structures. 
As a result, age dispersions in young star clusters have been interpreted 
either as real and indicating ``long'' time-scales for
star formation by some authors (e. g., Palla \& Stahler \cite{ps99}, \cite{ps00}), 
or as only apparent, due to observational errors and variability, episodic
accretion, and/or 
other physical effects unaccounted for by PMS
 evolutionary tracks 
(Hartmann \cite{hartie}; Baraffe et al.\ \cite{baretal}; Preibisch \cite{preiby};
Jeffries et al.\ \cite{jeffries}). 

Triggered star formation, i. e., the sequential birth of generations 
of stars, each originated
by feedback from the previous generation, is one of the ways which may naturally
lead to different but nearby bursts of star formation in the same region. 
The physical mechanisms proposed for triggering star formation are 
summarised in Deharveng et al.\ (\cite{deha}). 
On the other hand, numerical simulations of turbulent molecular clouds
(e. g., Bonnell et al.\
\cite{bonnell:11}; Dale \& Bonnel \cite{dale:11}) predict that large molecular
clouds can host several
clusters of roughly the same age in different sites, without invoking
triggered star formation. Thus, sequential star groups 
and stellar populations in a turbulent cloud cannot easily be observationally
recognised unless very accurate age measurements are possible.
Nevertheless, the observational evidence for
ongoing triggered star formation near massive stars has been growing recently (for an updated list,
see Deharveng \& Zavagno \cite{deha:zava}). Numerical simulations 
as well (e. g., Dale et al.\ \cite{dale:12b})
have shown that triggered star formation does occur in giant molecular clouds as a result
of ionising feedback (and molecular outflows) from massive stars, although this feedback
also tends to result in a globally lower star-formation efficiency.  

In the present paper, we test several widely-used observational tools 
based on multi-wavelength (from the near-infrared to the X-ray) observations
to try and
identify any young stellar generation in the star forming complex \object{NGC~6357}, 
to derive the star formation
history of the whole region. 
NGC~6357 is a massive star forming region containing young ($\sim 1$ Myr)
open star clusters interacting with the parental gas, bubble-like structures,
and pillars, one of which with a Young Stellar Objects (YSOs) at its apex.
The young open cluster \object{Pismis~24} is the most prominent of the 
young stellar clusters in the complex.
Both NGC~6357 and Pismis~24 are described in Sect.~\ref{intro:ngc6357}.
We also study the effects
of feedback UV radiation from massive stars on the currently forming generation
of stars, looking for any
evidence of triggered star formation both on the large and on the small scale.
Finally, we apply a method based on the $K$ luminosity function to derive
the initial mass function (IMF) of Pismis~24.

This paper is organised as follows. In Sect.~\ref{dat:red}
we describe observations and data reduction.
In Sect.~\ref{risultati},
we present large-scale, deep Spitzer/IRAC
photometry of NGC~6357, and briefly examine how YSOs are distributed in the region.
In the same section,
we then focus on Pismis~24 by using optical and infrared photometry,
complemented with X-ray high-angular resolution observations from the literature. 
We carefully address many critical issues (such as crowding effects on
the photometric completeness, contamination of cluster members by 
background stars, the reddening law, and infrared excess from
young objects), some of which are especially important for studying regions lying in the Galactic
plane and close in projection to the Galactic centre, but which have not been fully considered 
so far. 
We will also assess the limitations of the available YSO diagnostics,
and their effects on data interpretation. The IMF of Pismis~24
is derived in Sect.~\ref{imf:24}. The stellar populations in the other clusters
will be studied in 
forthcoming papers using recently obtained near-infrared (NIR) observations.
A tentative scenario for
star formation in the region is however discussed in Sect.~\ref{discussione}.
Finally, our conclusions are summarised in Sect.~\ref{conclusioni}. 

\subsection{NGC~6357 and the open cluster Pismis~24}
\label{intro:ngc6357}

\begin{figure}
   \centering
   \includegraphics[width=8cm]{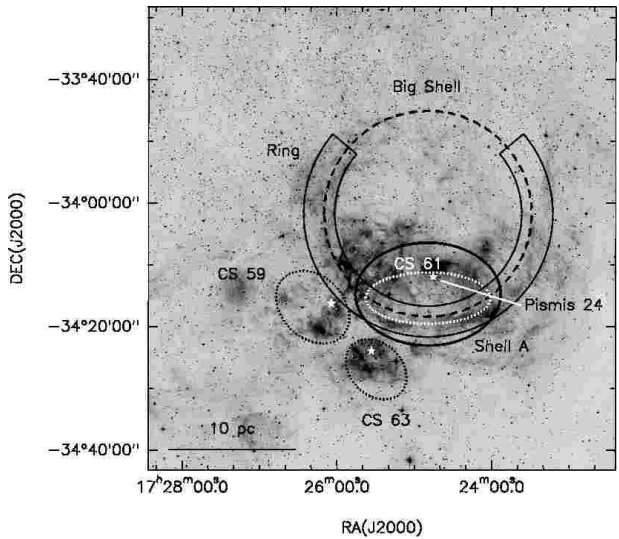}
      \caption{Large scale DSS red image (north up, east left) of the
	star forming region NGC~6357. The main structures described
	in the text are marked and labelled. The white stars mark
        the positions of known star clusters. The length scale is 
	drawn for an assumed distance of $1.7$ kpc. 
         \label{dss:red}}
   \end{figure}

NGC~6357 is a complex composed of giant molecular clouds, H\textsc{ii} regions
and open clusters, located at $l \sim 353\degr$,
$b \sim 1\degr$, at a distance of $1.7$ kpc
(see Sect.~\ref{dist&redd}). The large scale structure has been
studied in a number of papers and its most prominent 
components are highlighted in Fig.~\ref{dss:red}.
The gas distribution is outlined by a large optical shell opened to the north
(named {\em big shell} by Cappa et al.\ \cite{cappa11})
and three smaller cavities (namely \object{CWP2007 CS 59}, 
\object{CWP2007 CS 61}, and \object{CWP2007 CS 63}; Churchwell et al.\
\cite{churchwell}). Lortet et al. (\cite{lortet})
noted that the big shell is in low-excitation conditions
and ionisation-bounded, and suggested it is in fact a wind-driven bubble. 
Following Felli et al.\ (\cite{felli:90}), we will refer to the big shell
as {\it the ring}.
Three H\textsc{ii} regions are associated with the cavities:
\object{G353.2+0.9} (inside CWP2007 CS 61),
\object{G353.1+0.6} (inside CWP2007 CS 63), and \object{G353.2+0.7} 
(inside CWP2007 CS 59). 

The big shell is $\sim 60 \arcmin$
in diameter ($\sim 30$ pc at $1.7$ kpc) and is bordered by an {\em outer
shell} of giant molecular clouds with LSR velocities in the range
$-12.5$ to 0 km s$^{-1}$ (Cappa et al.\ \cite{cappa11}),
The total estimated gas mass of the outer shell amounts to $1.4 \times 10^{5}$ $M_{\sun}$. 
If the whole molecular gas structure is part of a large bubble, 
then the gas velocity range suggests an expansion rate of the order
of $\sim 10$ km s$^{-1}$. Given a radius of $\sim 15$ pc, the dynamical age
of this large shell would be $\sim 1.5$ Myr.

%
%

CS 61 is situated inside the southern half of the big shell, enclosed 
in an elliptical ring of NIR emission (see Fig.~\ref{3colori}) whose most prominent feature
is the H\textsc{ii} region G353.2+0.9 in the northern part. 
A few molecular clouds border the NIR emission, with LSR
velocities between $-7.5$ and 0 km s$^{-1}$ (Cappa et al.\ \cite{cappa11}), 
arranged in another shell structure
(shell A, labelled in Fig.~\ref{dss:red}). The total estimated mass of shell A is
$1.2 \times 10^{5}$ $M_{\sun}$. 

Pismis~24 (see Fig.~\ref{pismisJHK}) is a young open cluster inside CS 61, off-centred  
northward (as marked in Fig.~\ref{dss:red}). 
Several optical spectroscopic and photometric
observations have unveiled a number of coeval ($\sim 1$ Myr) O-type stars among its members
(Moffat \& Vogt \cite{mo:vo}, Neckel \cite{neckel}, Lortet et al.\ \cite{lortet},
Massey et al.\ \cite{massey01}, Russeil et al.\ \cite{russ:12}).
The two brightest stars were identified  
as an O3 If (\object{Cl Pismis 24 1}, a.k.a. HD~319718) and an O3 III 
(\object{Cl Pismis 24 17}) star
(Massey et al.\ \cite{massey01}), respectively. Pismis 24 1 was then resolved 
by HST imaging into two components,
Pismis 24--1SW and Pismis 24--1NE (the latter being a spectroscopic binary),
of $\la 100$ $M_{\sun}$ each (for a distance of $\sim 2.56$ kpc,
larger than that assumed in the present work;
Ma\'{i}z Apell\'{a}niz et al.\ \cite{maw}). 
We note that not necessarily these stars have already ended their main-sequence phase. Recent
simulations of a non-rotating 60 $M_{\sun}$ star have shown that this would display a supergiant
appearance (i. e., luminosity class I) already on the ZAMS
(Groh et al.\ \cite{groh}). 

Wang et al.\ (\cite{wang07}) obtained deep X-ray Chandra/ACIS observations
of Pismis~24, finding $\sim 800$ X-ray sources that cluster around
Pismis 24 1 and Pismis 24 17, mostly 
intermediate- and low-mass PMS cluster members. 
The radial distribution of X-ray source surface density
is characterised by a core $\sim 2\arcmin$ in radius superimposed on
a halo with a radius of at least $8\arcmin$. 
Optical ($RI$) photometry and spectroscopy 
(of a smaller sub-sample) with
VIMOS at the VLT (Fang et al.\ \cite{fang}) confirmed that these are PMS stars 
with a median age of 1 Myr and an age spread in the range $\la 0.1$ to
$\ga 10$ Myr. 
In addition, the IMF is consistent with that of the
Orion Nebula Cluster (\object{ONC}). Wang et al.\ (\cite{wang07})
estimated a number of members of $\sim 10000$ if the distance
is $2.56$ kpc. By combining 2MASS and Spitzer/IRAC photometry,
Fang et al.\ (\cite{fang}) derived a very low fraction 
($\sim 0.2$) of stars
with a circumstellar disk within $\sim 0.6$ pc from Pismis 24 1.
This is quite low for such a young
cluster, thus they suggested that this is observational evidence
of the effect of massive stars on the disk evolution around nearby,
less massive stars. 
The O-type stars in Pismis~24 are the ionising sources of G353.2+0.9
(Bohigas et al.\ \cite{bohi}; Giannetti et al.\ \cite{giannetti}). 
They are also 
responsible for the ionisation of the inner edge of shell A, 
originating a ring of H\textsc{ii} regions collectively referred to as \object{G353.12+0.86} by
Cappa et al.\ (\cite{cappa11}). 

Spitzer/IRAC photometry showed three large clusters of Class I and Class II sources
in NGC~6357 (Fang et al.\ \cite{fang}): one coinciding with Pismis~24, 
another coinciding with the open cluster AH03J1725--34.4 
towards CS 63 (Dias et al.\ \cite{dias},
Gvaramadze et al.\ \cite{gvaramadze}), and a third one towards CS 59. 
The most massive members of AH03J1725--34.4 are \object{N78 49}, \object{N78 50} and \object{N78 51}, 
that were classified as O5 to O9.5 (Neckel \cite{neckel}, Lortet
\cite{lortet}), although Damke et al.\ (\cite{damke})  
recently reported a new spectral
classification, namely O4III for N78~49 and O3.5V for N78~51, which suggests that 
AH03J1725--34.4
is roughly coeval with Pismis~24. N78~49 is the main ionising source of the
H\textsc{ii} region G353.1+0.6 (Massi et al.\ \cite{massi97}), 
located on the northern edge of CS 63, where a few giant molecular clouds
lie between CS 63 and CS 61 (Massi et al.\ \cite{massi97},
Cappa et al.\ \cite{cappa11}). 
Fang et al.\ (\cite{fang}) also found
two arcs of Class I and Class II sources, one encompassing CS 59 
and CS 61,
and the other in a symmetrically located position with respect to Pismis~24, which may be connected
with the cluster.   

A few authors have also suggested
that an older population of stars may exist in NGC~6357. 
Wang et al.\ (\cite{wang07}), by noting the off-centred position
of Pismis~24 inside CS 61, which might be inconsistent with a bubble 
originated by the energetic input from the stars of the cluster, and
Gvaramadze et al.\ \cite{gvaramadze}, by noting 
the age discrepancy between the very massive members of
Pismis~24 and the nearby ($\sim 4\arcmin$
from Pismis~24) older \object{WR93}. 

\subsection{Distance to Pismis~24 and extinction}
\label{dist&redd}

There is little doubt that the two most prominent H\textsc{ii} regions in NGC~6357,
(namely. G353.2+0.9 and G353.1+0.6)
are at the same distance. The gas velocity either from radio recombination lines or
from mm molecular lines does not change appreciably from one to the other
(Massi et al.\ \cite{massi97}). In particular, large-scale $^{12}$CO(1--0) emission
(Cappa et al.\ \cite{cappa11}) from the region displayed in Fig.~\ref{3colori}
confirms the visual impression that we are merely observing parts of a larger complex.
Remarkably, the nearby region \object{NGC~6334} also exhibits similar gas velocities and
stellar photometry yields a similar distance (Russeil et al.\ \cite{russ:12}).
Therefore, NGC~6357 as a whole could be part of an even larger galactic complex.

The distance to NGC~6357, and to 
Pismis~24 in particular, is still debated. As explained in Massi et al.\
(\cite{massi97}), the kinematical distance ($\sim 1$ kpc) based on the radial velocity
of the associated gas is uncertain and it is now clear
that it underestimates the actual distance. Neckel (\cite{neckel})
found $1.74 \pm 0.31$ kpc for NGC~6334 and NGC~6357 based on optical photometry
of stars. Lortet et al.\ (\cite{lortet}) found the same value from spectroscopic
observations of the most luminous stars in NGC~6357.
Based on a spectral analysis of the most massive members of Pismis~24, 
Massey et al.\ (\cite{massey01})
found a distance of $\sim 2.56$ kpc (distance modulus ${\rm DM} = 12$ mag), larger than the previous
values. 
On the other hand, all the most recent determinations point to a distance $\sim 1.7-1.8$
kpc (Fang et al.\ \cite{fang}, Russeil et al.\ \cite{russ:12}, Gvaramadze et al.\
\cite{gvaramadze}, Lima et al.\ \cite{lima}).
Reid et al.\ (\cite{reid14}) found an even lower distance to NGC~6334 
($1.34$ kpc) from trigonometric parallaxes of masers.
In the present work, we will assume a distance $1.7$ kpc (${\rm DM} = 11.15$ mag) following 
the most recent literature. 

Part of the discrepancy in the distance determination may arise due
to an anomalous reddening. 
Chini \& Kr\"{u}gel (\cite{ch:kr}) had already
noted that the region is seen through a dark cloud
and derived $R_{V} = 3.7 \pm 0.2$. 
Bohigas et al.\ (\cite{bohi}) found
$R_{V} = 3.5$, which agrees with $R_{V} = 3.53 \pm 0.08$ obtained by 
Russeil et al.\ ($\cite{russ:12}$). However,
Ma\'{i}z Apell\'{a}niz et al.\ (\cite{maw}) found $R_{V} = 2.9-3.1$ for
the two most massive stars of Pismis~24 through HST optical and 2MASS
NIR photometry.

Contrary to the distance, the average extinction towards Pismis~24 appears to be well constrained. 
Neckel (\cite{neckel}) and Lortet et al.\ (\cite{lortet})
found that the brightest members are affected by $A_{V}$ in the range 5--6 mag. Massey et al.\
(\cite{massey01}) found $E(B-V)$ between $1.6$--$1.9$ (i. e., the same $A_{V}$ range as above
if $R_{V} = 3.1$)
and a median $E(B-V) = 1.73$ for the most massive stars. Ma\'{i}z Apell\'{a}niz et al.\ (\cite{maw})
obtained $A_{V} =5.5$ mag for Pismis 24 1 and $A_{V} = 5.9$ mag for Pismis 24 17, 
using HST optical photometry (values increasing to
$A_{V} =5.87$ mag and $A_{V} =6.34$ mag, respectively, when including
2MASS photometry). Russeil et al.\ ($\cite{russ:12}$)
found $A_{V}$ in the range $5.01$--$6.39$ mag using multi-band photometry,
and values in agreement with those of Ma\'{i}z Apell\'{a}niz et al.\ (\cite{maw})
for the 2 most massive stars. 
Fang et al.\ (\cite{fang}) used optical spectroscopy and
$RI$ photometry for a large number of low-mass counterparts of X-ray sources
(hence, mostly cluster members) deriving an extinction range $3.2 < A_{V} < 7.8$.
Lima et al.\ (\cite{lima}) obtained an average $E(J - K_{s}) = 1.01$, i. e., $E(B - V) = 1.75$,
for the cluster stars.

\section{Observations and data reduction}
\label{dat:red}

\subsection{Near-Infrared imaging}
\label{nir:phot}

\begin{figure*}
   \centering
   \includegraphics[width=8cm]{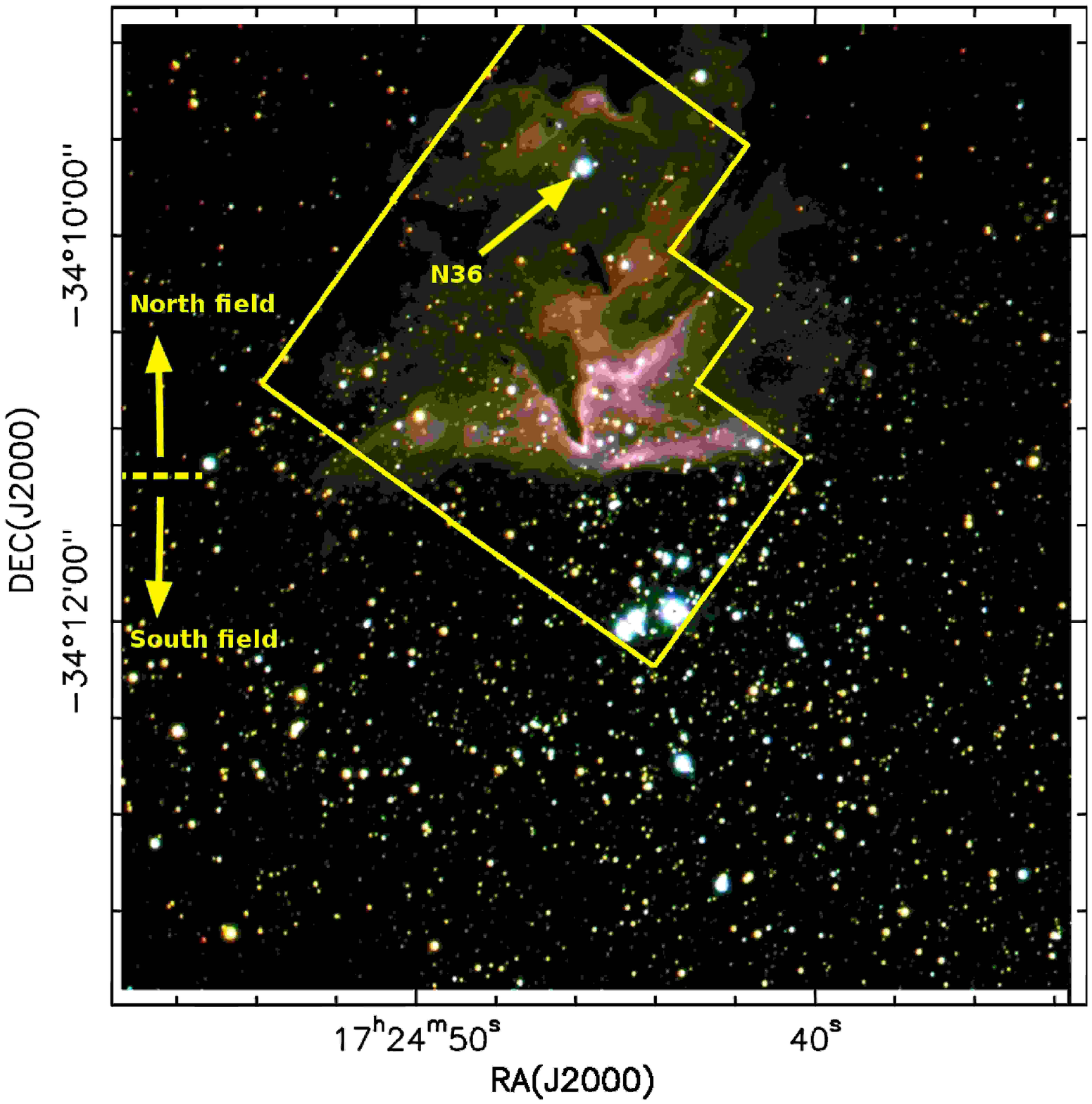}
   \includegraphics[width=8cm]{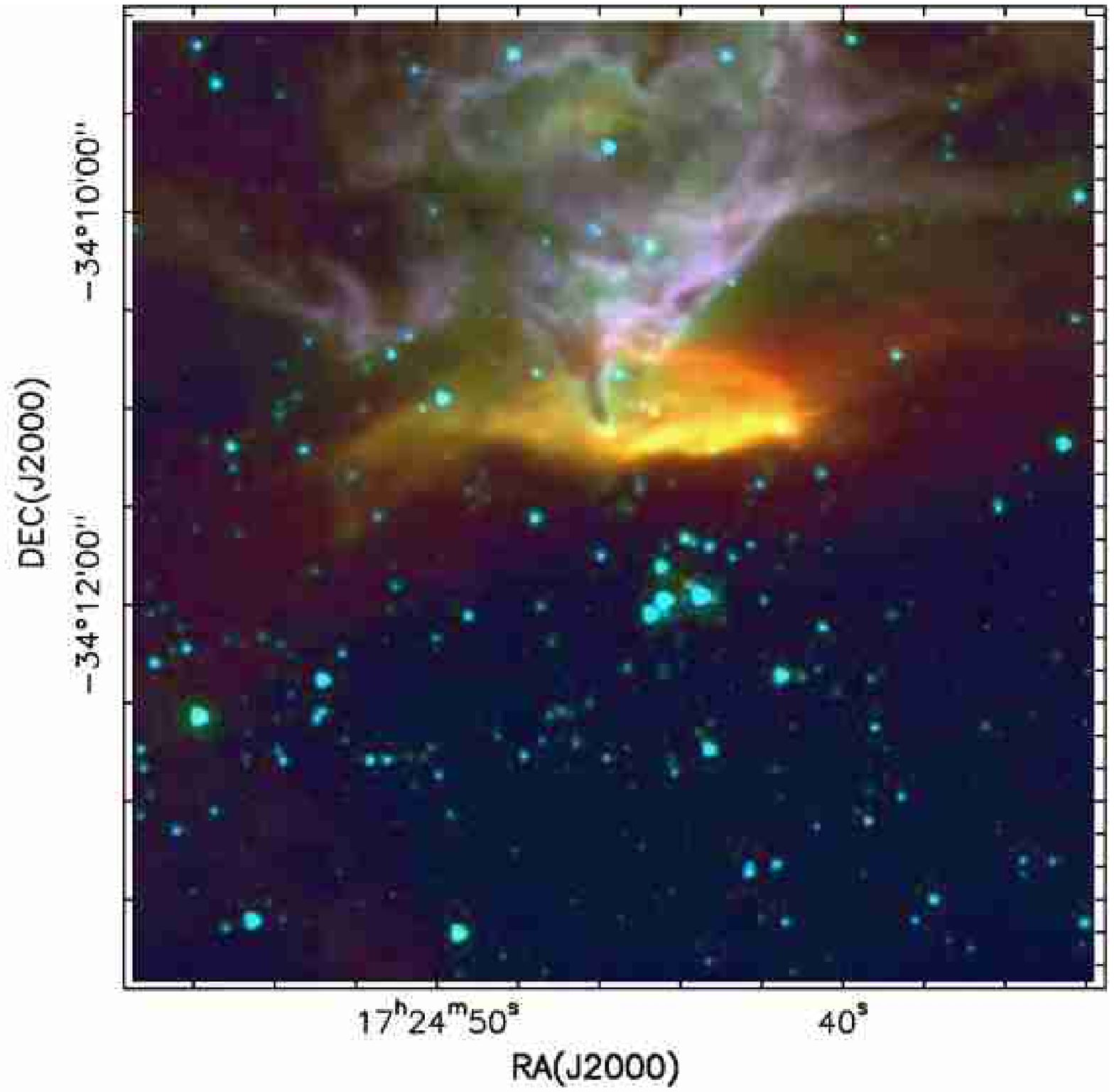}
      \caption{(left) Three-colour image
        ($J$ blue, $H$ green, $K_{s}$ red) of Pismis~24 and G353.2+0.9, obtained
        from the reduced SofI frames. Outlined, the HST/WFPC2 field.
        The border between the two sub-fields and the massive star N78~36
         are also indicated. (right) For comparison,
	three-colour image of the same area obtained by
 	combining the IRAC/Spitzer (short integration) frames 
	at $3.6$ $\mu$m (blue), $4.5$ $\mu$m (green) and
	$8.0$ $\mu$m (red).
(A colour version of this figure is available in the on-line edition.)	
         \label{pismisJHK}}
   \end{figure*}

The field towards G353.2+0.9 was imaged in the $JHK_{s}$ bands with SofI 
(Moorwood et al.\ \cite{moor}) at 
the ESO-NTT telescope (La Silla, Chile), during the night between
13 and 14 May, 1999.  The plate scale is $\sim$0.282\arcsec/pixel,
yielding an 
instantaneous field of view of $\sim 5 \times 5$ arcmin$^{2}$. For each band 
five pairs of on-source/off-source images were taken,
dithered according to a pattern of positions 
randomly selected in a box of side length $20\arcsec$. Each image consists of 
an average of 40 (80 at $K_{s}$) sub-exposures of $1.182$ s, resulting in a total
on-source integration time of $\sim 4$ min.\ ($\sim 8$ min. at $K_{s}$). 
The raw images were crosstalk corrected, flat fielded (using dome-flats), sky subtracted, 
bad-pixel corrected, registered 
and mosaicked using the special procedures developed for SofI and standard
routines in the IRAF\footnote{IRAF is distributed by
the National Optical Astronomy Observatory,
which is operated by the Associated Universities for Research in
Astronomy, Inc., under cooperative agreement with the National Science
Foundation.} package. 
The seeing in the final combined
images is $\sim$ $0\farcs9$. A three-colour image ($J$ blue, $H$ green, $K_{s}$ red)
from the reduced set of frames is shown in Fig.~\ref{pismisJHK}.
The field covered is $\sim 5\arcmin \times 5\arcmin$ in size, approximately centred at
RA(2000)$= 17^{h} 24^{m} 45.4^{s}$, DEC(2000)$= -34\degr 11\arcmin 25\arcsec$.
The $K_{s}$ image was already used by Giannetti et al.\ (\cite{giannetti}) in
their Fig.~2, where the structures visible in the image are labelled.
To calibrate the photometry, the standard star S279-F
(Persson et al.\ \cite{perss}) was observed in the same night. 

%
%

Photometry was carried out on the final images by using DAOPHOT routines in IRAF.
We selected a
$\sim 1$~(PSF-)FWHM aperture and sky annuli $\sim 2$~FWHM both in radius and width,
with the modal value as a background estimator.
From this, PSF-fit 
photometry was then performed with ALLSTAR. The results in the three bands were matched together
by using a radius of 3 pixels ($\sim 1$~FWHM). The obtained limiting magnitudes 
(at a signal to noise ratio of 3) are $J \sim 19.7$, $H \sim 19$ and $K_{s} \sim 18.5$.
In total, we found 6500 NIR sources detected at least in the $K_{s}$ band.

We compared our photometry with that from 2MASS, which is both less sensitive and less resolved
than ours. By computing $\Delta{m} = {\rm mag(SofI)}-{\rm mag(2MASS)}$ we found averages $0.05 \pm 0.32$ mag at $K_{s}$
($0.03 \pm 0.17$ mag for $K_{s} < 12$), $0.09 \pm 0.23$ mag at $H$
($0.05 \pm 0.09$ mag for $H < 12$), and $0.04 \pm 0.30$ mag at $J$ ($0.02 \pm 0.09$ mag
for $J < 12$).
The sources appear slightly fainter in the SofI photometry, as expected due to
its better angular resolution. Such large dispersions in $\Delta{m}$ 
are usually found in young, crowded stellar
fields when comparing photometry of very different sensitivity and resolution. However,
we note that $\Delta{m}$ 
exhibits a similar spread as that of the 2MASS photometric uncertainties in the same band. 
Therefore, we can conclude that 
our SofI photometry is consistent with that from 2MASS 
within errors. We exploited this to add 9 2MASS sources 
(including Pismis 24 1, Pismis 24 17, and \object{N78 36}) to our photometry list,
which are saturated in at least one of the SofI bands. 

To estimate the completeness limits, we examined the histograms of number of sources
as a function of magnitude. This was complemented with experiments of synthetic stars added to 
the images. To account for the different levels of extinction in the imaged field, this 
has been divided into two sub-fields:
a northern one (encompassing the molecular
gas region, thus more extincted) and a southern one (much less extincted,
containing Pismis~24). For the sake of simplicity, the two sub-fields have been 
separated by a line of constant declination (DEC[2000]$= -34\degr 11\arcmin 14\arcsec$,
see Fig.~\ref{pismisJHK}) 
bordering the southern edge of the structure named ``bar'' 
(see Fig.~2 of Giannetti et al.\ \cite{giannetti}). 
As expected, we found different completeness limits in the northern field, 
dominated by diffuse emission, and in the southern
one, dominated by source crowding. We were able to retrieve $\sim 80$ \% 
of the artificial stars 
at $K_{s} \sim 16.5$ in the northern field, and at $K_{s} \sim 15.8$ in the southern field.
We only carried out the test in the $K_{s}$ band, 
However, we estimate that the completeness limits in the other bands 
can be obtained by adding the following value to the $K_{s}$ completeness 
limit:
$0.5-1$ mag at $H$ and $1-1.5$ mag at $J$.

\subsection{HST/WFPC2 optical data}

We searched the HST archive for images of Pismis~24 suitable for photometry. We retrieved
WFPC2\footnote{See http://www.stsci.edu/hst/wfpc2} observations in 
the bands F547M and F814W from programme 9091 (P. I. Jeff Hester).
These filters can be easily transformed to the Johnson-Cousins $VI$ standard. Pismis~24 was observed 
on April 11, 2002. The images we used had an integration time of 500 s. To check the
transformations to the $VI$ standard, we also searched for WFPC2 images of clusters through the 
same bands
with in addition
$VI$ photometry from the ground in the literature. 
We found observations of the open cluster
\object{NGC~6611} from the same programme (9091) meeting this requirement, carried out on August 8, 2002.  
As known, the WFPC2 field of view is covered by four cameras, each of which
$800 \times 800$ pixels in size. Three of them (WFC)  are arranged in an L-shaped field and
operate at a pixel scale of $\sim 0\farcs1$, the fourth one (PC) operates at 
a pixel scale of $\sim 0.046 \arcsec$.

Removal of cosmic rays and photometry was performed using HSTphot
v1.1 (Dolphin \cite{dolphin}), a software package specifically developed for
HST/WFPC2 images. 
Quite a few stars in the two fields (i. e., Pismis~24 and NGC~6611) were 
rejected because saturated. HSTphot transformed
the photometry to the Johnson-Cousins standard by using the relations provided by
Holtzman et al.\ (\cite{holtzman}). Thus, stars without simultaneous detections in both bands 
were also discarded. Since the detection limit is $V \sim 25$, and $V - I > 4$
for most of the stars, these generally have $I < 21$.
In total, we obtained $VI$ photometry for 158 stars in Pismis~24 and 397 stars 
in NGC~6611, from all four chip fields (PC and WF). The HST/WFPC2 field of view is outlined in 
Fig.~\ref{pismisJHK}. Almost all the stars have photometric errors
$< 0.05$ in $I$ and $<0.1$ mag in $V$, although a few 
stars with $V > 24$ have photometric errors in $V$ up to $\sim 0.2$ mag.

We compared our results with the ground-based photometry in the
same bands. Unfortunately, no ground-based photometry is available for Pismis~24
in the $V$ band, although we could use that from Fang et al.\ (\cite{fang}),
obtained from VIMOS observations,
in the $I$ band. However, ground-based photometry in the $VI$ bands is
available for NGC~6611, obtained from WFI (at the 2.2 m telescope of ESO)
observations (Guarcello et al.\ \cite{guarcello}). 
Thus, we discovered an offset ($0.28 \pm 0.38$ mag) between our $V$ 
photometry of NGC~6611 and that from Guarcello et al.\ 
(\cite{guarcello}), with no apparent colour effects.
Nevertheless, a clear colour effect was found on our $I$ photometry of both clusters
(after correction, the r.m.s. of the difference in $I$ magnitudes is $\sim 0.25$ mag).
This is hardly surprising: Holtzman et al.\ (\cite{holtzman})
caution about their transformations being accurate only in the
range $-0.2 < V - I < 1.2$, whereas most of the sources towards Pismis~24 are well above
$V - I = 1$. Thus, we derived a more accurate transformation for $I$ through a linear
fit. Then, we corrected our $VI$ photometry to match the corresponding ground-based
photometry.  

\subsection{Spitzer/IRAC data}

The InfraRed Camera (IRAC, Fazio et al.\ \cite{fazioetal}) on board the
Spitzer Space Telescope, is equipped with four detectors operating at $3.6$, 
$4.5$, $5.8$ and $8.0$ $\mu$m, respectively. Each detector is composed
of $256 \times 256$ pixels with a mean pixel scale of $1\farcs22$, yielding a
field of view of $5\farcm2 \times 5\farcm2$. 
We retrieved all Spitzer/IRAC observations towards NGC~6357 from
the Spitzer archive. After a close examination of the available
data, we decided to use those from Program ID 20726 (P.I. Jeff Hester). The IRAC 
observations were
carried out in 2006, September 28, during the cryogenic mission.
The high dynamic range (HDR) mode was used, meaning that 
two images per pointing were taken,
a short exposure one ($0.4$ s) and a long
exposure one (10.4 s). This allows one to obtain unsaturated photometry of bright
sources from the short-exposure image and deeper photometry of faint sources
from the long-exposure one. 

\begin{figure*}
   \centering
   \includegraphics[width=14cm]{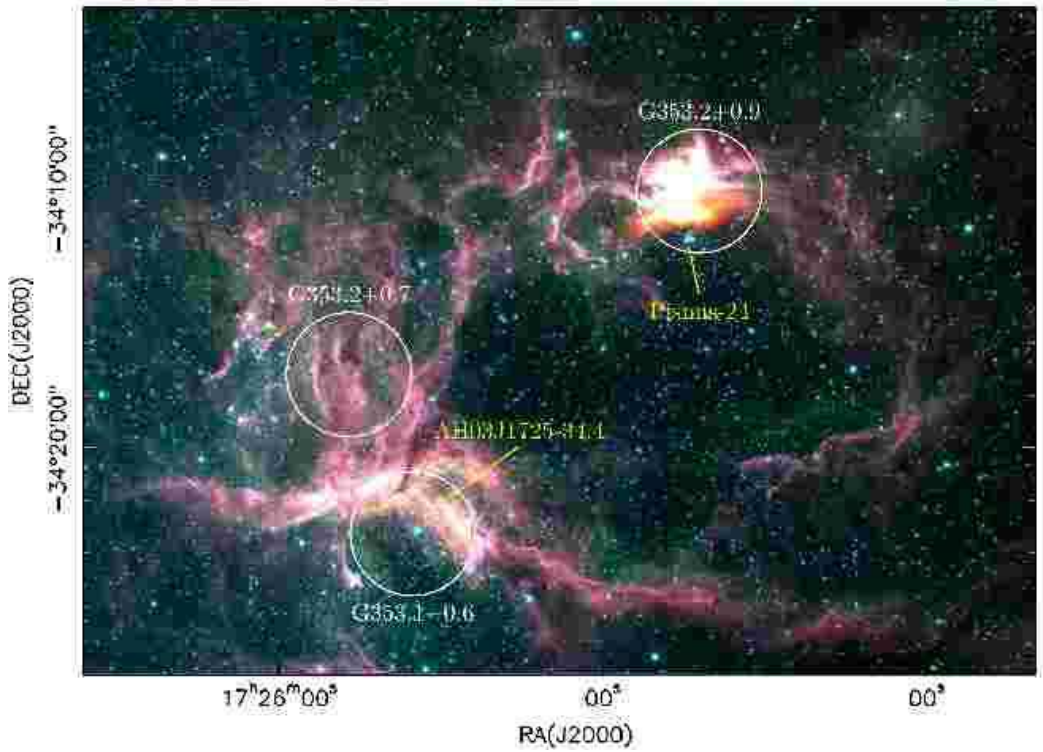}
      \caption{Three-colour image
	(3.6 $\mu$m blue, 4.5 $\mu$m green, 8.0 $\mu$m red) of NGC~6357, obtained
	from the short-exposure IRAC frames.
	The large circles mark the locations of the most
	prominent H\textsc{ii} regions. The clusters Pismis~24
	and AH03J1725--34.4 are also labelled.
(A colour version of this figure is available in the on-line edition.)
         \label{3colori}}
   \end{figure*}

Data reduction is detailed in App.~\ref{sp:ir:dr} and yielded
two mosaicked images (short-  and long-exposure)
per band, with a pixel size $0\farcs 6 \times 0\farcs 6$ (about half
the native pixel size).
We cropped each of them to a final field of view of
$\sim 38\arcmin \times 26\arcmin$ encompassing shell A and the three clusters 
in NGC~6357, fully covered at all bands.
A three-colour (3.6 $\mu$m blue, 4.5 $\mu$m green, 8.0 $\mu$m red) image
of the field, obtained from the final short-exposure IRAC frames,
is shown in Fig.~\ref{3colori}.

%
%

We performed aperture photometry on the final IRAC mosaics
(both long- and short-exposure)
by using DAOPHOT routines in IRAF. We adopted an aperture
radius of 4 pixels and a sky annulus 4 through 12 pixels from the
aperture centre (i. e., 2 and 2--6 native pixels, 
respectively), with the median value as a background estimator. These radii were chosen 
as small as possible to account for the variable diffuse emission in many 
areas of the region. We used the corresponding aperture corrections given in 
Table~4.7 of the IRAC Instrument Handbook (version 2.0.1).
Short-exposure and long-exposure photometry were
matched in each band, adding the brightest sources from the former to the faintest
sources from the latter.
The photometry is further detailed in App.~\ref{sp:ir:ph}, where we also show
that GLIMPSE II photometry and ours are consistent with each other,
and that the detection limits are similar, although 
GLIMPSE II photometry is much less deep and should consequently be less
sensitive. We attributed this to source crowding causing our photometry
to be confusion-limited
(NGC~6357 is in the Galactic plane, few degrees from the Galactic centre). 

After combining long- and short-exposure photometry, there remained 65875 sources
in the 3.6 $\mu$m band, 59160 in the 4.5 $\mu$m band, 47398 in the 5.8 $\mu$m band
and 13763 in the 8.0 $\mu$m band. We further merged the 4 lists of objects by adopting
a matching radius of 3 pixels ($1.8 \arcsec$). This is about equal to the FWHM of the Spitzer
PSF (Fazio et al.\ \cite{fazioetal}) at those wavelengths. 

Contamination due to PAH emission (and its cleaning) is discussed in App.~\ref{sp:ir:ph}.
To reduce any remaining effects from artefacts or false detections, 
in the following we will only consider sources
detected in at least the first two bands (i. e., 3.6 and 4.5 $\mu$m), 
and with photometric errors $<0.3$ mag. 
In addition, we will always discard detections with photometric errors
$\ge 0.3$ mag when adding other bands, unless otherwise stated. This means that multiple-band
detections will be discarded if the photometric error is $\ge 0.3$ in any of the bands. 
Nevertheless, we show in App.~\ref{compl:app} that the effect on source statistics is negligible.

The photometric completeness limit is highly variable, decreasing in magnitude
according to the number of bands where 
simultaneous detection is required, and depending on the position in the image. 
In addition, 
the long exposure images are mostly saturated over the  
areas of intense diffuse emission, particularly in the two upper wavelength IRAC bands (i. e., 5.8
and 8.0 $\mu$m),
and only the short exposure ones were suitable for photometry there.
The completeness limits are derived in
App.~\ref{compl:app} and listed in Table~\ref{compl:tab}.
In summary, we found that sources
detected in at least the first two bands should be almost complete down to $[3.6]=[4.5]=12.25$
in areas with at most faint diffuse emission, and $[3.6]=[4.5]=10.75$ in areas with intense
diffuse emission. 

\subsection{Matching of optical, NIR, MIR and X-ray detections}
\label{match:nmx}

We complemented the optical, NIR, and mid-infrared 
(MIR) observations with Chandra/ACIS-I X-ray observations from
Wang et al.\ (\cite{wang07}). These cover a field of view of $17\arcmin \times 17\arcmin$ 
centred at RA(2000)$= 17^{h} 24^{m} 42^{s}$, DEC(2000)$= -34\degr 12\arcmin 30\arcsec$,
with sub-arcsec angular resolution at the centre. Since we only use the detections in the
area covered by SofI, which lies at the centre of the ACIS field of view, no significant degradation
of the X-ray PSF is expected. The total integration time is $\sim 38$ ksec.

First, we merged our NIR source list with that from the X-ray observations by using IRAF routines 
and a matching radius of $\sim 1$ arcsec.
The optical, NIR-X and IRAC databases were then merged again with the same 
matching radius. We constructed a larger catalogue (without optical
data) with the sources falling in the whole SofI field (see Table~1), and
a smaller one with all the sources falling in the HST/WFPC2 field
(see Table~2).

Out of the 665 X-ray sources and 114 X-ray tentative sources found by
Wang et al.\ (\cite{wang07}), 337 sources and 52 tentative sources
fall in our SofI field. Of these, 
303 sources and 33 tentative sources match a detection in at least
the $K_{s}$ band. As for the unmatched
X-ray sources, 15 (plus 2 tentative) fall towards either of the two most
massive cluster members, where some SofI sources are saturated and others may well be
hidden inside the high-count wings of the O stars.  

Out of the 6500 $K_{s}$ sources in our larger catalogue, 
3308 were also detected at $J$ and $H$, 2322 only at
$H$, and 1660 match
a source of the IRAC catalogue.
Only 88 IRAC sources do not have
a match with a NIR source in  the Sofi field.
The smaller catalogue (HST field) contains 1140 NIR sources with at least a detection in
the $K_{s}$ band, 36 sources only detected in the X-ray, and 11 sources only detected 
in $VI$. The optical sources with a NIR counterpart are 147
(for conciseness, by NIR counterpart we will mean a source 
at least detected in the $K_{s}$ band). Of these, 62 have also
been detected in the X-ray. Finally, 29 X-ray sources have infrared counterparts but
not an optical one.
In the following sections we will always discard NIR detections with photometric errors
$> 0.3$ mag in any of the bands, as done for IRAC data, unless otherwise stated.

\section{Results}
\label{risultati}

\subsection{IRAC selection of Young Stellar Objects}
\label{irac:yso:guth}

Robust criteria to identify YSOs 
have become available in the literature based on IRAC colours.
We first removed contaminants from our catalogue of IRAC sources 
by following Gutermuth et al.\ (\cite{gutermuth}). As shown
in App.~\ref{app:cont}, the major source of contamination is PAH emission
at $5.8$ and $8.0$ $\mu$m, which can be associated with faint $3.6$ and $4.5$ $\mu$m
sources located in areas with strong diffuse emission. 
We removed these PAH contaminants from the list of all 
IRAC sources detected in at least the first three IRAC bands.

Then, we used the criteria of Gutermuth et al.\ (\cite{gutermuth}) to also remove
PAH galaxies, broad-line AGNs, and unresolved knots of shock emission from the
sources detected in all 4 bands. 
Since only shock emission can be identified based on the first three bands,
(all other contaminants also requiring a measurement at $8.0$ $\mu$m), extragalactic 
sources could not be filtered out 
of the sample of objects only detected in the first three bands. However, extragalactic
contaminants are probably less of a problem since we are observing through the galactic plane.
The very high reddening behind NGC~6357 should efficiently extinct most of the
background galaxies to below our detection limits.

More critical contamination may arise due to evolved background stars (i. e., AGB stars)
whose infrared colours can mimic those of YSOs (Robitaille et al.\ \cite{robi:agb}).
But these, unlike YSOs, should exhibit a more homogeneous
spatial distribution inside our field like any other type of background stars,
although still patchy due to extinction variations.

Finally, after contaminant cleaning we identified the Class I and Class II sources 
following the colour criteria of Gutermuth et al.\ (\cite{gutermuth}).
We show in App.~\ref{compl:app} and Sect.~\ref{sdIRACs}
(see also Table~\ref{mass:compl:tab}) that only the brightest and most massive Class II sources 
can be simultaneously
detected in all IRAC bands, particularly in the areas with intense diffuse emission.
The sample can be enlarged to include less massive young stars by 
requiring detections in the first three IRAC bands only, 
but the stellar mass corresponding to the completeness 
limit still remains quite high towards the areas of intense diffuse emission. 
Altogether, we found 50 Class I sources and 482 Class II sources out of
4560 sources with detections in all bands, and a further 14 Class I sources and
729 Class II sources out of 9114 sources with detections in the three lower-wavelength
bands only.
Their colours are shown in Fig.~\ref{fig:col-col}.
The number of YSOs classified from their colours in all four bands can be compared
with those found by Fang et al.\ (\cite{fang}) in a similar area, who used IRAC photometry 
in all four bands with a sensitivity comparable to that of our photometry.
They retrieved 64 Class I/flat sources and 244 Class II sources.
While the numbers of Class I sources found are consistent,
we found twice as many Class II sources as they did. This is not only due to the
different colour classification criteria adopted: by their criteria
(and using their cuts in photometric errors as well), we still found 438 YSOs. It is
likely that the much longer integration of our images yields more accurate measurements of faint
objects allowing more sources to get through the photometric error cuts. 

\subsection{Spatial (large-scale) distribution of IRAC sources}
\label{sdIRACs}

The completeness limits listed in Table~\ref{compl:tab} can be
converted into {\em mass} completeness limits if distance, mean age and average
extinction of the stellar population are known. 
We will assume that Pismis~24 is roughly representative of the 
stellar populations in most of the clusters of NGC~6357, which is confirmed by the results of 
Getman et al.\ (\cite{getman}) and Lima et al.\ (\cite{lima}).
Thus, we adopt a distance modulus of $11.15$ mag, and
an extinction $A_{V} \sim 5.5$ mag 
(see Sect.~\ref{dist&redd}).
Using the reddening laws of Rieke \& Lebofski (\cite{r&l}) and
Indebetouw et al.\ (\cite{indebetouw}), this translates into
$A_{3.6} \sim 0.34$, mag and  $A_{4.5} \sim 0.26$ mag.
We also assume that the IRAC $3.6$
$\mu$m band measurements are consistent with $L$-band measurements, so that
stellar masses can be derived from theoretical $L$
values. The latter were obtained by using the PMS evolutionary tracks 
(1 Myr old) of Palla \& Stahler (\cite{ps99}) and Siess et al.\ (\cite{siess}),
complemented with the colours of Kenyon \& Hartman (\cite{ke:ha}). 
In the evolutionary models of Palla \& Stahler (\cite{ps99}),
their birthline coincides with the zero age main sequence (ZAMS) for stars $\ga 8-10$ $M_{\sun}$, 
so we also used the colours 
of Koornneef (\cite{Koorn}), and the spectral types from 
Habets \& Heinze (\cite{h&h}) to convert the brightest [3.6]
values into masses. 
The estimated mass completeness limits are listed in Table~\ref{mass:compl:tab}.

 \addtocounter{table}{2}
\begin{table*}
\caption{Estimated [3.6] completeness limits and corresponding
stellar masses for given extinction values.
\label{mass:compl:tab}}
\centering                          
{\footnotesize
\begin{tabular}{ l l  l l l l l l l l l l}        
\hline\hline                 
Image area & \multicolumn{3}{c}{good detections$^{b}$ in} &
\multicolumn{3}{c}{good detections$^{b}$ in} & \multicolumn{4}{c}{good detections$^{b}$ in} \\
   & \multicolumn{3}{c}{4 bands} & \multicolumn{3}{c}{3 adjacent lowest bands} &
\multicolumn{4}{c}{2 adjacent lowest bands} \\
\hline                        
 &       & Naked & Class 
 &       & Naked & Class 
 &       & Naked & Class & Naked & Class \\
 &       & PMS & II 
 &       & PMS & II 
 &       & PMS & II & PMS  & II \\
 & [3.6] & $M$ & $M$ 
 & [3.6] & $M$ & $M$ 
 & [3.6] & $M$ & $M$ & $M$ & $M$ \\
 & (mag) & ($M_{\sun}$) & ($M_{\sun}$) 
 & (mag) & ($M_{\sun}$) & ($M_{\sun}$) 
 & (mag) & ($M_{\sun}$) & ($M_{\sun}$) & ($M_{\sun}$) & ($M_{\sun}$) \\ 
 & & \multicolumn{2}{c}{$A_{V}=5.5$ mag} 
 & & \multicolumn{2}{c}{$A_{V}=5.5$ mag} 
 & & \multicolumn{2}{c}{$A_{V}=5.5$ mag} & \multicolumn{2}{c}{$A_{V}=10-20$ mag} \\ 
\hline                        
Faint diffuse emission & $10.25$ & $6-10$ & $2.5$ 
        & $11.75$ & $3$ & $1.5$ & $12.25$ & $2$ & $1$ & -- & --- \\
Intense diffuse emission$^{a}$ & -- & -- & -- & -- 
        & -- & -- & $10.75$ & -- & -- & $\sim 6-10$ & $\sim 2.5-3$ \\
\hline                                   
\end{tabular}
}
\vspace*{1mm} {\footnotesize $^{a}$~Completeness limits unreliable when requiring good detections
   in more than two bands (see App.~\ref{compl:app}). $^{b}$ a detection with photometric
   error $< 0.3$ mag is referred to as a good detection}
\end{table*}

Towards areas with intense diffuse emission
(e. g., G353.2+0.9 north of Pismis~24), 
coinciding with local molecular clouds, 
both the reduced sensitivity and the higher
extinction must also be taken into account. Table~\ref{mass:compl:tab} lists
mass completeness limits, as well, for $A_{V}=10-20$ mag. 
Clearly, these areas may suffer from heavy incompleteness.

Due to their infrared excess, the mass completeness limit is bound to
be lower for Class II sources. One can obtain very crude estimates
of this as follows. First, we assume a spectral index 
${\rm d}\log \lambda F_{\lambda}/{\rm d}\log \lambda = -1$
(representative of Class II sources). 
Then, we derive
the flux in the $J$ band from the completeness [3.6] value and the
spectral index. Since the flux in the $J$ band is less affected
by the excess, we use the $J$-band magnitudes obtained and the same evolutionary tracks
as above to derive the corresponding stellar masses. 
The values Yielded 
are listed in Table~\ref{mass:compl:tab}, as well. For Class I sources,
estimates are much more difficult to make. As an example, the model in Whitney et al.\ 
(\cite{whitney}) of a Class I source with a $0.5$ $M_{\sun}$ central star
is always fainter than our faintest completeness limit (i. e., $[3.6] = 12.25$), whereas the one in
Whitney et al.\ (\cite{whitney2}) with $M_{*} = 0.5$ $M_{\sun}$ and $T_{*} = 4000$ K
would be brighter unless it is seen with the disk edge-on.

To map the surface density of young stars in the whole NGC~6357 region,
we counted the IRAC sources in squares 1 arcmin in size, displaced by 0.5 arcmin 
from each other both
in right ascension and in declination (i. e., a Nyquist sampling interval). 
We included all sources with simultaneous detection in at
least the 3.6 and 4.5 $\mu$m bands 
up to $[3.6]=12.25$. As shown in Table~\ref{mass:compl:tab},
this limit should allow us to retrieve
most of the young stars down to 2--3 $M_{\sun}$ outside the areas 
of the image with intense diffuse emission (e. g., away from G353.2+0.9 
and G353.1+0.6). 
We also tried a 2D binned kernel density estimate as done in Sect.~\ref{redde:law},
which gave similar results although yielding a slightly smoother distribution.
In this case, the routine dpik returns $33\arcsec - 46\arcsec$ as optimal bandwidths,
which justifies our choice of a 1-arcmin sampling size.

\begin{figure}
   \centering
   \includegraphics[width=8cm]{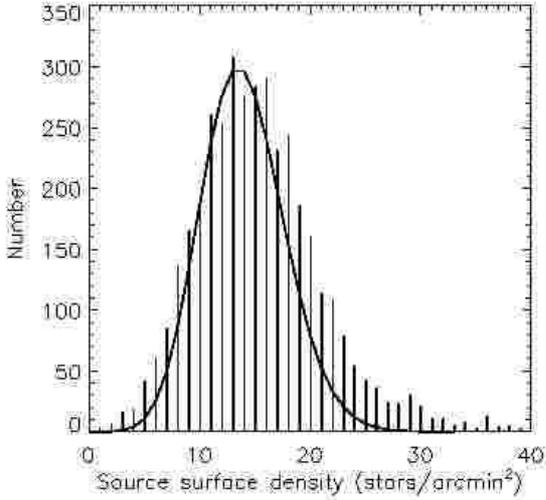}
      \caption{
Histogram of the statistics of the surface distribution of IRAC sources
       detected at least in the two lower-wavelength bands,
      counted in squares 1 arcmin in size, displaced each other by 0.5 arcmin 
       in right ascension and/or declination (i. e., a Nyquist sampling interval).
	Overlaid (full line), a Poisson curve with mean $=14$
        stars/arcmin${^2}$.
         \label{stat:hist}}
   \end{figure}

The area imaged is large enough to enable us to derive a meaningful average for the
source surface density of background stars. The count statistics is shown
as a histogram in Fig.~\ref{stat:hist}, obtained by sorting all 1-arcmin size squares.
It resembles a Poisson statistics but exhibits an excess towards high counts,
produced by non-random clustering.
The peak is roughly fitted by a Poisson curve with mean equal to 14 stars/arcmin${^2}$,
also shown in Fig.~\ref{stat:hist},
that we assume to be the mean surface density of field stars ($\sim 14 \pm 4$ 
stars/arcmin${^2}$).  

%
%

\begin{figure*}
   \centering
   \includegraphics[width=18cm]{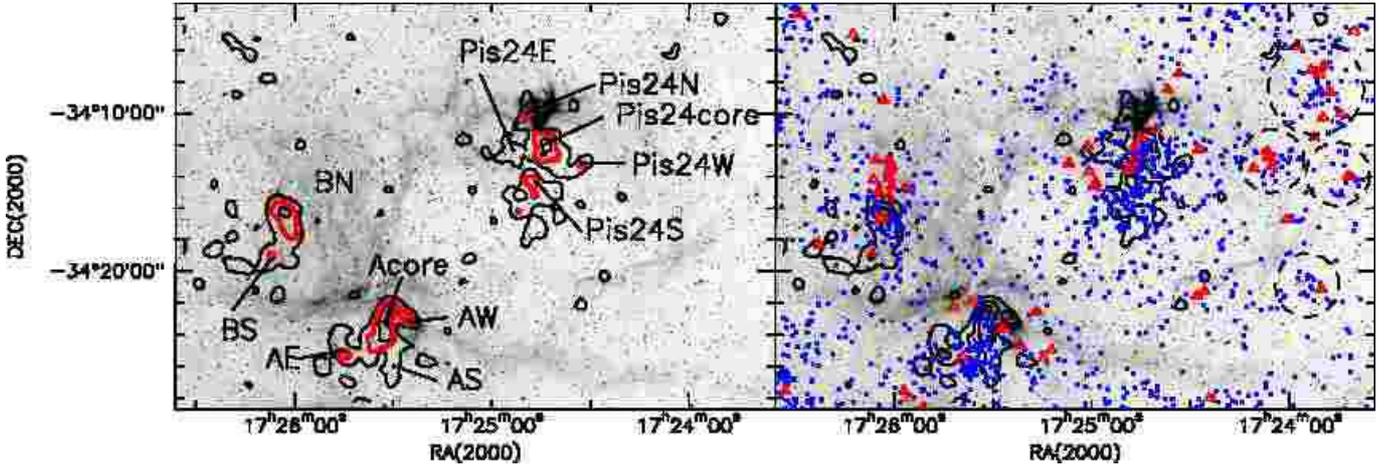}
      \caption{(left) Contour map of the surface density of IRAC
	sources, computed as explained in the text. 
	Only sources detected in at least the first two Spitzer/IRAC
        bands, with photometric errors $< 0.3$ mag,
	and up to $[3.6]=12.25$ are considered. The contours
	are: 25 stars/arcmin${^2}$, 36 stars/arcmin${^2}$
	 (red line), 47 stars/arcmin${^2}$, ranging in steps
         of $3 \sigma$ (11 stars/arcmin${^2}$) from the estimated 
         average surface density (14 stars/arcmin${^2}$) of field stars
         plus $3 \sigma$, overlaid on the image at $3.6$ $\mu$m (grey-scale). 
	Also labelled, the tentatively identified sub-clusters
	(A is also known as AH03J1725--34.4).
	(right) Same as left, but with the positions of identified
        Class II sources (full blue squares) and Class I sources
 	(open red triangles) superimposed. Other YSO concentrations
        are enclosed in dashed-line circles.
(A colour version of this figure is available in the on-line edition.)
         \label{map:irac}}
   \end{figure*}

The surface density
of IRAC sources obtained is shown in Fig.~\ref{map:irac}. The lowest contour
is equal to the mean field star density (14 stars/arcmin${^2}$) plus a $3\sigma$ 
of the fitted Poisson distribution (11 stars/arcmin${^2}$), 
and the contour step is $3\sigma$. The three larger clusters
(Pismis~24, AH03J1725--34.4, hereafter A, 
and the one roughly located towards G353.2+0.7,
hereafter B)
stand out well above a $3\sigma$ fluctuation of the field star statistics.
They are associated with the three cavities identified by Churchwell et al
(\cite{churchwell}; see Sect.~\ref{intro:ngc6357} for details),
close in projection to high density gas
clumps  where H\textsc{ii} regions (G353.2+0.9, G353.1+0.6, and
G353.2+0.7) are produced by ionisation from their massive members.
Thus, the three clusters are situated towards the brightest parts of 
(or outside) the 
ring-like structure (i.\ e., in the north and east)
surrounding CS 61, as evident in Fig.~\ref{3colori}

Each cluster appears further subdivided in several sub-clusters at
the 1 arcmin resolution scale. We provide a tentative list of their properties
in Table~\ref{irac:clust:tab}, including approximate positions based on those of the 
local peaks of surface density. Number of sources and size are computed with
respect to the $3 \sigma$ level above the average field star surface density.

%
%

We named ``cores'' the maxima of surface density towards the three clusters.
As shown in Fig.~\ref{map:irac} and Table~\ref{irac:clust:tab},
both Pismis~24 and AH03J1725--34.4 stretch
to the edge of the bright diffuse emission areas (where most of the molecular gas
is also located). 
Checking the effects of contamination, we found that by removing from the counts all sources with 
detections in 3 and 4 bands identified as contaminants, lower
surface densities are obtained towards the areas with diffuse emission. Consequently, none
of the contours in the modified map cross these regions, and sub-clusters
Pis24 E, Pis24 N, BS, and AW almost disappear. Both contamination and decreased sensitivity
make it difficult to identify clustering towards these bright areas. 
However, the remaining sub-clusters can still be retrieved in the modified map 
suggesting they are real local structures. This is further confirmed by the distribution
of the YSOs, i. e., all Class I and Class II sources
identified using at least the first three IRAC bands (shown
in Fig.~\ref{map:irac}, as well). 
In fact, Kuhn et al.\ (\cite{kuhn}), using the MYStIX database, essentially retrieved 
our subclusters Pis24 core (their NGC 6357 A), Pis24 S (NGC 6357 B), 
Acore (NGC 6357 C),
AS (NGC 6357 D), and AE (NGC 6357 E). In addition, Lima et al.\ (\cite{lima}), using
VVV NIR photometry, also noted our subcluster Pis24 W (VVV CL164), but suggested 
that it is a much older cluster ($\sim 5$ Gyr) unrelated with NGC 6357. Out of the clusters
studied by Lima et al.\ (\cite{lima}), our Acore and AS roughly coincide with their BSD 101 and 
ESO 392-SC 11, respectively.

%
%
\begin{table*}
\caption{Main parameters of the clusters of IRAC sources. Surface
   densities are calculated for a distance of $1.7$ kpc.
\label{irac:clust:tab}}
\centering                          
\begin{tabular}{ c c c c c c c c c c}        
\hline\hline                 
Designation & \multicolumn{6}{c}{Approx.\ peak position} & Number & Mean surface & Size \\
            & \multicolumn{3}{c}{RA(2000)} & \multicolumn{3}{c}{DEC(2000)} &
                                   of sources ($^{c}$) & density &   \\
        & ($h$) & ($m$) & ($s$) & ($\degr$) & ($'$) & ($''$) &
             & (pc$^{-2}$) & (arcmin$\times$arcmin) \\
\hline                        
Pis24 core & 17 & 24 & 43 & $-34$ & 12 & 05 & $120 \pm 16$ & $98$ & $2 \times 2$ \\
Pis24N$^{a}$     & 17 & 24 & 49 & $-34$ & 10 & 19 & $37 \pm 10$ & $76$ & $0.5 \times 2$ \\
Pis24S     & 17 & 24 & 48 & $-34$ & 14 & 17 & $137 \pm 19$ & $70$ & $3 \times 4$\\
Pis24E$^{a}$     & 17 & 24 & 54 & $-34$ & 12 & 28 & $42 \pm 11$ & $57$ & $2 \times 1$ \\
Pis24W$^{b}$     & 17 & 24 & 32 & $-34$ & 13 & 16 & $23 \pm 7$ & $94$ & $1 \times 1$ \\
Pis24 (global) & & & & & & & $359 \pm 30$ & $77$ & $6 \times 10$ \\
Acore & 17 & 25 & 33 & $-34$ & 23 & 41 & $134 \pm 17$ & $91$ & $2 \times 4$ \\
AS    & 17 & 25 & 31 & $-34$ & 25 & 57 & $38 \pm 10$ & $78$ & $1 \times 2$ \\
AE    & 17 & 25 & 45 & $-34$ & 25 & 22 & $55 \pm 12$ & $75$ & $1 \times 2$ \\
AW$^{a}$    & 17 & 25 & 25 & $-34$ & 22 & 51 & $35 \pm 10$ & $72$ & $1 \times 1$ \\
A (global) & & & & & & & $262 \pm 25$ & $82$ & $5 \times 5$ \\
BN    & 17 & 26 & 03 & $-34$ & 16 & 21 & $119 \pm 17$ & $81$ & $2 \times 3$ \\
BS$^{a}$    & 17 & 26 & 07 & $-34$ & 19 & 08 & $78 \pm 15$ & $64$ & $2 \times 1$ \\
B (global) & & & & & & & $197 \pm 23$ & $73$ & $2 \times 5$ \\
\hline                                   
\end{tabular}

\vspace*{1mm} $^{a}$~identification less certain; $^{b}$~coinciding with VVV CL164
of Lima et al.\ (\cite{lima}), possibly a much older cluster unrelated with NGC 6357;
$^{c}$~the estimated average
              field star surface density (by the surface area) has been subtracted  
\end{table*}
%
%

Clearly, the YSOs concentrate towards 
the three clusters. An underlying diffuse population of Class II sources
is also visible, although this could be 
contaminated by background evolved stars displaying the same colours.
The arc-like distributions of YSOs claimed by Fang et al.\ (\cite{fang}),
symmetrical to the centre of the largest bubble, 
show up south-east of Pismis~24
(roughly overlapping both our cluster A and our cluster B) and north-west of it (four circled areas
in Fig.~\ref{map:irac}). 
The Class I sources appear to avoid the largest peaks of IRAC source surface
density in the area; rather they tend to be found towards the molecular clouds 
(compare with Fig.~\ref{map:irac});
for example, east of Pis24 core, north of Acore, and north of BN.
The four encircled areas showing small YSO concentrations also
lie towards molecular gas clumps. Only few Class I and Class II sources
are found towards the H\textsc{ii} regions G353.2+0.9 and G353.1+0.6, but this is 
probably due to the incompleteness. We can therefore conclude that star
formation appears to be in progress in the molecular gas associated with the 
clusters.

%
%

\subsection{NIR stellar population of Pismis~24 and reddening law}
\label{redde:law}

\begin{figure}
   \centering
   \includegraphics[width=9cm, height=12cm]{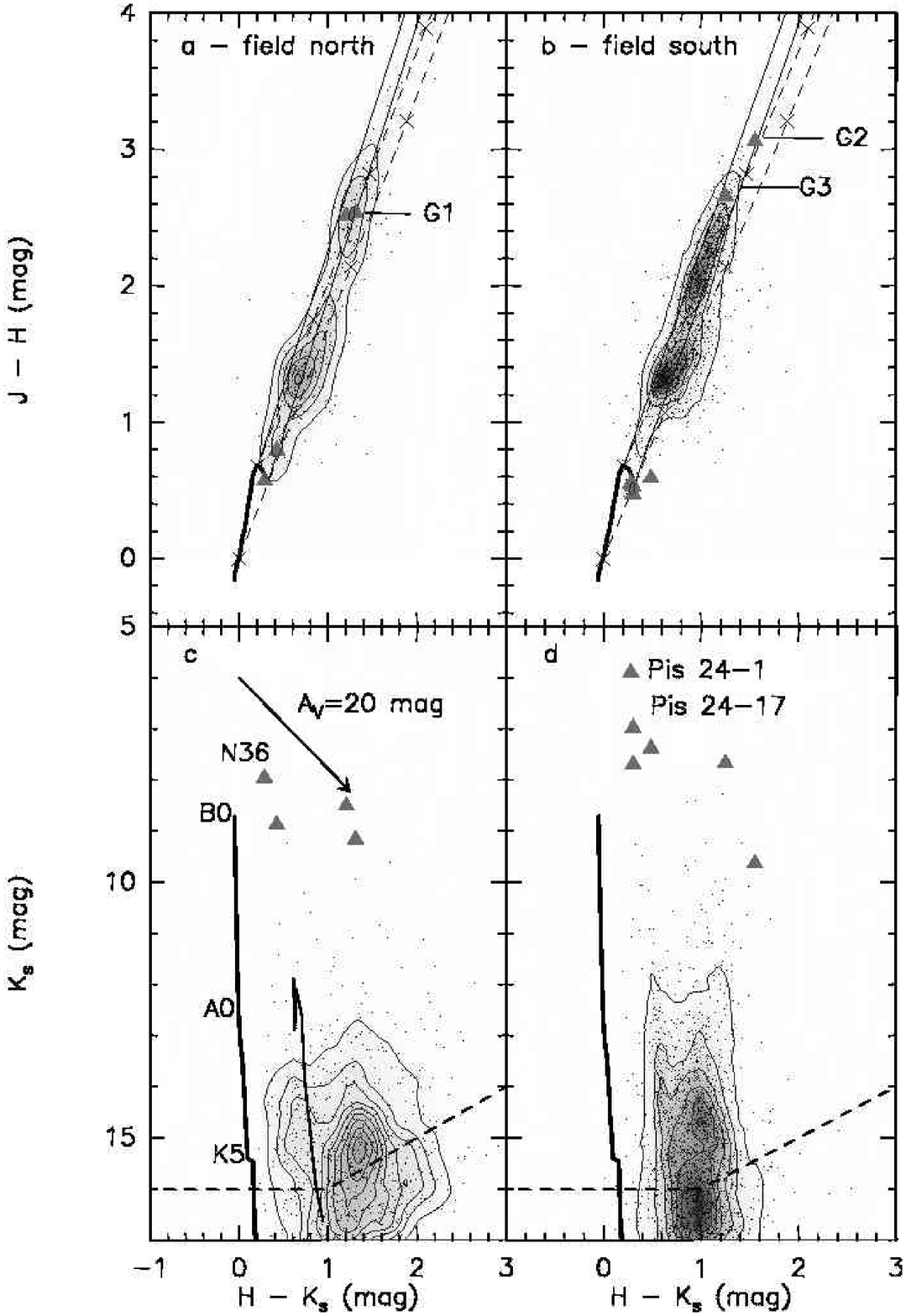}
      \caption{{\bf (a)} SofI $J-H$ vs.\ $H-K_{s}$ diagram for the northern field.
	{\bf (b)} SofI $J-H$ vs.\ $H-K_{s}$ diagram for the southern field.
        {\bf (c)} SofI $K_{s}$ vs.\ $H-K_{s}$ diagram for the northern field.
       {\bf (d)} SofI $K_{s}$ vs.\ $H-K_{s}$ diagram for the southern field.
	Only sources with $K_{s} < 16$ and photometric errors $< 0.3$ mag
	have been selected for the colour-colour diagrams, whereas
        all sources with photometric errors $< 0.3$ mag in $H$ and $K_{s}$
        have been selected for the colour-magnitude diagrams. 
        The datapoint density after smoothing is shown with contours in all panels.
        The grey triangles mark the sources whose colours have been taken from 2MASS. 
        A few of these sources, associated with Pismis~24, are also indicated,
        although those labelled G1, G2, and G3 are the
        candidate background giant stars discussed in the text.  
        The main sequence locus is drawn in all panels with a thick solid line
	(using the colours from Koornneef \cite{Koorn}, complemented with
         the absolute magnitudes from Allen \cite{all76} for panels c,d). 
        Spectral types are labelled next to the ZAMS in panel c.
	The dashed lines in panels a,b are reddening paths with crosses
	every $A_{V} = 10$ mag (from $A_{V} = 0$ mag), according to Rieke \& Lebofsky (\cite{r&l}), 
	while the thin solid lines are the reddening paths according to 
	the extinction law derived by Straizys \& Laugalys (\cite{StraLau}). 
        The dashed line in panels c,d marks the completeness limit.
       The arrow in panel c spans a reddening of $A_{V} = 20$ mag.
       Also shown as a full line parallel to the ZAMS in panel c is the sequence of PMS stars 1 Myr old
       from the evolutionary tracks of Palla \& Stahler (\cite{ps99}), for
       $A_{V} = 10$ mag, and masses in the $0.1-6$
       $M_{\sun}$ range. A distance of $1.7$ kpc is assumed for ZAMS and isochrones. 
         \label{nir:ccd}}
   \end{figure}

The members of Pismis~24 can be clearly isolated in the
NIR colour-colour diagrams (CCD) 
shown in Fig.~\ref{nir:ccd}a,b of the sources found with SofI.
Only sources with $K_{s} < 16$ 
(roughly the completeness limit in the southern field, as discussed
in Sect.~\ref{nir:phot}) 
have been selected. We have overplotted both the main
sequence locus (thick solid lines, Koornneef \cite{Koorn}) and the reddening
band of the main sequence (dashed lines) according to the extinction law derived 
by Rieke \& Lebofsky (\cite{r&l}). The latter usually fits well the
reddening of NIR sources whose photometry is obtained in the SofI filters 
(e. g., Massi et al.\ \cite{massi06}). 
 
A smoothing of the datapoints (contours in Fig.~\ref{nir:ccd}) unveils
two different stellar populations in the CCD. To this purpose, we computed a
2D binned kernel density estimate using the routine bkde2D from the
library KernSmooth in the package R
(R core team \cite{erre}). The bandwidth was estimated using the routine dpik, which is based 
on a direct plug-in methodology (see Wand \& Jones \cite{w&j}). 
Two different
datapoint concentrations can be distinguished along the main sequence reddening band, roughly
separated by $H - K_{s} = 1$ and $J - H =2$. The less reddened concentration 
is roughly centred at $A_{V} \sim 5$ mag from the main sequence and extends
up to $A_{V} \sim 10-15$ mag, i. e., it lies in the reddening interval of the cluster
(see Sect.~\ref{dist&redd}). 
The two features are easily recognised in the magnitude-colour
diagrams ($K_{s}$ vs.\ $H-K_{s}$, hereafter CMD) of Fig.~\ref{nir:ccd}c,d.
The smoothed distribution using a 2D binned kernel density estimate is also shown.
Here, the less reddened concentration appears as a branch
extending from the massive stars Pismis 24 1 and Pismis 24 17 down to the
completeness limit, displaying a larger colour spread in the northern field.

The two concentrations are clearly separated in the CCD by a sort of gap, which
is more evident in the northern field. This suggests that the gap is caused by the molecular
gas associated with NGC~6357 and allows one to discriminate between background
stars (the more reddened concentration) and a nearby population
of mostly cluster members (the less
reddened concentration). The foreground extinction can be easily estimated
from the distance between the ZAMS and the outer envelope of sources
facing the ZAMS in
Fig.~\ref{nir:ccd}d, ranging from $A_{V} \sim 5.7$ mag to $A_{V} \sim 7.6$ mag.
This is consistent with the average reddening measured for Pismis~24 (e. g., Massey et al.\
\cite{massey01}, Fang et al.\ \cite{fang}).

Interestingly, the more reddened datapoint concentration extends above the extincted main
sequence band in the CCD. This is also evident using less deep 2MASS photometry 
and the extinction law of Indebetouw et al.\ (\cite{indebetouw}), which is based on 
large-scale 2MASS photometry. A possible explanation is that the extinction
law in the area has a higher $E(H-K_{s})/E(J-H)$ ratio than $1.72$,
derived by Rieke \&
Lebofsky (\cite{r&l}). In fact, as shown in Fig.~\ref{nir:ccd}a,b (thin solid lines),
the steeper law found by  Straizys \& Laugalys (\cite{StraLau}),
with $E(H-K_{s})/E(J-H) = 2$, would make the main sequence reddening
band encompass the more reddened source concentration. This would also
imply a larger fraction of sources with a NIR excess (YSOs) associated with the cluster.
Given the many claims of an anomalous $R_{V}$ in the region
(see Sect.~\ref{dist&redd}), we have also tried the extinction laws
tabulated in Fitzpatrick (\cite{fitz}) as a function of $R_{V}$. 
However, our conclusion is that there is no need for an extinction
law significantly steeper than that of Rieke \&
Lebofsky (\cite{r&l}), which we conservatively adopt
in this work. The shift of the redder clump is mostly due to a predominance of
background giant and supergiant stars, whose locus lies
slightly above that of the main sequence in the CCD, whereas main sequence
stars become too faint to be detected and this somewhat depopulates the main sequence reddening band
(e. g., a G0 V star would be fainter than $K_{s} = 16$ for $A_{V} = 15$ mag even
at the distance of Pismis~24). 
This is confirmed by
three stars which are bright enough to be saturated, despite being
very reddened ($A_{V} \sim 20$ mag). They are labelled
in Fig.~\ref{nir:ccd} as G1, G2, and G3, among the sources whose NIR magnitudes have
been taken from 2MASS.
We checked that they lie far from the brightest stars of Pismis~24,
two of them being at opposite sides in the images. Their reddening identifies them as
background stars, and their intense brightness points towards them being giant or
supergiant stars. 
As such, their correct location in the CCD is above the extincted main sequence band,
which is more consistent with the extinction law by Rieke \& Lebofsky 
(\cite{r&l}).

%
%

%
%

Figure~\ref{nir:ccd}c,d also show how deep our observations are. The completeness 
limit $K_{s} = 16$ can be converted into a mass limit by using the adopted 
distance modulus $11.15$ mag and assuming
a PMS stellar population 1 Myr old. From the evolutionary tracks of Palla 
\& Stahler (\cite{ps99}), the corresponding mass completeness limit is $0.2$ $M_{\sun}$
for $A_{V} = 10$ mag (appropriate for Pismis~24). We showed in Sect.~\ref{nir:phot}
that the completeness limit in the northern field
is $K_{s} = 16.5$; by adopting this value we find masses of 
$0.4$ $M_{\sun}$ for
$A_{V} = 20$ mag (appropriate for young stars embedded in the molecular clouds).
In a scenario of sequential star formation, one may expect 
a population of younger
stars in the molecular gas than the (triggering) cluster members, 
hence intrinsically brighter
and less massive at the completeness limit. On the other hand, if the
region were farther away then the mass at the completeness limit would be
accordingly larger (e. g., at $\sim 2.56$ kpc, $K_{s} = 16$ and $A_{V} = 10$ mag
would correspond to a $\sim 0.4$ $M_{\sun}$ star).

\subsection{X-ray source population of Pismis~24}
\label{SDXS}

It has long been known that X-ray surveys are very efficient in
revealing populations of weak-line T Tauri stars (or Class III sources),
given that these PMS stars are much brighter in X-rays than their main sequence 
counterparts (see e. g., Feigelson \& Montmerle \cite{fm:99}).
Classical T Tauri stars (or Class II sources) are also strong
X-ray emitters, but these can be efficiently recognised based on their
infrared excess, as well. So, the distribution of X-ray sources in
star forming regions usually reflects that of PMS stars with and without a
prominent circumstellar disk. 

\begin{figure}
   \centering
   \includegraphics[width=8cm]{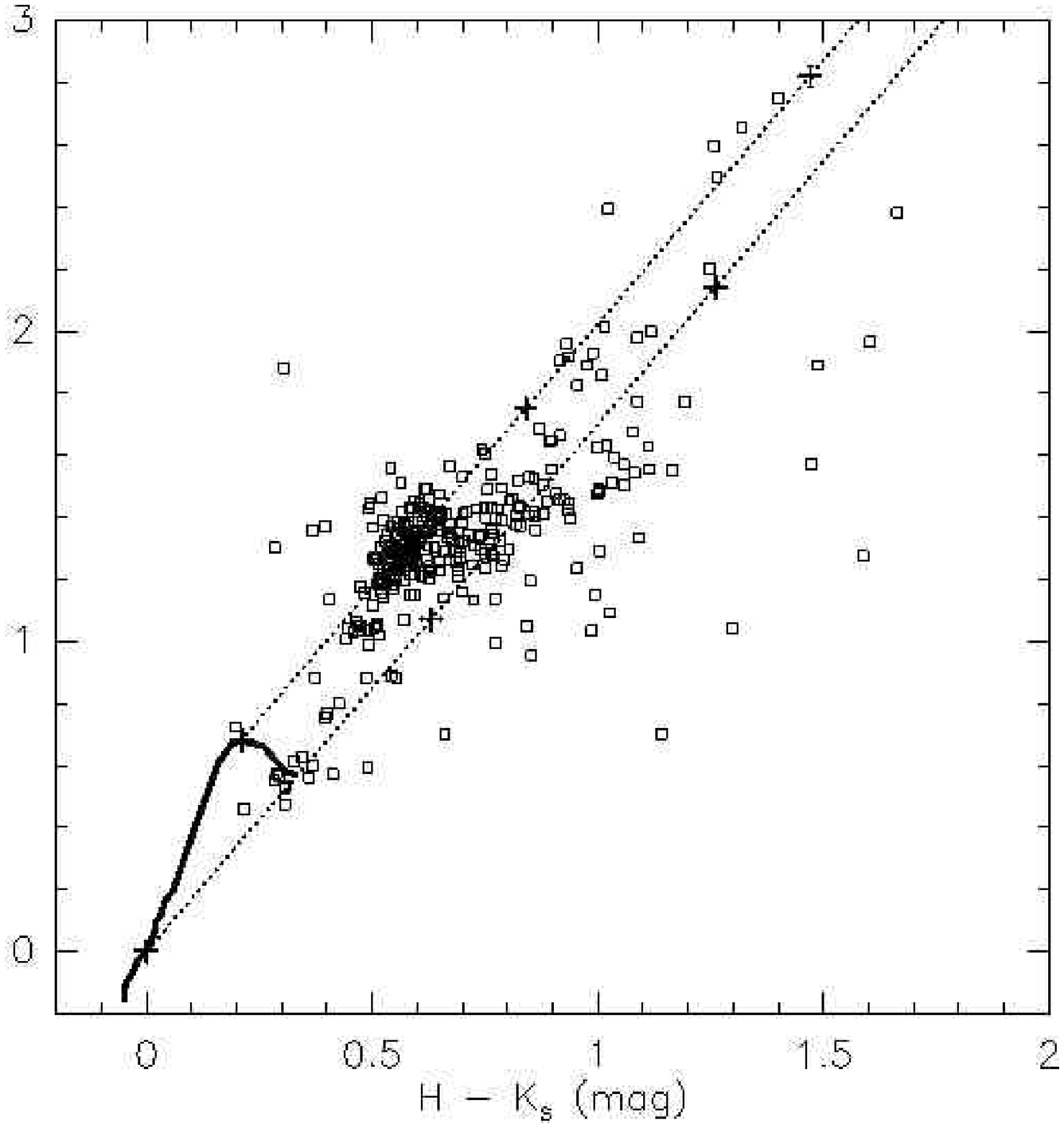}
      \caption{SofI $J-H$ vs.\ $H-K_{s}$ diagram for the 
	NIR counterparts of the X-ray sources detected towards Pismis~24. 
        Only sources with photometric errors $< 0.3$ mag
        have been selected. The thick solid line is the main sequence
        locus (using the colours from Koornneef \cite{Koorn}).
        The dotted lines are reddening paths with crosses
        every $A_{V} = 10$ mag, following Rieke \& Lebofsky (\cite{r&l}).
         \label{xray:nir}}
   \end{figure}

The high spatial resolution
X-ray survey of NGC~6357 by Wang et al.\ (\cite{wang07}) provides
a clear view of how Class II/III sources are distributed inside Pismis~24.
The $JHK_{s}$ colours of the NIR counterparts of the X-ray
sources (see Sect.~\ref{match:nmx}) can be used to check their nature as PMS stars. 
The $J - H$ vs.\ $H - K_{s}$ diagram in Fig.~\ref{xray:nir} confirms this and
is consistent with Fig.~6 of Wang et al.\ (\cite{wang07}),
although we retrieve many more NIR counterparts from our deeper SofI
photometry. In particular, 278 (out of 303) $K_{s}$ counterparts
(plus 25 out of 33 $K_{s}$ counterparts of tentative X-ray sources) were also 
detected in the $J$ and $H$ bands with photometric errors $< 0.3$ mag.
Most datapoints populate the reddened main sequence region
(which occurs for Class III source, as well), between 
$A_{V} \sim 5$ and 10 mag, and some extend below it, exhibiting
a NIR excess. This is consistent with a population of PMS stars.
We note that the reddening law of Rieke \& Lebofsky (\cite{r&l})
fully accounts for the colours of this population, further supporting 
its adoption (Sect.~\ref{redde:law}).

To estimate the fraction of X-ray emitting sources with a NIR excess,
we counted all datapoints more than 1$\sigma$
(where $\sigma$ is the photometric error of each source)
below the line defined by the reddening path from a main sequence M8
star (according to the colours from Koornneef \cite{Koorn}). Such MS stars
(at the distance of the cluster) are too faint to be detected in our
SofI image, so most of the objects below this line must be young stars with a colour
excess.  Out of the 303 X-ray emitting sources with detections
in $JHK_{s}$, 47 ($\sim 15$ \%) exhibit
a colour excess and can be classified as Class II sources. 
However, we will show in Sect.~\ref{age:24} that the fraction
of NIR counterparts with a colour excess is actually higher than
derived from the colour-colour diagram of Fig.~\ref{xray:nir}.
Wang et al.\ (\cite{wang07}) estimate that the contamination from extragalactic
sources is less than 2-4 \% and that from foreground stars less than 1-2 \%.
We can therefore confirm that the X-ray sources from Wang et al.\ (\cite{wang07}) 
falling in our SofI field mostly represent a population of Class II/Class III sources belonging
to the cluster.

%
%

The sensitivity of the X-ray observations can also be estimated following 
Wang et al.\ (\cite{wang07}). They quote an on-axis detection limit of
3 counts ($0.5-8$ keV) and derived a corresponding absorption corrected luminosities in
the $0.5-8$ keV band $\log(L_{X}/[{\rm erg\, s^{-1}}]) \sim 30.2$ at $d = 2.56$ kpc,
which scales to $\log(L_{X}/[{\rm erg\, s^{-1}}]) \sim 29.9$ at $1.7$ kpc.
These can be converted into stellar masses using the empirical
relationship found by Preibisch \& Zinnecker (\cite{p&z}) for a $\sim 1$ Myr old
PMS star population, obtaining $M \sim 0.7 M_{\sun}$ (at $1.7$ kpc).
We note that from the best fit of Flaccomio et al.\ (\cite{flaccomio}) to CTTSs we obtained similar
results. This is what one can expect in the southern field.
To assess the effects of reddening in the northern field, we repeated their
calculations using PIMMS\footnote{http://asc.harvard.edu/toolkit/pimms.jsp}, but 
assuming an extinction $A_{V} = 20$ mag. In this case,
we obtained $\log(L_{X}/[{\rm erg\, s^{-1}}]) \sim 30.6$ and $M \sim 1.8 M_{\sun}$ ($d = 1.7$ kpc).

The completeness of the NIR counterparts of the X-ray sources has been estimated in App.~\ref{compl:nir-irac-x}
to be at $K_{s} \sim 13$ {\em in the southern field}. Assuming a distance of $1.7$ kpc 
and adopting the evolutionary tracks of Palla \& Stahler (\cite{ps99}) for
1 Myr old PMS stars, this corresponds to a mass completeness limit of
$\sim 2 M_{\sun}$ for $A_{V}$ between 5--10 mag, which is consistent with what was found by  
Wang et al.\ (\cite{wang07}).

\subsection{Spatial distribution of young stars and YSOs towards Pismis~24}
\label{SDNIRS}

\begin{figure}
   \centering
   \includegraphics[width=8cm]{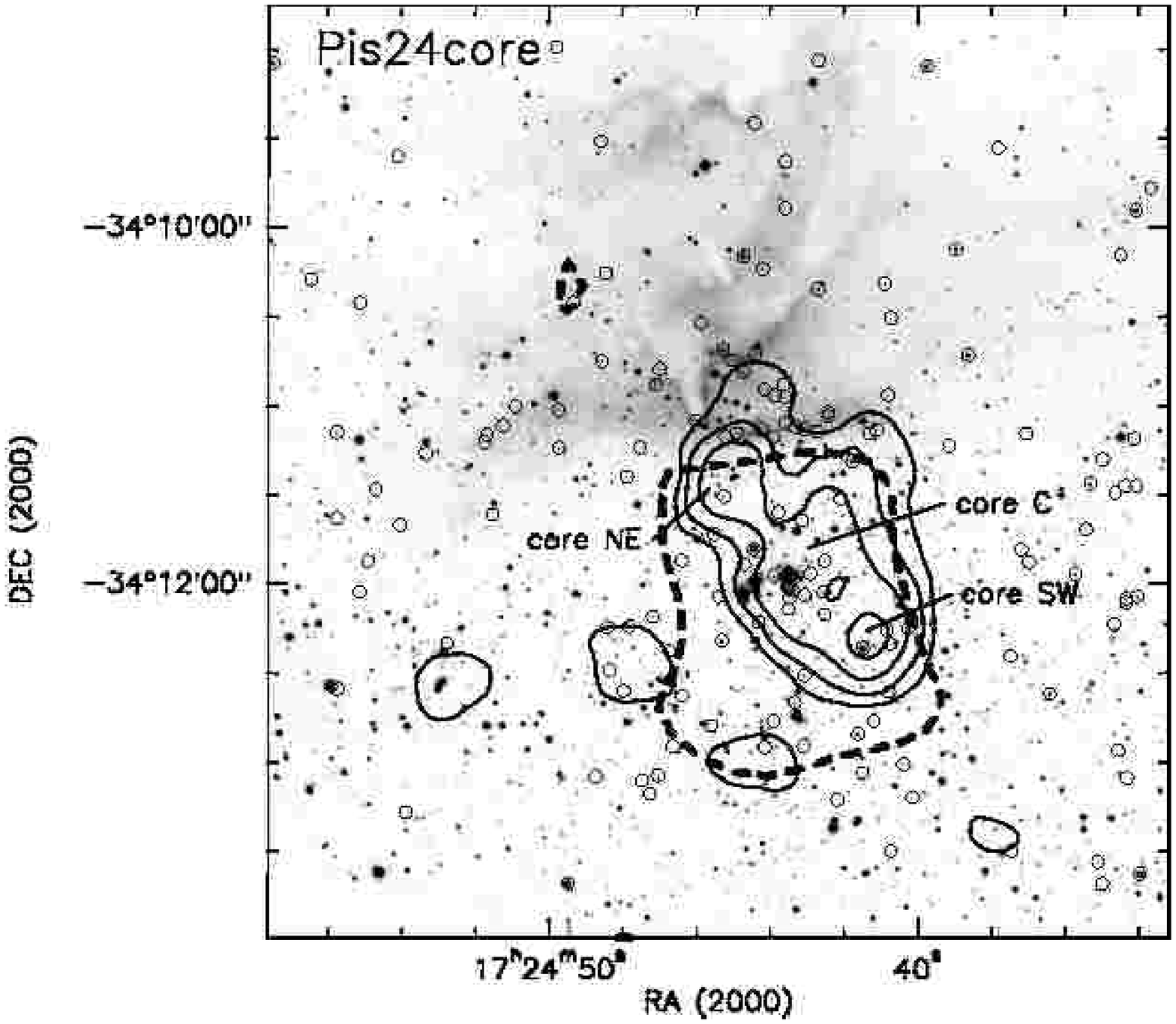}
      \caption{Contour map of the surface density of NIR sources from the
        SofI data, overlaid
	with the $K_{s}$ image of Pismis~24 and G353.2+0.9. Contours range
        from 240 stars arcmin$^{-2}$ (the average for field stars
        plus 2$\sigma$, using the values found in the southern field) 
        in steps of 30 stars arcmin$^{-2}$ ($\sim 1 \sigma$).
        Also overlaid:
        (dashed contour) the surface density of IRAC 
        sources (level 36 stars arcmin$^{-2}$ of Fig.~\ref{map:irac})
        delineating Pis24 core; (open circles) the
	location of sources with a NIR excess (see the text for
 	the selection criteria). All data used in figure have 
    	$K_{s} < 16$ and photometric errors $<0.3$ mag either in $K_{\rm s}$
         (surface density) or in all $JHK_{\rm s}$ bands (NIR excess). 
         \label{nir:map}}
   \end{figure}

To derive the surface density of NIR sources from the SofI data
towards Pismis~24 and G353.2+0.9,
we followed the same method as in Sect.~\ref{sdIRACs}. We used squares 
of side $\sim 29\arcsec$ (100 pixels), i. e., half the size
of the squares used for the IRAC sources, 
and counted sources up to $K_{s} = 16$ over the whole $K_{s}$ image. 
To estimate the average surface density
of field stars, we constructed the histograms of number of squares as a function of counts
per square
separately for the northern and the southern fields. In both cases, we obtained 
quasi-Possonian distributions with an excess frequency of high counts. 
There is actually a slightly higher frequency of lower counts,
as well,
particularly in the northern field, which is probably caused by
varying extinction. The average field star surface density was estimated from the peaks 
of the distributions and is $\sim 180$ stars arcmin$^{-2}$ in the southern field, and
$\sim 80$ stars arcmin$^{-2}$ in the northern one. The standard deviation in the southern field
is $\sim 30$ stars arcmin$^{-2}$. In Fig.~\ref{nir:map}, a contour map of the surface density
of NIR sources is overlaid on the $K_{s}$ image, starting from 240 stars arcmin$^{-2}$
(i. e., the average of field stars plus $2\sigma$, using the values found in
the southern field). Clearly, we retrieve the
Pis24 core found from the IRAC data and discussed in Sect.~\ref{sdIRACs}, although it appears
slightly smaller in $K_{s}$. This is not unexpected, given that the IRAC field is much
larger and hence allows a better determination of the average surface density of field stars.
Conversely, even the outer parts of the SofI field do probably include cluster members
and all bins are therefore biased. However, this will mostly affect the statistics of
cluster members rather than the peak locations.
Heavy reddening could distort the surface density of NIR sources, as well. This
may well occur north of the bar (i. e., in the northern field) where most of the associated
molecular gas is distributed (Massi et al.\ \cite{massi97}, Giannetti et al.\ \cite{giannetti}),
but the southern field is expected to be much less affected and the most prominent
surface density peaks are expected to outline real structures there.

The core of Pismis~24 appears elongated in a NE-SW direction, with three smaller sub-clusters:
a central one (core C) including the massive stars Pismis 24 1 
and Pismis 24 17 
(although they are a bit
off-centre, there is a local peak of density towards Pismis 24 1, not shown in the figure), 
a small compact concentration of stars between core C and the elephant trunk (core NE),
and the highest peak of surface density, which lies in the south-west (core SW). 
Lima et al.\ (\cite{lima}) suggest that core NE (their VVV CL 164) is a sub-cluster
of Pismis~24, as well. They also indicate a further sub-cluster
(their VVV CL 166), roughly located towards one of the regions in our image
with counts $2\sigma$ above the average field star density.

The sources with a NIR excess are marked by small circles in Fig.~\ref{nir:map}.
These have been selected as all sources with $K_{s} < 16$ 
lying in the colour-colour
diagram on the right of the reddening line passing through the colours of main sequence 
late M stars (roughly coincident with the lowest of the pair of lines 
drawn in Fig.~\ref{nir:ccd}a,b as dashed lines).
These NIR-excess objects are distributed all over the field, 
but their number decreases in the northern field and they tend to
concentrate towards both the Pis24 core and the H\textsc{ii} region.
Background AGB stars can mimic the colours of YSOs and may then contaminate the field
(Robitaille et al.\ \cite{robi:agb}), but they are not expected to
exhibit strong clustering, although they would as well tend to avoid the 
heavily extincted areas. Most of the sources with a NIR excess then
represent real YSOs.
 
%
%

We estimated the number of cluster members in the core by counting the sources (up to
$K_{s} = 16$) within the contour corresponding to 240 stars arcmin$^{-2}$. 
We subdivided this area in three parts roughly coinciding with the
sub-clusters. The results
are listed in Table~\ref{pis24:clust:tab}. The average density of field stars has been
subtracted and the errors given are obtained by just propagating the Poissonian ones.
The total number of members is larger than that found by counting
IRAC sources (see Table~\ref{irac:clust:tab}).
Although this is obvious, due to the differences in the mass completeness limits
and resolutions between the two datasets, we would have expected to retrieve much more 
core members from the $K_{s}$ image. This will be discussed in Sect.~\ref{star:pop}.

%
%
\begin{table*}
\caption{Main parameters of the core of Pismis~24 (Pis24 core). 
\label{pis24:clust:tab}}
\centering                          
\begin{tabular}{ c c c c c c c c c }        
\hline\hline                 
Designation & \multicolumn{6}{c}{Approx.\ peak position} & Peak & Number \\
            & \multicolumn{3}{c}{RA(2000)} & \multicolumn{3}{c}{DEC(2000)} &
                                   surface density$^{a}$ & of cluster members$^{b}$\\
        & ($h$) & ($m$) & ($s$) & (${\degr}$) & ($'$) & ($''$) & (pc$^{-2}$) & \\
\hline                        
core NE & 17 & 24 & $45.0$ & $-34$ & 11 & 24 & 565 & $34 \pm  9$ \\ 
core C & 17 & 24 & $43.4$ & $-34$ & 12 & 00 & 565 & $116 \pm 19$ \\ 
core SW & 17 & 24 & $41.3$ & $-34$ & 12 & 17 & 835 & $56 \pm 12$ \\ 
core (global) & & & & & & & & $290 \pm 24$ \\ 
\hline                                   
\end{tabular}

\vspace*{1mm} $^{a}$~the average field star surface density of $\sim 180$ stars/arcmin$^{2}$
   has been subtracted. $^{b}$~the global number is derived inside the $2\sigma$ 
   contour from the background, whereas a higher contour is used for the 
   sub-clusters. 
\end{table*}
%
%

More specifically, various samples of different
indicators can be used to derive the spatial distribution of YSOs
in Pismis~24. In fact, given the nature of the region and the many
connected effects already discussed, and the different completeness
limits in the various bands, different indicators trace different
types of sources with inhomogeneous results. 
For example, as discussed in Sect.~\ref{SDXS}, X-ray sources 
trace T Tauri stars, 
whereas in principle, we can use the IRAC colours
to discriminate Class I and Class II sources following
Gutermuth et al.\ (\cite{gutermuth}).

NIR photometry can be used to identify stars with colour excess typical of
circumstellar disks. For this purpose we used the
criterion discussed in Sect.~\ref{SDXS}, based on the 
$J - H$ vs.\ $H - K_{s}$ diagram.  Our SofI photometry is much deeper than the IRAC one,
but it has been known that stars with circumstellar disks may not exhibit a colour
excess in $JHK_{s}$ CCDs (Haisch et al.\  \cite{haisch}) and
a fraction of YSOs is likely to be missed.
To make the most out of NIR and IRAC photometry, 
we combined the NIR fluxes  
with that in the IRAC 4.5 $\mu$m band. The first two IRAC bands 
are the most sensitive of the four 
and are much less affected by problems of saturation than the two 
longer wavelength bands.
We adopted the criteria of Winston et al.\ (\cite{wins07}) to identify
YSOs by using $J - H$ vs.\ $H - [4.5]$ or, when $J$ is not available,
$H - K_{s}$ vs.\ $K_{s} - [4.5]$ diagrams (not shown). These will be
referred to as sources with a $JH(HK_{s})[4.5]$ excess. 

We checked that the YSOs identified either through IRAC colours alone
or through combined IRAC-$JHK_{s}$ colours fall in the expected regions
of a $J - H$ vs.\ $H - K_{s}$ diagram, i. e., either in the 
band of the reddened main sequence or to the right of it. 
As a whole, we identified in the SofI field: 5 Class I
sources and 99 Class II sources (from IRAC colours only), 514 sources
with a $JH(HK_{s})[4.5]$ excess, and 390 sources with a $JHK_{s}$ excess.

The completeness degree of each of the four indicator samples obtained 
is assessed in App.~\ref{compl:nir-irac-x}. We show not only that it is 
different in the southern and northern fields, as expected, but
that it is also very different depending on the chosen indicator.
This might bias any conclusion if not taken into account properly. 
All things considered, X-ray sources seems to be the less biased tracer of YSOs,
displaying a fair trade-off between sensitivity and background contamination.

\begin{figure*}
   \centering
   \includegraphics[width=14cm]{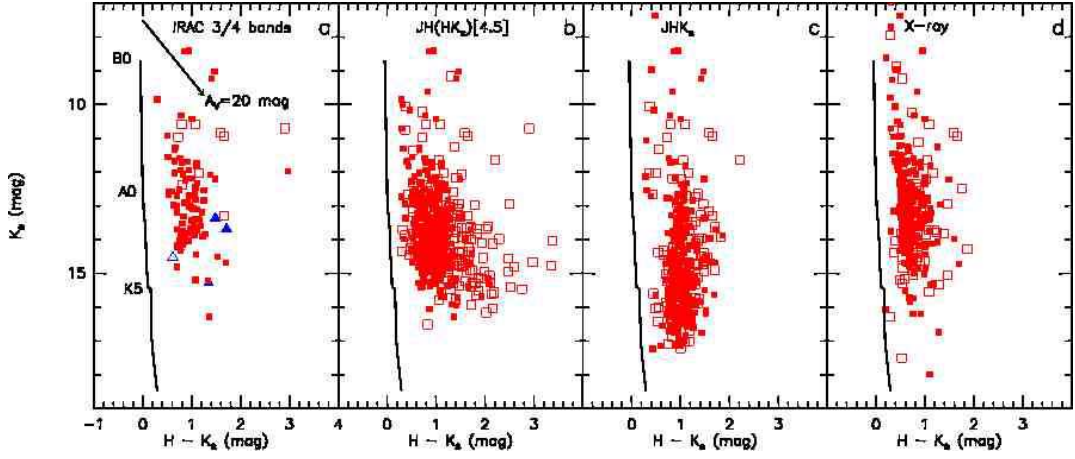}
        \caption{ $K_{s}$ vs.\ $H - K_{s}$ for the NIR counterparts of the YSOs
	towards Pismis~24
	identified according to four different indicators, as explained in
        the text. Open symbols always refer to sources in the northern field, whereas
	full symbols refer to sources in the southern field.
	{\bf (a)} YSOs identified on the basis of their IRAC colours (triangles:
	Class I; squares: Class II); {\bf (b)} YSOs identified on
	the basis of their combined $JH(HK_{s})[4.5]$ colours; {\bf (c)} YSOs
        identified on the basis of their $JHK_{s}$ colours;
        {\bf (d)} NIR counterparts of X-ray sources.
	The thick solid vertical line marks the main sequence
        locus (using absolute magnitudes from Allen \cite{all76} and
        colours from Koornneef \cite{Koorn}) for a distance modulus of $11.15$ mag
        ($d = 1.7$ kpc).
	The arrow in {\bf a} shows a reddening of
        $A_{V} = 20$ mag according to Rieke \& Lebofsky (\cite{r&l}).
	Spectral types are labelled on the ZAMS.
(A colour version of this figure is available in the on-line edition.)
         \label{yso:mc:d}}
   \end{figure*}

The different properties of the four indicator samples used
can be demonstrated by plotting their $HK_{s}$ counterparts
in a $K_{s}$ vs.\ $H - K_{s}$ diagram (Fig.~\ref{yso:mc:d}).
The first thing to note is that 
some sources spread to redder colours in the northern field.
Then, we note that the various samples exhibit different
sensitivity limits at $K_{s}$.
In particular, the IRAC photometry misses 
a large fraction of sources, especially the fainter ones.
On the other hand, the SofI photometry yields the largest number of faint
sources, but misses a fraction of the reddest sources compared to
the combined $JH(HK_{s})[4.5]$ selection, due to the extinction
affecting the $J$ band. 
Most of the X-ray selected sources have $H - K_{s} \la 1$, whereas
both $JHK_{s}$- and $JH(HK_{s})[4.5]$-selected YSOs spread to $H - K_{s} > 1$.
We note that 100 out of the 303 X-ray sources with a $JHK_{s}$
counterpart in the SofI field (see Sect.~\ref{match:nmx}) exhibit
an excess in the $JH(HK_{s})[4.5]$ CCDs,
whereas we only retrieved 47 X-ray emitting sources with a
NIR excess through the $JHK_{s}$ CCD.

By comparing $VI$ and $JHK_{s}$ photometry as well
(see Sect.~\ref{age:24}), one finds that the number of X-ray 
sources displaying a NIR excess must be higher than that.
By plotting the position of the $JH(HK_{s})[4.5]$-selected
YSOs with $H - K_{s} > 1.2$ (i. e., the colours of the background 
group of sources in the CCDs diagrams of Fig.~\ref{nir:ccd}) one also finds
they are anti-correlated both with the
cluster and the H\textsc{ii} region, confirming they
are mostly background evolved stars with a NIR excess. 
 
%
%

\begin{figure*}
   \centering
   \includegraphics[width=12cm]{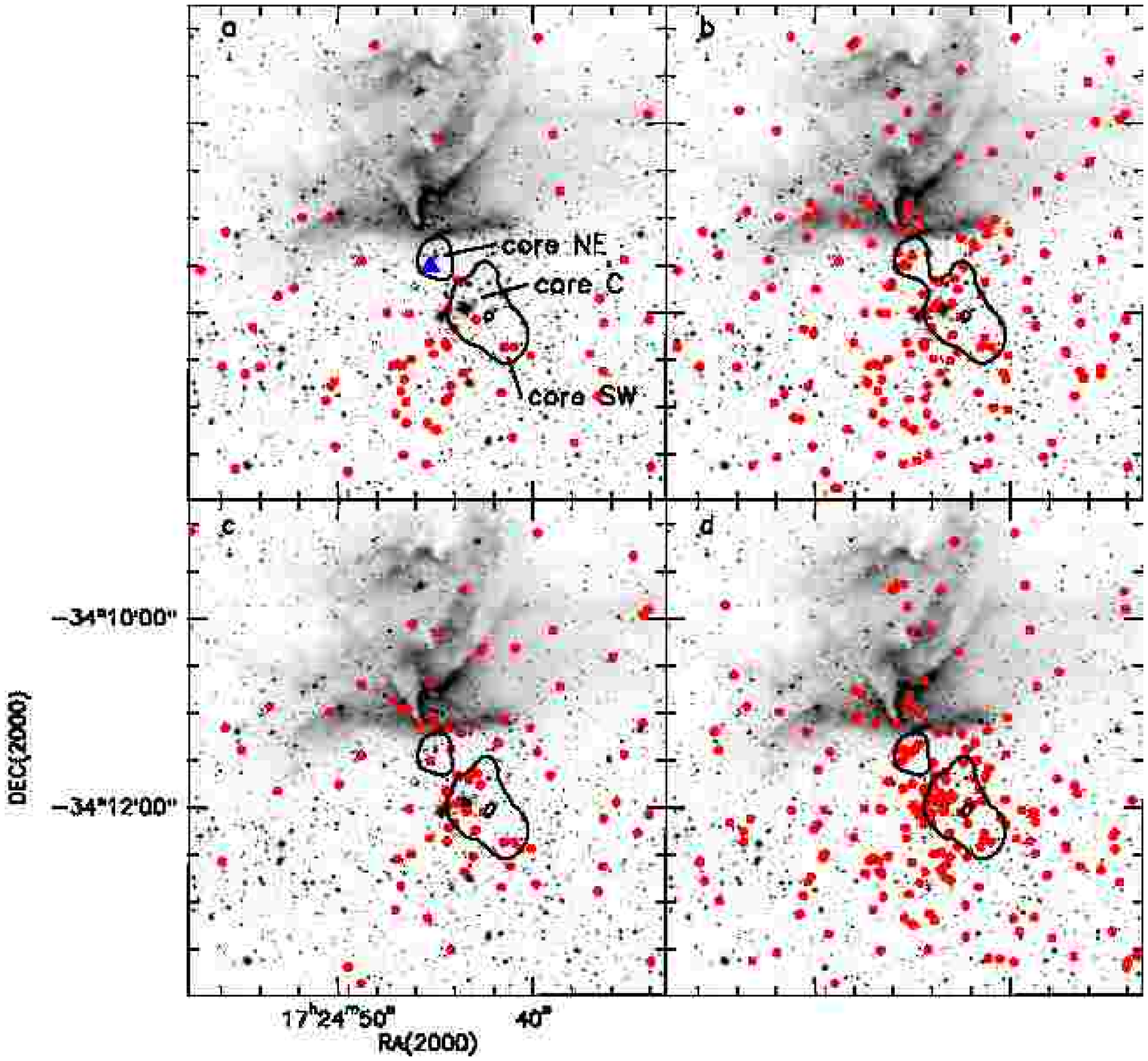}
        \caption{Spatial distribution of the NIR counterparts of the YSOs
        identified according to four different indicators, as explained in
        the text, overlaid with the $K_{s}$ image. Also drawn, a contour
	of the surface density of $K_{s}$ sources. Only objects
        with $K_{s} \le 13.5$ (unlike Fig.~\protect\ref{nir:map}, where
         $K_{s} \le 16$ has been used) have been selected.
        {\bf (a)} YSOs identified on the basis of their IRAC colours:
        red symbols for Class II sources, a large blue triangle marks 
        the only Class I source; {\bf (b)} YSOs identified on
        the basis of their combined $JH(HK_{s})[4.5]$ colours; {\bf (c)} $JHK_{s}$
        sources exhibiting a colour excess; {\bf (d)} X-ray sources.
(A colour version of this figure is available in the on-line edition.)
         \label{yso:dist:map}}
   \end{figure*}

In Fig.~\ref{yso:dist:map} we plot the spatial distribution of the $K_{s}$
counterparts of the YSOs identified through the four indicators discussed above.
To avoid biases due to the different sensitivity limits
and completeness levels, 
only sources with $K_{s} \leq 13.5$ are plotted,
so that all groups of objects are as 
homogeneously complete as possible 
(see App.~\ref{compl:nir-irac-x} and Fig.~\ref{yso:mc:d}).
A sample selection based on $K_{s}$ is prone
to contamination from sources with an infrared excess, so
using a band where excess emission is fainter, such as $J$, would be
more suitable. Unfortunately, $J$ would miss a significant
fraction of YSOs.
IRAC-selected Class I and Class II sources are mostly distributed
in the southern field, avoiding the more
extincted and brighter areas due to the sensitivity problem, 
unlike sources with a NIR excess in at least one of the
$JHK_{s}$ and $JH(HK_{s})[4.5]$ colours. 
More YSOs are retrieved when
adding the IRAC 4.5 $\mu$m band to the SofI $JHK_{s}$ bands, as expected.
Many more sources with a NIR excess are detected in the northern field
compared to IRAC-selected YSOs, particularly towards the
H\textsc{ii} region. Generally, these spread all over the SofI field,
although they tend to concentrate south of the Pis24 core. A slight 
increase in source surface density occurs towards the core. 
On the other hand, X-ray sources show clear and distinct surface density 
increases towards
core NE, core SW, core C and the H\textsc{ii} region. Core C appears to
be further composed of two parts aligned south-east to north-west.
 
%

\subsection{$K_{s}$ Luminosity Function of Pismis~24}
\label{klf:sect}

$K$ luminosity functions (KLFs) are valuable tools in constraining 
the IMF of young star clusters (Lada \& Lada
\cite{L&L}, and references therein). Our SofI field is large enough to
contain most of the cluster core members, so in principle it allows us to construct a 
KLF representative of the cluster. Unfortunately, nearby control field images are not available to
correct for field star contamination as usually done. 
It would even be difficult to find a suitable field, given that 
we are nearly looking towards the Galaxy centre. 
Hence, we must resort to an alternative method in order to minimise
contamination from field stars. This means using either {\em colour} intervals
or the NIR counterparts of the X-ray sources to select the cluster population.

We followed Massi et al.\ (\cite{massi06}) to produce
a dereddened KLF from sources with detections in the $H$ and $K_{s}$ bands.
Each datapoint is shifted along the reverse reddening direction in the ($H-K_{s}$, $K_{s}$) space
to a locus representative of the magnitudes and colours of young unextincted PMS stars. We adopted 
the locus they proposed without NIR excess correction
(hereafter the pseudo-sequence) and shifted it to account for
a distance of $1.7$ kpc. We have also modified 
the upper part of the locus by adding one more segment
(essentially a vertical line with $H-K_{s}=-0.05$
coinciding with the ZAMS; Koornneef \cite{Koorn}) to account for
the most massive cluster members. 
Given that the cluster age is rather well constrained
to $\la 3$ Myr (see Sect.~\ref{age:24}) the method is
particularly suited to obtain a reliable dereddened cluster KLF
(as shown by Massi et al.\ \cite{massi06}).

A few different selection criteria can be chosen to identify the cluster members. 
We showed in Sect.~\ref{redde:law}
that $H - K_{s} \la 1$ and $J - H \la 2$ enclose the cluster population
in the NIR CCDs of Pismis~24. Alternatively, one can use the pseudo-sequence as a reference
and pick all the sources inside a well-defined extinction interval. In principle one
could choose the extinction range found in the visible, $3.2 < A_{V} < 7.8$ (Fang et 
al.\ \cite{fang}). However, Fig.~\ref{nir:ccd}c,d suggests that 
the cluster population may span a wider range in $A_{V}$. 
Therefore, we will use the NIR counterparts of the X-ray sources
to refine the extinction cuts. 

%
%

\begin{figure}
   \centering
   \includegraphics[width=8cm]{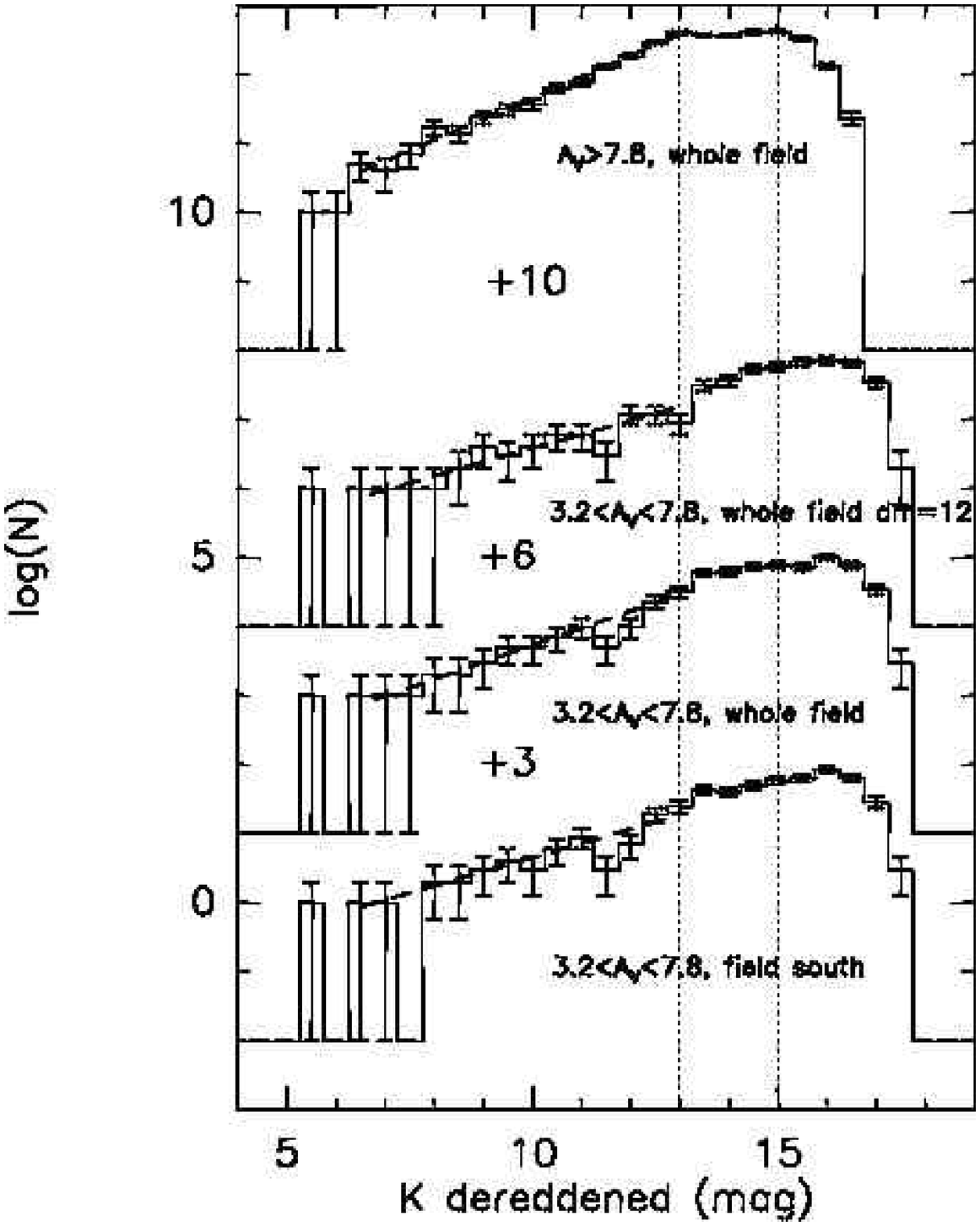}
 \caption{Dereddened KLF (see text) for sources in the 
  $H - K_{s}$ colour interval corresponding to a reddening interval
  $3.2 < A_{V} < 7.8$ (see text) from the mean locus (pseudo-sequence) of young stars.
  From bottom up: sample selected from the southern field
  only (pseudo-sequence at $1.7$ kpc), from the whole SofI field (pseudo-sequence at $1.7$ kpc), 
  and from the whole SofI field (using a pseudo-sequence shifted to a distance of $2.56$ kpc). 
  The KLF at the top has been obtained for sources with $A_{V} > 7.8$ mag picked from the whole field
  (pseudo-sequence at $1.7$ kpc).  The two vertical dotted lines mark the completeness
  limit for the KLF at the top ($K_{s} = 13$) and that for the other three KLFs ($K_{s} = 15$).	
         \label{KLF:show}}
   \end{figure}

Nevertheless, the reddening interval $3.2 < A_{V} < 7.8$ sets up an extinction-limited sample 
whose contamination from field stars is expected to be negligible, and we used it
to perform a few statistical tests.
In Fig.~\ref{KLF:show} we show four dereddened KLFs obtained by changing the most 
critical parameters (sampling field sub-region, distance, extinction interval).
A simple $\chi^{2}$ test confirms the visual impression from Fig.~\ref{KLF:show}
that the KLFs obtained are not statistically different if sources are selected 
from the whole SofI field or the southern field only, and proves that the dereddened KLFs do not 
change if the selected sources are further required to also have a valid $J$ detection.
By fitting a linear relation to the logarithmic KLFs between $K_{s} = 6.5$ and $K_{s} = 13$,
one obtains a slope of $0.21 \pm 0.05$ ($\chi^{2}$ $=1.25$), if the sources are picked from
the southern field only, and $0.20 \pm 0.09$ ($\chi^{2}$ $=0.80$), if a
pseudo-sequence shifted to $2.56$ kpc
is used. Thus, even a large systematic error in distance do not change the shape of the KLF. 
Finally, the $\chi^{2}$ test indicates only a marginal difference between the KLFs obtained from
sources in the $3.2 < A_{V} < 7.8$ and that obtained
from sources in the $A_{V} > 7.8$ mag extinction range; i. e., 
the effect of contamination of the KLF by field stars cannot be derived 
in a clear-cut, statistical way.  

A rough estimate of the completeness limit for the KLF is (dereddened magnitude) $K_{s} = 15$. 
This can be derived from the $K_{s}$ vs.\ $H-K_{s}$
diagram  as the dereddened magnitude of the point of intersection of the completeness
limit locus (dashed line in Fig.~\ref{nir:ccd}c,d) with a pseudo-sequence
extincted by a value equal to the upper limit in the sampling $A_{V}$ range.
It can be seen in Fig.~\ref{KLF:show} that the
limiting magnitude obtained lies roughly 1--2 mag below the KLF peak.

%
%

\begin{figure}
   \centering
   \includegraphics[width=6cm]{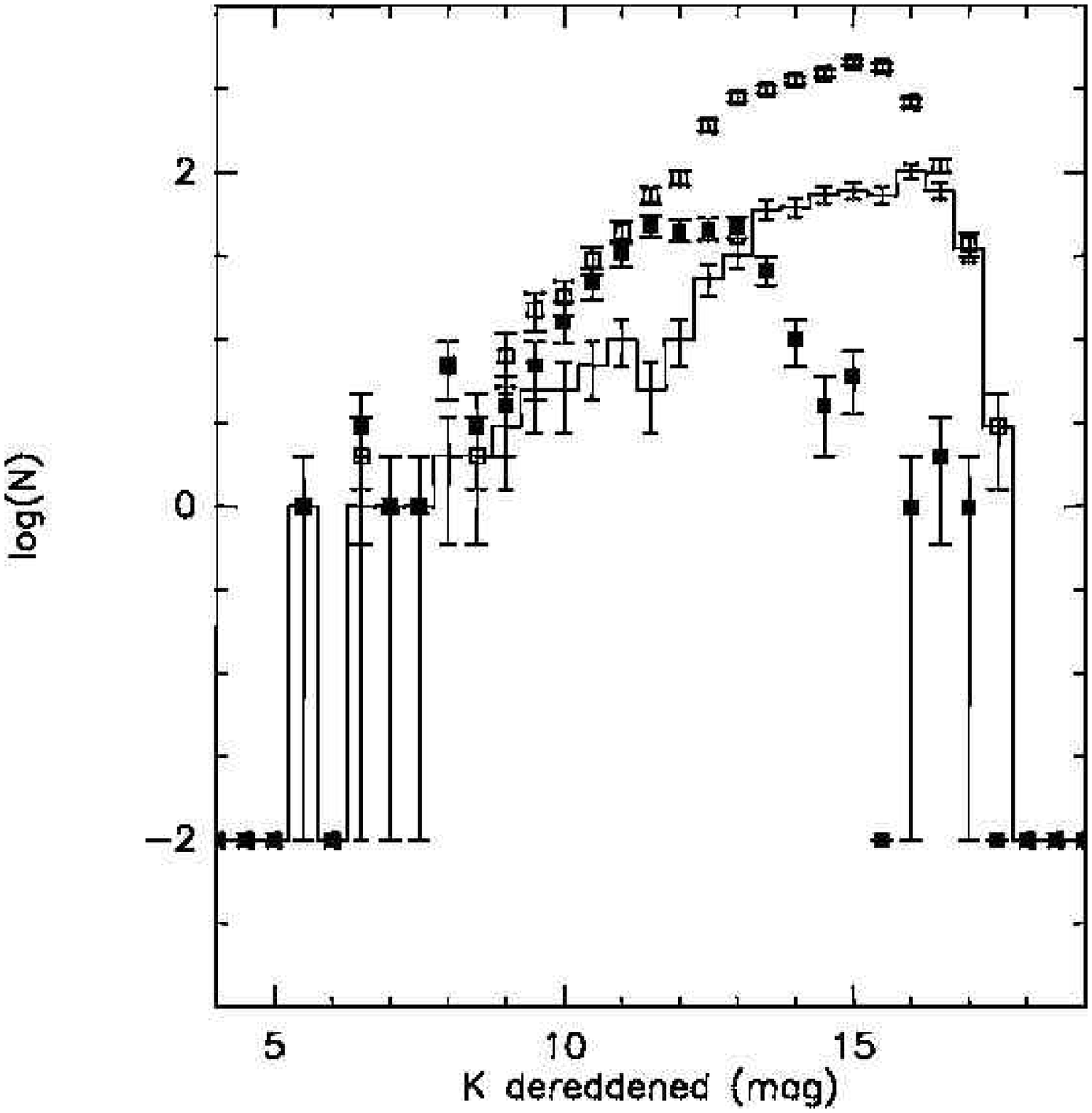}
 \caption{Dereddened KLF (see text) for sources in the  
  $H - K_{s}$ colour interval corresponding to a reddening interval
  $3.2 < A_{V} < 7.8$ (see text) from the mean locus (pseudo-sequence) of young stars 
  (histogram), for $K_{s}$ counterparts of X-ray sources
  (full squares), and for sources in the reddening interval $3.2 < A_{V} < 15$ 
  (open squares). All are obtained with	
  the pseudo-sequence at $1.7$ kpc and from the whole field. Error-bars
  are derived from Poisson statistics. 
         \label{newklf:oldklf}}
   \end{figure}

The most critical issue in selecting a representative sample of cluster members
by using a reddening interval is finding a suitable upper limit that allows
including most of the population with only small contamination. As discussed
in Sect.~\ref{SDXS}, the X-ray sources are mostly cluster members and their NIR
counterparts can be used to construct a KLF completely independent of 
any assumption of a reddening interval.
A comparison between the dereddened X-ray counterpart 
KLF and that from sources with $3.2 < A_{V} < 7.8$ is shown in Fig.~\ref{newklf:oldklf}
(pseudo-sequence at $1.7$ kpc, data from the whole field).
They are clearly different and a $\chi^{2}$ test 
on the segment from $K_{s}$ $6.5$ to
to $11.5$ yields a significance level of $0.35$, confirming they are only marginally 
consistent with each other.
On the other hand, Fig.~\ref{newklf:oldklf} shows that the dereddened KLF 
obtained from sources in the wider range $3.2 < A_{V} < 15$ looks more consistent with
that from X-ray counterparts, which is confirmed by a $\chi^{2}$ test on
the rising part of the distributions.
The stars in the extinction range $3.2 < A_{V} < 15$ from the pseudo-sequence fall
inside the less extincted datapoint concentration in the CCDs of Figs.~\ref{nir:ccd}a,b
that mostly features cluster members. Thus, they make up a natural extension of
the sample of X-ray counterparts, picking sources down to a fainter completeness limit
in $K_{s}$.

As a final remark, we caution against the effects of the NIR excess of 
a fraction of cluster members. 
The selected extinction interval does not account for the 
 NIR excess exhibited by a number of YSO.
These YSOs are both brighter in $K_{s}$ and look 
as if they were more extincted than their central stars  actually
are, hence they look more massive than they are. As discussed by Massi et al.\ (\cite{massi06}), 
this will result in a steeper KLF.

%
%

\subsection{Age of Pismis~24 and masses of stars from the optical HST data}
\label{age:24}

Following Massi et al.\ (\cite{massi06}), if the age and star formation history
of Pismis~24 are known then
its KLF can be converted into an IMF.
So far, the age of Pismis~24 has been estimated in two ways: from the 
age of the most massive stars and from the age of the intermediate to low-mass stars.
Massey et al.\ (\cite{massey01}) compared the most massive cluster stars
with isochrones in a Hertzsprung-Russell diagram finding a median age of $1.7$ Myr and a 
very likely coevality. A similar exercise was done by Fang et al.\ (\cite{fang}),
who found that the age of the 6 most massive stars is 
between 1--$2.7$ Myr if they are assumed to be coeval. 
On the other hand, Fang et al.\ (\cite{fang}) obtained optical spectra
and $RI$ photometry for 151 stars with an X-ray counterpart,
deriving their spectral type and extinction. As expected, once dereddened
these stars spread out well above the ZAMS locus in an $R$ vs.\ $R-I$ diagram,
confirming their PMS nature.
By comparing their location with PMS isochrones from Dotter et al.\ (\cite{dotter}),
Fang et al.\ (\cite{fang}) showed that their age distribution peaks at 1 Myr.
Finally, Lima et al.\ (\cite{lima}) obtained an older age of 5 Myr by fitting
Padova isochrones in a $J$ vs.\ $J - K_{s}$ diagram.

\begin{figure*}
   \centering
   \includegraphics[width=16cm]{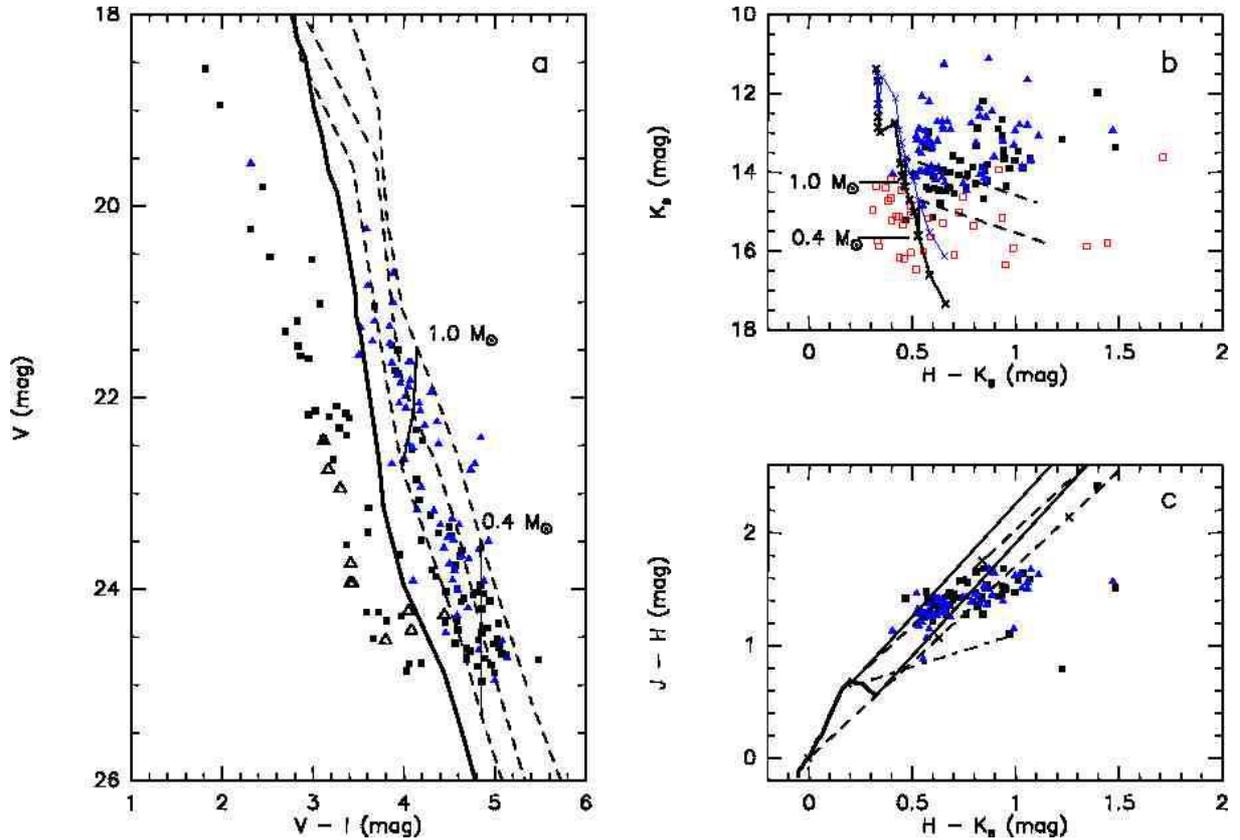}
 \caption{
   {\bf (a)} $V$ vs. $V-I$ for stars towards Pismis~24 from HST/WFPC2 images.
  The full line marks the ZAMS, the dashed lines are isochrones for
  (from the right) 1, 3, and 10 Myr old PMS stars obtained from the
  evolutionary tracks of Palla \& Stahler (\cite{ps99}). These have been
  reddened by $A_{V} = 5.5$ mag and scaled to $1.7$ kpc.
  The loci of 1 and $0.4$ $M_{\sun}$ are labelled near the 1 Myr isochrone.
  Large open triangles are stars without a NIR counterpart, whereas small
  blue full triangles are stars with X-ray detection.
  {\bf (b)} $K_{s}$ vs.\ $H - K_{s}$ for the
   NIR infrared counterparts of the optical stars (red open squares:
  stars on the left of the optical ZAMS; full squares: stars on the right
  of the optical ZAMS; blue full triangles: stars on the right of the
  optical ZAMS also detected in X-ray emission). The full lines are the
  isochrones for (from the right) 1 and 3 Myr old PMS stars obtained from the
  evolutionary tracks of Palla \& Stahler (\cite{ps99}), same reddening and distance
  as above.  The loci of 1 and $0.4$ $M_{\sun}$ are also labelled
  and the dashed lines are their reddening paths (for 1 Myr old stars,
  up to $A_{V} = 10$ mag, according to Rieke \& Lebofsky \cite{r&l}).
  {\bf (c)} $J - H$ vs.\ $H - K_{s}$ for the
   NIR infrared counterparts of the optical stars (same symbols as above,
   but stars on the left of the optical ZAMS are omitted). The full line
   is the unreddened MS, the dashed lines are reddening paths with crosses
   every $A_{V} = 10$ mag, following Rieke \& Lebofsky (\cite{r&l}).
   Also shown as full lines, the reddening paths according to
   the extinction law derived by Straizys \& Laugalys (\cite{StraLau}).
   The dot-dashed line marks the locus of Classical T-Tauri Stars
   (Meyer et al.\ \cite{meyer}).
(A colour version of this figure is available in the on-line edition.)
   \label{v:vi}
      }
   \end{figure*}

We used our $VI$ photometry from the HST/WFPC2 images to check the age derived by the
other authors. 
The results are shown in the $V$ vs. $V-I$ diagram of Fig.~\ref{v:vi}a,
along with the ZAMS locus and isochrones for PMS stars 1, 3, and 10 Myr
old (for a reddening $A_{V} = 5.5$ mag and a distance of $1.7$ kpc).
For the ZAMS, we have adopted $V-I$ colours from Ducati et al.\
(\cite{ducati}) for B stars (transformed to the Johnson-Cousins standard),
and from Bessel (\cite{bessel}, \cite{bessel91}) for later than B stars. 
We note that $V-I$ colours from Ducati et al.\
(\cite{ducati}) and Bessel (\cite{bessel}) show a discrepancy of $\sim 0.1$ mag for 
B7 stars, although they are in good agreement for A0 stars.
The PMS sequence of the cluster is easily identified above the ZAMS, and  
the stars fall in the age interval 1--10 Myr with an average of about $3$ Myr.
The 11 optical sources without a NIR counterpart (and an X-ray detection, either)
mostly lie to the left of the ZAMS, as expected for field stars.

To further investigate the age dispersion, we show in Fig.~\ref{v:vi}b the
$K_{s}$ vs.\ $H - K_{s}$ diagram for the NIR counterparts of the optical
stars, along with the isochrones for PMS stars 1 and 3 Myr old (same reddening
and distance as above). The NIR counterparts of the  PMS stars with X-ray emission 
are shown as full triangles. By comparing the data 
with the isochrones, the PMS stars appear more massive in the NIR than at 
optical wavelengths. This discrepancy already appears if one compares the masses found
by Fang et al.\ (\cite{fang}) and by Wang et al.\ (\cite{wang07}). The latter
authors found more massive counterparts for the X-ray sources from 2MASS NIR data.
We tried estimating the PMS star extinction by shifting their datapoints along the reddening
vector to the 1 and 3 Myr loci in the $K_{s}$ vs.\ $H - K_{s}$ diagram.  
When turned into optical extinction,
this causes the stars to move to the upper corner of the $V$ vs. $V-I$ diagram
of Fig.~\ref{v:vi}a,
to the left of the ZAMS. Although a photometric error of $0.1$ mag
in the NIR would result in an error of $\sim 1$ mag in $A_{V}$, 
the shift is much larger for most of the objects.
This is evidence
that the dispersion in the magnitude-colour diagram of Fig.~\ref{v:vi}b
is due not only to differential extinction, but also to an infrared excess. 
Although only part of the stars exhibit
an excess in the $J - H$ vs.\ $H - K_{s}$ diagram
(i. e., lie below the reddening band of the MS; 
see Fig.~\ref{v:vi}c), a comparison between the NIR sources with X-ray counterparts
and the locus of Classical T-Tauri Stars
(Meyer et al.\ \cite{meyer}), also drawn in Fig.~\ref{v:vi}c, confirms
that their spread in the CCD is strongly affected by NIR excess emission. 
Figure~\ref{v:vi}c also shows that the steepest of the extinction
laws we considered, that of Straizys \& Laugalys (\cite{StraLau}),
would only marginally increase the fraction of sources with a NIR  excess.

\begin{table*}
\caption{Statistics of the detection of HST sources in infrared bands
and X-ray.
\label{HST:dete}}
\centering                          
\begin{tabular}{ c c c c c c c }        
\hline\hline                 
Identification & $VI$ & $VIHK_{s}$ & $VIHK_{s}[4.5]$ & $VIHK_{s}[4.5]$ & $VIHK_{s}[4.5]$ & 
     $VIHK_{s}[4.5]$ \\
w.r.t. ZAMS  & sources & sources  & sources & with  X-ray & with NIR exc. & with X-ray and NIR exc. \\
\hline
Cluster members ($V<23$) & 42 & 42 & 21 & 17 & 11 & 10 \\
Cluster members ($V \ge 23$) & 71 & 70 & 23 & 9 & 20 & 7 \\
All cluster members & 113 & 112 & 44 & 26 & 31 & 17 \\
Field stars ($V<23$) & 27 & 24 & 2 & 0 & 2 & 0 \\
Field stars ($V \ge 23$) & 18 & 11  & 3 & 0 & 3 & 0 \\
\hline
\end{tabular}
\end{table*}

The issue can be better tackled by adding the IRAC 4.5 $\mu$m measurements to $JHK_{s}$,
using the criteria of Winston et al.\ (\cite{wins07}) to discriminate
sources with a NIR  excess, as done in Sect.~\ref{SDNIRS}.
We found 49 HST stars with a detection in the 4.5 $\mu$m band, only 5 of which
lie to the left of the ZAMS in Fig.~\ref{v:vi}a. A statistics of the
detection of the optical sources in infrared bands and X-ray is shown in
Table~\ref{HST:dete}. The stars on the left of the optical ZAMS are referred 
to as field stars, whereas those on the ZAMS or to
the right of it are referred to as cluster members. Clearly, 50 \% of the HST
cluster members with $V < 23$ were detected at 4.5 $\mu$m, whereas
this fraction drops to 32 \% for $V \ge 23$. This is obviously a signature of
incompleteness in the IRAC bands; in fact, Table~\ref{mass:compl:tab} lists mass completeness
limits in the 3.6 and 4.5 $\mu$m bands of $2 M_{\sun}$
for naked PMS stars and $\sim 1 M_{\sun}$ for Class II sources.
Out of the 44 HST cluster members with detection at 4.5 $\mu$m,
70 \% exhibit a NIR excess, a fraction increasing to 87 \% 
if considering only those with $V \ge 23$.
Hence, at least 27 \% of all the optically detected 
cluster members exhibit a NIR excess. Since PMS star emission is hardly affected
by disk emission in the optical bands, the sample of HST cluster members is not
biased towards stars with excess emission in the NIR, although limited by extinction.
Therefore, this fraction may well be a lower limit for the fraction of cluster
members as a whole with a circumstellar disk. 

Figures~\ref{v:vi}b,c suggest that a large fraction of $JHK_{s}$ sources with X-ray detection
must also exhibit a strong NIR excess (having ruled out heavy extinction).
In fact, 59 \% of the 
HST cluster members with a detection in the 4.5 $\mu$m band also have an X-ray counterpart.
Of these, 17 (65 \%) exhibit a NIR excess.  
Thus, a large number of cluster members must be 
associated with a circumstellar disk and are therefore quite young; 
the mass mismatch between optical and NIR CMDs would even increase
by assuming an age of 10 Myr.
Therefore, we think that the cluster age can be constrained to 1--3 Myr and the
age dispersion shown in the $V$ vs. $V-I$ diagram is not real. 
No clear age patterns can be noticed by selecting stars falling in limited areas
of the field, although those near the massive stars show a tendency to lie close
the 3 Myr isochrone (however with a number of objects quite small).

The nature of the optical cluster members undetected in 
X-ray can be deduced from Fig.~\ref{v:vi}a.
At the high mass end, X-ray emitting sources are more numerous than
at the low mass end. The X-ray activity in PMS stars is known to be variable
(see e. g., Flaccomio et al.\ \cite{flaccomio}), so
a fraction of low-mass stars in a phase of increased X-ray activity were detected.
This clarifies why Fang et al.\ (\cite{fang}) found $RI$ counterparts of X-ray sources
down to $< 0.4$ $M_{\sun}$, whereas the estimated mass detection limit in X-ray is higher
($> 0.7$ $M_{\sun}$; see Sect.~\ref{SDXS}). The optical cluster members undetected in X-ray
are therefore lower-mass stars in a phase of lower X-ray activity.

Unlike intermediate to low-mass stars, the cluster massive members are
Main Sequence (MS) stars.
For MS isochrones the time is measured starting from the ZAMS,
while PMS tracks use different zero-age reference isochrones; e. g.
Palla \& Stahler (\cite{ps99}) adopted their birthline. 
Classical PMS evolutionary
tracks usually start from the Hayashi tracks rather than the birthline. 
Tognelli et al.\ (\cite{pisa}) followed this approach and showed that
their tracks intersect the birthline before 1 Myr for almost all stellar masses.
In particular, for $M < 1.5$ $M_{\sun}$ this happens before $0.5$ Myr.

Birthlines, like that of Palla \& Stahler (\cite{ps99}),
necessarily join the ZAMS for stellar masses somewhere around $10$ $M_{\sun}$,
hence cannot be used as a zero-age isochrone for massive stars. 
So the results obtained from massive (MS) and intermediate to low-mass (PMS) stars
simply suggest that most of the massive stars were already on the ZAMS
when most of the intermediate and low-mass stars were still emerging from
the birthline or were still contracting along a PMS track, 
within $\sim 1$ Myr from each other. 

%
%

\subsection{The Initial Mass Function of Pismis~24}
\label{imf:24}

To convert dereddened $K_{s}$ magnitudes in masses, we used the PMS evolutionary
tracks computed by Palla \& Stahler (\cite{ps99}), as in Massi et al.\ (\cite{massi06}).
Since the relationship between $K_{s}$ and stellar mass depends on age, 
we used isochrones for
2 different ages, 1 and 3 Myr respectively. As shown in the previous section,
these encompass the range of mean ages derived for the cluster. 
At these ages, Palla \& Stahler
(\cite{ps99}) predict that intermediate mass stars are already on the 
ZAMS, so
the isochrones have been complemented with a ZAMS model,
as explained in Massi et al.\ (\cite{massi06}). 
We have further extended the ZAMS relationship up to O3--4 V
stars by assuming these to have the same $V-K$ colour as O6--9 V stars
(given by Koornneef \cite{Koorn}). This should yield a
good approximation for the mass of the OB members even though
they have already departed from the ZAMS, also considering that
the masses of early O stars (the most evolved ones)
are highly uncertain (see Weidner \& Vink \cite{WV10}).  

In addition, 
to test the effects on the IMF of adopting different models
for massive stars to convert brightness into mass,
we also used the synthetic tracks 
of Lejeune \& Schaerer (\cite{basel}) for main sequence stars
from $0.8$ to 100 $M_{\sun}$, in particular 
the ones for solar metallicity ($Z=0.02$), age 1 Myr and increased mass
loss from high-mass stars. 
The $K_{s}$ magnitudes of OB stars were, therefore, converted into {\it initial} 
rather than current masses.
These tracks also allowed us to test the effect of using a MS $K_{s}$-mass
relationship for all stellar masses,
as opposed to our adopted mixed MS-PMS isochrones.

\begin{figure}
   \centering
   \includegraphics[width=8cm]{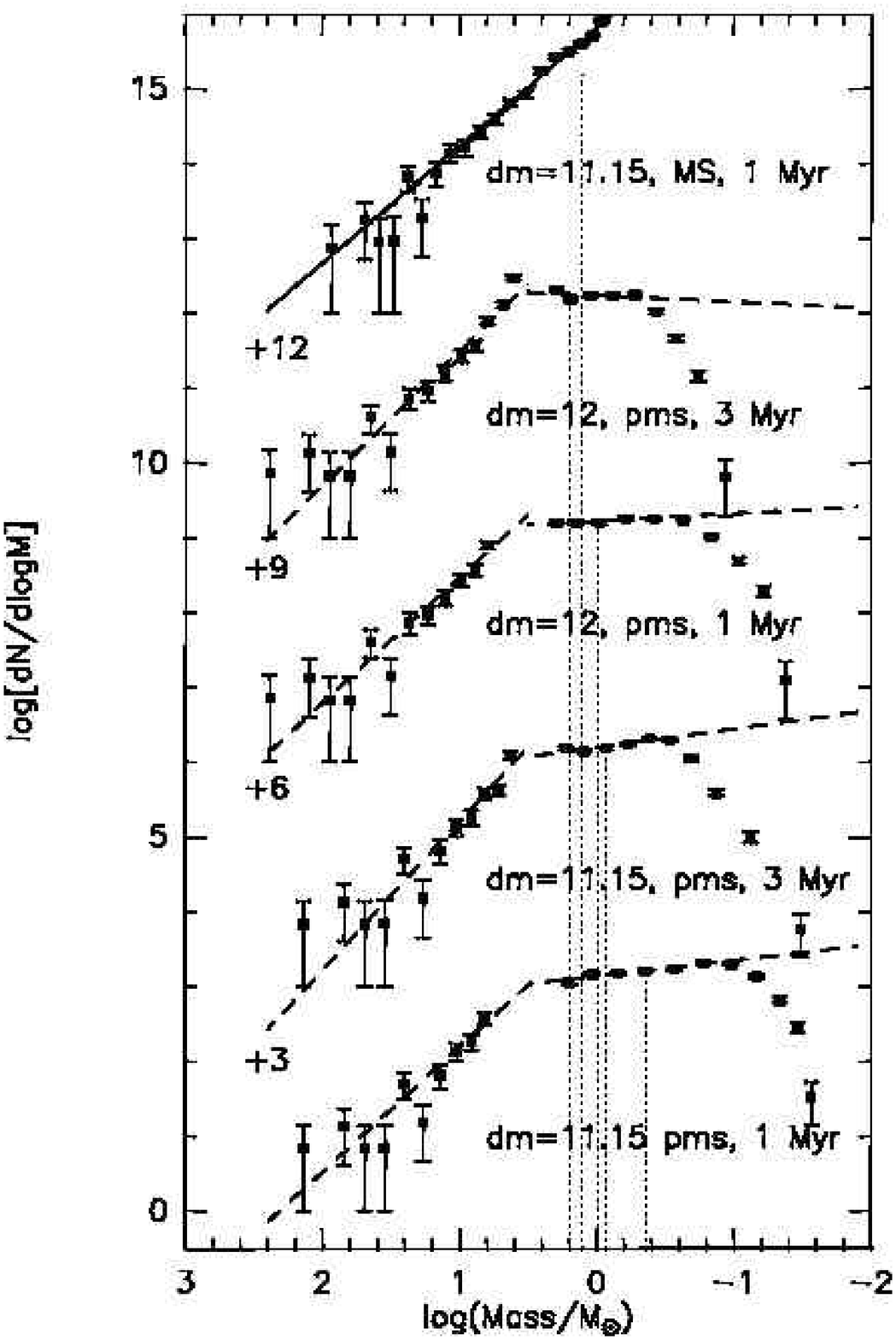}
 \caption{Logarithmic initial mass functions obtained from the dereddened KLFs,
	each shifted by the amount indicated on the left. The assumed
        distance moduli, ages, and synthetic tracks (either PMS for low-mass
	stars, or MS for all stars) are also labelled.  
        The dashed lines show the two-segment fits to the IMFs (for the
        IMF on the top, the full line is a fit to the whole distribution,
        which cannot be distinguished in the figure from the fit to the high-mass
        end only ).
	The vertical dotted lines mark the mass for each distribution
        at the completeness limit
        of the corresponding dereddened KLF.
         \label{imf:pile}}
   \end{figure}

Finally, 
we note that a small range of intermediate
masses is necessarily degenerate when using the PMS tracks of Palla \& Stahler (\cite{ps99}), 
given that ZAMS stars and PMS stars
of different masses overlap and exhibit the same $K_{s}$ brightness in that range
(lying around a value of a few solar masses). Thus, we cannot
perform any mass conversion in that limited range and this explains the gaps
in Fig.~\ref{imf:pile}.  
%
%

We show in Fig.~\ref{imf:pile} a set of IMFs obtained by sampling the whole SofI field
in the reddening interval $3.2 < A_{V} < 15$ with respect to the pseudo-sequence.
Given that we use 2MASS magnitudes for
the brightest stars (saturated in our SofI images), massive binary stars (such as Pismis 24 1;
see Ma\'{i}z Apell\'{a}niz et al.\ \cite{maw}) are not resolved. 
Thus the IMFs obtained are primary star IMFs. On the other hand, 
this allows us to make a direct comparison with most of the IMFs of clusters
found in the literature, which are usually primary star IMFs, as well. 
We also show an IMF derived by assuming a distance of $2.56$ kpc, 
which allows  us to assess the effect on the IMF of a systematic error 
on distance.

All IMFs obtained using
PMS tracks for low-mass stars exhibit a change of slope around $M=2.5$ $M_{\sun}$. 
This flattening does not depend on the sampling reddening interval, since 
it is also displayed by the IMFs derived from the KLFs of sources in the 
reddening interval $3.2 < A_{V} < 7.8$.
We fitted the logarithm of the IMF from the PMS plus ZAMS tracks with 
two-segment functions; the slope of each one directly yields $\Gamma$ (where d$N$/d$\log M$ $\sim
\Gamma$; i. e., for a Salpeter IMF, $\Gamma = -1.3$) in the fitting range. 
The brightest star (Pismis 24 1) is always omitted
from the fits to the $M > 2.5$ $M_{\sun}$ part of the IMF, to avoid any biases due to its 
overestimated mass (for comparison, Ma\'{i}z Apell\'{a}niz et al.\ \cite{maw} found $97$ $M_{\sun}$,
which in turn is an upper limit because of the double nature of this star). Adding this point 
results in a slightly flatter IMF with $\Delta\Gamma$ within $0.1-0.2$ in all cases, 
which is inside the errors.  

We list the results of our two-segment fits to the IMFs
in Table~\ref{tab:imfit}, along with the fitting mass range and the reduced
$\chi ^{2}$. 
It has been shown (see, e.\ g., Maschberger \& Kroupa \cite{mushy})
that fitting straight lines to a histogram with bins of constant width 
leads to a biased result. Thus we
also derived $\Gamma$ for the high-mass    end of the IMF by adapting the 
modified Maximum Likelihood estimator proposed by Maschberger \& Kroupa (\cite{mushy}),
along with their estimator for the upper limit of a truncated power law, in the hypothesis
that the actual IMF is a power-law truncated at high masses. These result are also 
listed in Table~\ref{tab:imfit} and, again, we do not include the brightest star
in the computations. Its inclusion does not change significantly 
the obtained $\Gamma$, 
but of course it leads to an increase in the upper mass limit, by a factor $\sim 2$.
To show how the choice of the KLF affects the results, Table~\ref{tab:imfit}
also lists fits to the dereddened KLF from stars selected 
in the reddening interval $3.2 < A_{V} < 7.8$. 

A larger distance than that we have adopted yields higher stellar masses
but does not affect $\Gamma$ significantly.
For masses $\la 2$ $M_{\sun}$ the IMF flattens,
except when using only MS tracks to convert brightness into mass.
The flatter segment of the IMF extends beyond the completeness limit. 
In any case, by adopting the KLF of sources sampled in the range $3.2 < A_{V} < 7.8$, the completeness
magnitude can be increased by $\sim 1$ mag, thus decreasing the corresponding mass
value.
The IMFs associated show a similar flattening, so there is no doubt
that this is real, since this more conservative KLF is still expected to provide a representative
sample of low-mass members.

%
\begin{table*}
\caption{Two-segment fits to the IMFs.
\label{tab:imfit}}
\centering                          
\begin{tabular}{ c c c c c c c c c c c}        
\hline\hline                 
Distance & Age & $\Gamma_{1}$ & Mass & Reduced &
             $\Gamma_{2}$ & Mass & Reduced & $\Gamma_{3}^{b}$ & Mass & Upper \\
 modulus & & & range & $\chi^{2}$ & & range & $\chi^{2}$ & & range & mass  \\
 (mag) & (Myr) & & ($M_{\sun}$) & & & ($M_{\sun}$) &
                  & & ($M_{\sun}$) & ($M_{\sun}$) \\
\hline                        
\multicolumn{11}{c}{Star selection: reddening interval $3.2 < A_{V} < 15$} \\
\hline                        
$11.15$ & 1 & $-1.7 \pm 0.2$ & 6--80 & $1.1$ & $-0.21 \pm 0.03$ & $0.1$--$1.5$ &
		$0.7$ & $-1.9 \pm 0.2$ & 6--77 & 78 \\
$11.15$ & 3 & $-2.0 \pm 0.1$ & 4--80 & $2.4$ & $-0.25 \pm 0.05$ & $0.4$--$1.7$ & 
		$1.6$ & $-2.0 \pm 0.2$ & 4--77 & 77 \\
$11.15$ & 1$^{a}$ & $-1.5 \pm 0.1$ & 3--80 & $0.4$ & -- & -- &
                -- & $-1.8 \pm 0.1$ & 2--53$^{c}$ & 53 \\
$11.15$ & 1$^{a}$ & $-1.57 \pm 0.04$ & 0.6--80 & $0.8$ & -- & -- & -- &
                --  & -- & -- \\
12 & 1 & $-1.7 \pm 0.1$ & 6--124 & $1.3$ & $-0.10 \pm 0.04$ & $0.4$--2 &
		$0.5$ & $-1.8 \pm 0.2$ & 6--138 & 138 \\
12 & 3 & $-2.1 \pm 0.1$ & 4--124 & $3.8$ & $0.08 \pm 0.07$ & $0.8$--2 &
		$2.8$ & $-2.1 \pm 0.1$ & 4--138 & 138 \\
\hline                        
\multicolumn{11}{c}{Star selection: reddening interval $3.2 < A_{V} < 7.8$} \\
\hline                        
$11.15$ & 1 & $-1.2 \pm 0.3$ & 6--80 & $0.1$ & $-0.3 \pm 0.1$ & $0.2$--$1.5$ &
		$0.9$ & $-1.3 \pm 0.4$ & 6--77 & 79 \\
$11.15$ & 3 & $-1.1 \pm 0.2$ & 3--80 & $0.1$ & $-0.3 \pm 0.2$ & $0.4$--$1.7$ & 
		$0.1$ & $-1.2  \pm 0.2$ & 4--77 & 79 \\
$11.15$ & 1$^{a}$ & $-1.0 \pm 0.2$ & 5--80 & $0.07$ & -- & -- & 
                -- & $-0.9 \pm 0.2$ & 4--53$^{c}$ & 54 \\
$11.15$ & 1$^{a}$ & $-1.4 \pm 0.1$ & 1--100 &
                $1.2$ & --  & -- & -- & --  & -- & -- \\ 
12 & 1 & $-0.8 \pm 0.2$ & 5--160 & $0.3$ & $-0.3 \pm 0.2$ & $0.5$--2 &
		$0.2$ & $-1.0 \pm 0.2$ & $6.5$--138 & 142 \\
12 & 3 & $-1.1 \pm 0.2$ & 3--160 & $0.5$ & $-0.2 \pm 0.2$ & $0.6$--2 &
		$0.3$ & $-1.1 \pm 0.2$ & 4--138 & 140 \\
\hline                                   
\end{tabular}

\vspace*{1mm} $^{a}$~Main Sequence tracks of Lejeune \& Schaerer (\cite{basel});
$^{b}$~modified Maximum Likelihood estimator (Maschberger \& Kroupa \cite{mushy}). 
$^{c}$~the upper limit is the mass of the actual second most massive stars, whereas
for $\Gamma_{1}$ the midpoint of the bin containing the second most massive star is used.
\end{table*}

To conclude, the cluster IMF appears to be scarcely affected by the choice of age and distance,
the most significant changes being due to the choice of the underlying KLF and
of the evolutionary tracks. Table~\ref{tab:imfit} shows 
that by increasing the upper limit of the
extinction range to select the cluster members from $A_{V} = 7.8$ mag to
$A_{V} = 15$ mag one obtains a steeper IMF at the high mass end. 
Although using the MS tracks of Lejeune \& Schaerer (\cite{basel})
to convert $K_{s}$ magnitudes into masses is a useful exercise, it is not appropriate
for the low-mass cluster members since, as demonstrated in this work, they are
mostly PMS stars. Its most notable effect is then on the MS high-mass part of the IMF,
where nevertheless it only produces a slightly flatter distribution.
This also allows one to assess the effect of an age spread, rather than 
a single burst of star formation, on the IMF: including 
a fraction of older PMS stars results in steepening the low-mass end of the IMF.
On the other hand, the inclusion of cluster members with a NIR excess
is likely to have steepened the IMF as a whole and, probably,
also to have shifted the turning point towards a higher mass. Since stars with a NIR excess
shift redwards in the $K_{s}$ vs.\ $H - K_{s}$ diagram, lowering the upper end of the reddening interval to
$A_{V} = 7.8$ mag should result in filtering out a large fraction of these objects. Thus,
as a rule of thumb the actual slope of the
high mass end of the IMF should be bracketed by the values derived for the two extinction
ranges we have considered. 

Our data point to a Salpeter-like IMF, truncated to an upper
mass limit of 50--100 $M_{\sun}$. A flattening occurs at the low mass end, with 
a turning point somewhere between 1--3 $M_{\sun}$. If taken at face value, this 
flattening is
consistent with the IMF prescribed by Scalo (\cite{scalo}),
 and the field
star IMF of Kroupa et al.\ (\cite{kroupa}). It is also consistent
with the IMF of young clusters given by Chabrier (\cite{chabry}),
although our data are unsuitable to investigate the log-normal
part of this distribution. 

\section{Discussion}
\label{discussione}

\subsection{Star formation on the smaller scale: substructures in Pismis~24}

Our NIR images allow us to study in more depth the core of Pismis~24. 
The SofI field of view covers an area containing what we have called, in Table~\ref{irac:clust:tab},
 Pis24 core,
Pis24N, and Pis24E, whereas  Pis24W and Pis24S lie outside. Star \object{N78 36} is visible in the
northern part of the images along with the small associated cavity
(more evident in the IRAC three-colour image of Fig.~\ref{pismisJHK}). 
However,
only Pis24 core stands out in the contour map of $K_{s}$ source surface density 
(Fig.~\ref{nir:map}). Apart from some doubts on Pis24N, both Pis24N and Pis24E are
anyway characterised by a much lower surface density of IRAC sources than Pis24 core. Thus,
they probably cannot be separated statistically from the background star counts in the
smaller SofI field. If so, the value adopted for the surface density of field stars is
overestimated and, consequently, the number of NIR sources associated with the cluster 
is underestimated. This is confirmed by the radial profile derived by Lima et
al.\ (\cite{lima}) for Pismis 24, which has a radius of $\sim 3\arcmin$.

One critical issue is whether the multiple density peaks of 
the core (see Fig.~\ref{nir:map} and Table~\ref{pis24:clust:tab})
represent real substructures or are rather the effect of variable
extinction.
Although the bulk of molecular gas lies in the SofI northern field,
a few molecular clumps have been found projected towards the core
(e. g., CO clumps E, F, and G of Massi et al.\ \cite{massi97},
and CS clump D3 of Giannetti et al.\ \cite{giannetti}).
However, some of these clumps appear red-shifted and those authors suggest that
this is because these may lie in the rear part of the complex, accelerated away
by rocket effect. In addition, the multiple density peaks 
are also clearly visible in the long exposure SPITZER images at $3.6$ and
$4.5$ $\mu$m, where extinction is much lower than in the $K_{s}$ band
(e. g., compare the SofI and IRAC images in Fig.~\ref{pismisJHK}).
They are also well outlined by X-ray sources in Fig.~\ref{yso:dist:map}.
Thus, the sub-clustering of the core of Pismis~24 is probably real.
Sub-cluster core NE (Fig.~\ref{nir:map}) lies between the elephant trunk and the massive
stars, and all these structures are roughly aligned along the elephant trunk axis.
Figure~\ref{yso:dist:map}
highlights a large number of YSOs towards the H\textsc{ii} region,
as well.


Whether this source distribution is the result of sequential star formation
started by the massive stars in core C, with the youngest generation
being the closest to the molecular gas, should be proven by demonstrating a corresponding
age sequence. However, no clear trends are obvious from the HST
photometry (see Sect.~\ref{age:24}). Alternatively, the fraction
of cluster members with a circumstellar disk is believed to be
an indicator of the cluster age (Haisch et al.\ \cite{haisch}).
This fraction is derived by counting the number of members
exhibiting a NIR excess. As already noted,
not all the stars with a circumstellar disk exhibit a colour excess 
in a $J - H$ vs. $H - K_{s}$
diagram, so some data in at least one longer wavelength band are needed. 
We used  our NIR-IRAC combined photometry 
($JHK_{s}[3.6][4.5]$) and determined the number of NIR sources up to
$K_{s} = 13.5$ which exhibit a NIR excess either in $JHK_{s}$ or
in $JH(HK_{s})[4.5]$ (or both). The upper limit in $K_{s}$
was chosen to minimise incompleteness effects, following the 
results of App.~\ref{compl:nir-irac-x}. 
Furthermore, we ruled out all sources with $J - H > 2$ in order 
to keep background giants out.

We counted the NIR sources distributed in five 
different areas: a rectangular one ($\sim 0.74 \times 0.31$ pc$^{2}$)
including the H\textsc{ii} region (G353.2+0.9), and four circular ones.
Two of the circles coincide with sub-clusters core NE (radius $\sim 0.11$ pc) 
and core SW (radius $\sim 0.14$ pc), respectively
(see Fig.~\ref{nir:map}), whereas sub-cluster core C has been further subdivided in
two parts, one centred on the massive stars (SE, radius $\sim 0.21$ pc), 
the other including the north-western tip (NW, radius $\sim 0.16$ pc).
The results are listed in Table~\ref{subcl:counts}.

\begin{table*}
\caption{Source counts inside selected areas in the SofI field of view.
\label{subcl:counts}}
\centering                          
\begin{tabular}{ l c c c c c c c}        
\hline\hline                 
Sub-cluster & Centre or & Radius or & Total number & X-ray & IRAC-derived & Sources with & Fraction \\ 
	   & south-eastern & side lengths & of NIR & NIR &
                Class II& a $JHK_{s}$ and/or & of sources \\
 & corner & & sources & counterparts & sources & a $JH(HK_{s})[4.5]$ excess & with c/s disk\\ 
\hline                        
H\textsc{ii} & 17:24:47.50 -34:11:14.8 & $90\arcsec \times 38\arcsec$ & 22 & 11 & 0 & 13 & $0.59 \pm 0.21$ \\ 
core NE  & 17:24:45.27 -34:11:25.7 & $50\arcsec$ & 13 & 8 & 0 & 5 & $0.38 \pm 0.20$ \\
core C SE & 17:24:43.85 -34:12:01.7 & $90\arcsec$ & 43 & 34 & 6 & 19 & $0.44 \pm 0.12$ \\ 
core C NW & 17:24:41.23 -34:11:29.1 & $70\arcsec$ & 18 & 12 & 1 & 6 & $0.33 \pm 0.16$ \\
core SW & 17:24:41.34 -34:12:21.5 & $60\arcsec$ & 13 & 6 & 2 & 5 & $0.38 \pm 0.20$ \\
\hline                                   
\end{tabular}
\end{table*}

The small number of IRAC-selected Class II sources in Table~\ref{subcl:counts}
obviously reflects the lower sensitivity of this photometry
(see Sect.~\ref{SDNIRS}), given that $K_{s} = 13.5$ samples 
mostly intermediate-mass stars and only the brightest of the lower-mass Class II sources.
The fraction of sources with a circumstellar disk that we find is always larger than the
27 \% obtained in Sect.~\ref{age:24}, 
which is plausible (taking into account that 27 \% may be a lower limit).
The results do not depend much on the source selection criteria. 
In fact we obtained only slightly smaller values by either including sources up to $K_{s} = 15.5$
or adding the further criterion $0.2 \leq H - K_{s} \leq 1$.
Therefore, the fraction obtained should not be affected significantly by lower mass
stars whose brightness raises them above the adopted $K_{s} = 13.5$ limit due to their NIR excess.
In conclusion, the fractions of star-disk systems
in the five areas are consistent with each other within the errors and point
to a global value between $0.3$--$0.6$, which would imply ages between 2--4 Myr
(Haisch et al.\ \cite{haisch}), consistent with the inferred age of Pismis~24.

If the derived fractions were taken at face value, the area towards the H\textsc{ii} region would be 
younger than sub-cluster core C SE, which harbours the most massive stars in Pismis~24.
Sub-cluster core C SE in turn would appear younger than the surrounding areas.
However, a few sources of bias, as well, must be considered. 
The most massive stars, towards core C SE, are saturated in the SofI image 
and a few close-by fainter
NIR sources could not be measured (we remind the reader, e. g., of the 15 X-ray 
sources without a NIR
counterpart located close to the two most massive stars). 
On the other hand, the extinction gets higher towards the H\textsc{ii} region
and the sampling of 4.5 $\mu$m sources  gets worse.

The fraction of disk-star systems listed in Table~\ref{subcl:counts} is consistent with
those derived by Fang et al.\ (\cite{fang}). However, we do not find evidence of a
significant decrease towards the two most massive stars as they do. 
As we have shown,
the computation has to accurately take into account the issues of completeness
and contamination in the different bands.   

More insight into the differences between sub-cluster populations can be gained
by examining in detail all the sub-regions. This is done in App.~\ref{ind:so}
along with a discussion on a few individual objects. O-type  stars are only contained in
the south-eastern part of sub-cluster core C (except possibly one in core SW),
whereas the most massive stars in the other sub-clusters 
are at most early B stars.
Thus, there are no intense ionising sources inside the H\textsc{ii} region,
as instead was suggested by Felli et
al.\ (\cite{felli:90}), who misplaced two NIR sources (see App.~\ref{ind:so}).
This conclusively rules out the scenario proposed
by those authors, where the ionisation of the H\textsc{ii} region was due to embedded sources
lying in the ionised gas.
The extended cluster halo underlying the sub-clusters may be composed of low-mass stars
drifting out of the core. For typical dispersion velocities of $\sim 2$
km s$^{-1}$, one in fact expects to find stars up to $\sim 4\arcmin$ from the core
of Pismis~24 in $\sim 1$ Myr.

Assuming that the sub-clustered structure is real, one might ask 
whether this can reflect inhomogeneities that were already present 
in the parental molecular cloud. 
Parker \& Meyer (\cite{Par:Mey}) performed numerical simulations of
clusters of 1000 members, showing that sub-structured distributions with a radius of 1 pc
collapse to a central concentration after 1 Myr if they are initially subvirial
(i. e., with a virial ratio $T/|V| < 1$).
However, they remain sub-structured even after 5 Myr if they are originally supervirial
(i. e., with a virial ratio $T/|V| > 1$),
as expected in case of fast gas removal. Since these supervirial clusters expand quickly,
the presence of substructures within $\sim 1$ pc in the Pis24 core is consistent with
a scenario in which
a young ($< 5$ Myr) stellar population formed in different nearby gas clumps,
quickly dispersed by the ionising photons from the most massive stars. These clumps 
should have had sizes of the order of few $0.1$ pc. 

\subsection{Star formation on the smaller scale: stellar population of Pismis~24}
\label{star:pop}

Although the massive members of Pismis~24 are main sequence stars, its low-mass members
are PMS stars. This is typical of very young ($\sim 1$ Myr) clusters; for example 
the Orion Nebula Cluster (Hillenbrand \cite{hillen}), the centre of NGC~3603 (Stolte
et al.\ \cite{stolte}), or NGC~602 (Gouliermis et al.\ \cite{gouliermis}). Whether O-type stars are
the last stars to form in a cluster is still an open issue (Tan et al.\ \cite{tan}). 
This point of view is not at odds with the PMS nature
of the lower-mass stars, provided that the growth of massive stars towards the ZAMS is
very fast and takes only a small fraction of the cluster age. Russeil et al.\
(\cite{russ:12}) estimated that the starless and protostellar phases
in the development
of massive stars in NGC~6334--NGC~6357 have statistical lifetimes of
$\sim 0.01$ Myr and $0.1$ Myr, respectively. If so, high-mass stars can form, grow and
start their main-sequence phase while already-formed low-mass stars are still
contracting towards the ZAMS. 

As for the stellar content of Pismis~24, our derived IMF includes 11 (if 
$3.2 < A_{V} < 7.8$) to 22 (if $3.2 < A_{V} < 15$) stars with $M > 14-15$ $M_{\sun}$
(i. e., early B to early O stars), irrespective of the adopted age or evolutionary
tracks. Massey et al.\ (\cite{massey01}) list 8 earlier than O9 stars in the cluster.
Wang et al.\ (\cite{wang07}) review the total number of massive stars quoted in
the literature and list 17 known earlier than B0 stars plus 13 candidate O stars based
on their X-ray luminosity. Of these, 11 plus 4 lie inside the SofI field, which is
consistent with the number from our derived IMF. Interestingly, most of the remaining 
high-mass stars are located south of Pis24 core, and may be associated with
Pis24 S (see Fig.~\ref{map:irac}). 

The number of stars in the cluster core down to $0.1$ $M_{\sun}$ can be derived from either the 
surface density of $3.6$ $\mu$m
sources (see Table~\ref{irac:clust:tab}) or the surface density of $K_{s}$ sources
(see Table~\ref{pis24:clust:tab}), by assuming the mass completeness limits obtained
for an age of 1 Myr and the IMF prescribed by Scalo (\cite{scalo}). 
We found $\sim 2300$ stars from the counts at $3.6$ $\mu$m and
$\sim 350$ stars from the counts in the $K_{s}$ band. The difference is probably due
to both contamination from Class II sources mimicking intermediate-mass star 
fluxes at $3.6$ $\mu$m, and an overestimate
of the average field star density in the much smaller SofI field.
Alternatively, the total number of members can be estimated from the high-mass 
end of our derived IMF.
Considering that we have found 28 (for $3.2 < A_{V} < 7.8$) to 85 (for $3.2 < A_{V} < 15$)
stars down to $\sim 6.5$ $M_{\sun}$ (assuming an age of 1 Myr), a Scalo IMF yields
$\sim 3600$ to $\sim 11000$ members in the Sofi field. This is consistent with
$\sim 10000$ (in fact $\sim 5000$ if the distance is shortened to $1.7$ kpc) found
by Wang et al.\ (\cite{wang07}) by scaling their derived X-ray luminosity function 
(obtained from sources on a larger area than our SofI field) to that of the ONC.
Thus, the stellar mass of Pis24 core amounts to $\sim 2-6 \times 10^{3}$
$M_{\sun}$ for a Scalo IMF. 
Lima et al.\ (\cite{lima}) actually derive a lower mass, $533 \pm 50 M_{\sun}$, which would
imply that most of the cluster mass is in massive stars. Incidentally, this is 
the lowest mass estimate quoted 
in the literature; nevertheless it highlights how sensitive this numbers are to
the decontamination technique adopted.

\subsection{Star formation on the smaller scale: dynamical properties of Pismis~24}

One interesting property of Pismis~24 is the possible early ejection of massive stars
from the cluster.
In particular, we note that the O-type star \object{Cl Pismis 24  13}/N78~36
(see Fig.~\ref{pismisJHK}) is located on the opposite 
side of sub-cluster core C SE with respect to the O stars \object{Cl Pismis 24 2} 
and \object{Cl Pismis 24 3}, 
all roughly aligned along a line passing through core C SE. The ejection of massive stars
from clusters may happen due to three-body encounters, in which massive binaries are 
involved (see, e.\ g., Zinnecker \& Yorke \cite{zin:yo}). 
Massive stars located on a line roughly centred on the cluster core may
indicate momentum conservation in one of these episodes. This should become clear
by measuring their radial velocities. In addition, Gvaramadze et al.\
(\cite{gvaramadze}), in a systematic search for runaway OB stars associated
with NGC~6357, listed four other distant B0 to O V stars that may have been 
ejected from Pismis 24, 
three more B0 to O V stars that may have been ejected from AH03J1725-34.4,
and one O6 V star possibly ejected from either of them. The Wolf-Rayet star
WR93, $4\arcmin$ east of Pis24 core, could have been ejected, as well
(this would cause an age inconsistency according to some authors,
but would not be an issue according to Massey et al.\ \cite{massey01}).
This early ejection of massive stars would indicate both
that the early phases of the two clusters were highly dynamically unstable,
and a high fraction of pristine binaries among massive stars. By the way,
the two most massive members of Pismis~24 are known to be binary stars.   

\subsection{Star formation on the smaller scale: from gas to stars}

To infer the properties of the gaseous environment from which Pismis~24
originated, we start by noting that, from Fig.~\ref{nir:map}, the most
massive members of Pismis~24 lie at a projected distance
$\sim 1\farcm\ 4$ ($\sim 0.8$ pc) from the ionisation front in the north. The gas facing
these stars (with pillars, see Fig.~\ref{pismisJHK}) 
reminds one of the inner wall of the cavities in the simulations by
Mellema et al.\ (\cite{mellema}) and Arthur et al.\ (\cite{arthur}).
Pismis~24/G353.2+0.9 is also somewhat reminiscent of other young open clusters 
interacting with nearby gas such as
\object{NGC~602}/N90 in the \object{SMC} (e. g., Gouliermis et al.\ \cite{gouliermis}).

The molecular gas north of Pismis~24 
is likely to be the remnant of a larger complex where the cluster was born.
Giannetti et al.\ (\cite{giannetti}), summarising both their own and previous observations,
confirmed that the molecular gas is being eroded and pushed aside by the intense UV field.
Hence, Pismis~24 has quickly cleared off its parental cloud, whose gas must have been ejected to
the south through a champagne flow. 
A sketch of the primordial environment of the cluster could
be the following. Let us assume that the H\textsc{ii} region has evolved
in a medium of constant density with a radius equal to the distance between the most
massive stars and the ionisation front ($\sim 1$ pc). This would encompass
all three sub-clusters of Fig.~\ref{nir:map}. If we take the age of the cluster
($\sim 1$ Myr) as the age of this idealised H\textsc{ii} region, we can use Eq.~(12:20)
of Spitzer (\cite{spitzer}) to derive the Str\"{o}mgren radius ($\sim 0.02$ pc). 
In turn, from the Str\"{o}mgren radius 
we can derive the average medium density ($n_{\rm H} \sim 10^{6}$ cm$^{-3}$).
Such a cloud would have had a mass $\sim 10^{5}$ $M_{\sun}$.
Gendelev \& Krumholz (\cite{genkru}) showed that the ionisation front
expands only slightly faster in a blister-type H\textsc{ii} region compared with a
spherically-symmetric one, so this does not affect the final result significantly.  
Using the total stellar mass derived in Sect.~\ref{star:pop}, the star formation in this cloud
would have occurred with an efficiency of $\sim 2-6$ \%. This efficiency 
is of the same order of that usually measured in molecular clouds as a whole
(Padoan et al.\ \cite{padoan}), but
could be even larger if the stars formed in several clumps reflecting the current
sub-clustered structure, hence with sizes less than a few $0.1$ pc. In this case,
the expansion of the H\textsc{ii} region could be delayed to match the cluster age 
by increasing the average clump density (embedded inside a lower density gas),
by invoking UV radiation absorbed by dust (which would imply lower gas densities),
or by gas inflow (Dale \& Bonnel \cite{dale:11}).
The remaining gas must have been quickly cleared via champagne flows. 
Some of the properties of the molecular clouds in the north
should still reflect those of the cluster parental cloud. Giannetti et al.\
(\cite{giannetti}) found that the molecular gas is distributed in clumps
$\sim 0.2$ pc in size, with average densities $n(\rm H_{2}) \sim 10^{3} -
10^{5}$ cm$^{-3}$ and masses $\sim 10$ to few $100$ $M_{\sun}$
(once scaled to our assumed distance).  
The densest and most massive of them may be similar to 
those where the cluster stars formed.

Are there any hints that star formation in parts of Pismis~24 has been
triggered by the feedback from the massive stars?  Numerical simulations
show that photo-ionisation from massive stars inside molecular clouds
generally leads to a slight decrease in the {\it global} star formation efficiency,
with gas expulsion partly compensated for by triggered star formation
(e. g., Dale et al.\ \cite{dale:12a}, \cite{dale:12b}). Nevertheless,
they show that triggered star formation does occur, and amounts to a few \%
of the total cloud mass (see also Walch et al.\ \cite{walch:11}). 
Dale et al.\ (\cite{dale:12b}) find that the association of young stars with
gas structures such as shells or pillars, as occurs in NGC~6357,
is a good but not conclusive indication of triggered star formation.
We could not detect significant age differences in different areas of Pis24 core
based on the fraction of stars with a circumstellar disk, that could have pointed
to sequential star formation as well. 
However, this fraction appears to be at least marginally higher towards the H\textsc{ii} region 
where other indications of triggered
star formation have been found. Figure~\ref{yso:dist:map} shows that
young stars  are embedded in the H\textsc{ii} region. 
The elephant trunk with the young star IRS 4 at its tip
is another sign of possible triggering. As discussed in App.~\ref{ind:so}, this region
hosts less massive stars than sub-cluster core C 
which may be an effect of photo-ionisation (Dale et al.\ \cite{dale:12b}).
Giannetti et al.\ (\cite{giannetti})
found that many of the gas clumps in the north are virialised and/or exhibit
possible compression from outside, which may lead to further star formation.


The elephant trunk with IRS 4 at the tip inside G353.2+0.9
suggests that the 
so-called ``radiation-driven implosion''
(see Bisbas et al.\ \cite{bisbas:11} and references therein) may be under way.
When an ionisation front crosses a dense clump, it generates a shock
which compresses the clump making it collapse under some conditions.
Westmoquette et al.\ (\cite{west:10}) in fact found evidence of a D-shock
propagating into the elephant trunk.  The
numerical simulations by Bisbas et al.\ (\cite{bisbas:11}) nicely show
how pillars with newly-born stars at the tip can be produced by this
mechanism. Interestingly, they found that more intense ionising fluxes
should disrupt less dense clumps rather than make them collapse. 
However, these authors considered clumps whose masses 
($15$ $M_{\sun}$ at most) are much less than
that of the elephant trunk (a few $100$ $M_{\sun}$ according to Giannetti
et al.\ \cite{giannetti}), so their results cannot be easily compared
with our observations. Alternatively, Gritschneder et al.\ (\cite{grit})
showed that pillars may naturally arise in turbulent clouds exposed to
UV radiation, which acts to enhance the initial turbulent density
distribution.
In conclusion, although we cannot conclusively prove that triggered star
formation is in progress towards G353.2+0.9, this seems quite likely.

\section{Conclusions}
\label{conclusioni}

We have retrieved archival IRAC/Spitzer data and studied the young star population
in the star forming region NGC~6357, located in the Galactic
plane ($d \sim 1.7$ kpc). We complemented these data with
new $JHK_{s}$ observations, archival HST/WFPC2 images, and CHANDRA 
X-ray observations available in the literature,
and focused on the stellar population (age, IMF, evidence of multiple generations of stars)
in the young open cluster Pismis~24, which is part of NGC~6357.
Since the region is only few degrees from the Galactic Centre
in projection, we had to
carefully assess the various effects that can bias the results, in
particular when comparing photometry from different instruments
at different wavelengths. We showed that the IRAC/Spitzer photometry is
confusion-limited and, at longer wavelengths, suffers from saturation
in the areas with intense diffuse emission (i. e., from the UV-illuminated
surface of molecular clouds). We also showed that the $JHK_{s}$ photometry,
IRAC/Spitzer photometry and high-resolution X-ray observations do
not sample the stellar population homogeneously, but are complete only
down to different stellar masses. In the case of IRAC/Spitzer photometry,
the completeness gets even worse when including either of the bands at $5.8$
and $8.0$ $\mu$m.
Having considered all these effects, we obtained the following results:
   \begin{enumerate}
      \item The Spitzer/IRAC photometry on the large scale shows another two stellar
	clusters associated with NGC~6357, i. e., the cluster
	AH03J1725--34.4 (associated with the H\textsc{ii} region G353.1+0.6),
	and the cluster we named ``B'' (associated with the H\textsc{ii} region G353.2+0.7). 
	Class II sources (identified from IRAC colours) also cluster towards 
	these three sites, suggesting a stellar population
	of very young stars in all three cases. All have sizes
	of $\sim 2.5$ pc and exhibit sub-clustering on scales down to
	$0.5$--1 pc. 
	We estimated total numbers of stars of a few
        hundreds in each cluster down to the completeness limit. 
 	If the cluster ages are roughly the same
        (assumed $\sim 1$ Myr), the stars are counted down to $\sim 2$ $M_{\sun}$
	and all three clusters must contain similar total numbers
	of members. 
      \item Both Pismis~24 and AH03J1725--34.4 lie inside cavities where
        they appear off-centred, interacting with molecular clouds
	to the north. Cluster B lies inside a cavity, as well. 
      \item We discussed briefly the NIR extinction law towards Pismis~24,
	concluding that the one derived by Rieke \& Lebofsky (\cite{r&l}) 
	should be appropriate.  The $J - H$ vs.\ $H - K_{s}$ diagram
	indicates a large population of extincted giant stars, which
	is expected given that we are looking towards the innermost parts
	of the Galaxy. 
      \item The NIR photometry shows that the core of Pismis~24,
        itself one of several other substructures, contains
        three surface density peaks, which are likely to represent
        real sub-clusters.  All the 
	most massive members lie in the same sub-cluster, whereas 
	only intermediate- to low-mass members populate the
 	other sub-clusters.
      \item Optical HST photometry confirms a cluster age of $\sim 1-3$ Myr.
      However, we found that many of the NIR counterparts of the optical stars
      (those with X-ray detection in particular) appear more massive
      than from optical photometry. This points to a large fraction of stars  
      exhibiting a NIR excess.
       \item We derived from their NIR excess (evaluated including the
        IRAC $4.5$ $\mu$m band) the fraction of stars with a circumstellar
        disk and this is in the range $0.3$--$0.6$, consistent
        with an age $\la 4$ Myr. We found no decrease of this fraction towards
        the massive stars.
       \item We derived the IMF towards the core of Pismis~24 by using
	$K$ luminosity functions. We minimised contamination by 
	field stars by selecting $HK_{s}$ sources in two different
        extinction ranges with respect to a pseudo-sequence of young stars, 
	i.  e. $3.2 < A_{V} < 7.8$ (derived from 
	optical spectroscopy) and $3.2 < A_{V} < 15$ (derived
	from the NIR counterparts of X-ray sources). We found
	an IMF consistent with a Salpeter one down to $\sim 2.5$
	$M_{\sun}$. For $M \la 2.5$ $M_{\sun}$, the IMF flattens.
      \item Assuming the IMF prescribed by Scalo (\cite{scalo}), we
	estimated that Pismis~24 contains $\sim 3600$ to $\sim 11000$
	stars ($2-6 \times 10^{3}$ $M_{\sun}$), depending on the extinction 
        interval used to select the cluster members. 
      \item There are indications of triggered star formation towards
	the H\textsc{ii} region G353.2+0.9, where the UV radiation from
	the core of Pismis~24 interacts with the molecular cloud.
   	We confirmed the results of Giannetti et al.\ (\cite{giannetti}), 
        who ruled out the scenario proposed by
	Felli et al.\ (\cite{felli:90}) in which the ionising sources
	of G353.2+0.9 are embedded massive stars inside the H\textsc{ii} region
	itself. In fact, we did not find any stars earlier than B-type in the area.  
      \item There are indications of an early ejection
	of massive stars from the core of Pismis~24, suggesting 
	a dynamically unstable environment after gas expulsion.
      \item We proposed a picture of the molecular gas where
	the core of Pismis~24 formed. It must have been a dense (average $n[{\rm H_{2}}]
	\sim 5 \times 10^{5}$ cm$^{-3}$) cloud, $\sim 1$ pc in radius.
	Or, most probably, a group of denser clumps
          ($n[{\rm H_{2}}] > 5 \times 10^{5}$ cm$^{-3}$) $\sim 0.1$ pc
	in radius, embedded in a lower density gas ($n[{\rm H_{2}}] < 5 \times 10^{5}$ 
        cm$^{-3}$). 
   \end{enumerate}

\begin{acknowledgements}
This work is based in part on observations made with the 
Spitzer Space Telescope, which is operated by the Jet Propulsion Laboratory, 
California Institute of Technology under a contract with NASA.
It is also partially based on observations made with the NASA/ESA Hubble Space Telescope, 
obtained from the Data Archive at the Space Telescope Science Institute, which is operated 
by the Association of Universities for Research in Astronomy, Inc., under NASA contract 
NAS 5-26555. These observations are associated with program \# 9091 (P. I. Jeff Hester). 
This publication makes use of data products from the Two Micron All Sky Survey, which is a 
joint project of the University of Massachusetts and the Infrared Processing and Analysis 
Center/California Institute of Technology, funded by the National Aeronautics and Space 
Administration and the National Science Foundation.
FM and EdC acknowledge partial funding by the PRIN INAF 2009 grant CRA 1.06.12.10 (“Formation and
early evolution of massive star clusters”, P.I. R. Gratton).
EdC is also grateful for financial support from PRIN INAF 2011
("Multiple Populations in Globular Clusters: their role in the Galaxy
assembly", P.I.: E. Carretta).
\end{acknowledgements}

\Online

\begin{appendix}
\section{Spitzer/IRAC data reduction}
\label{sp:ir:dr}

We downloaded all the basic calibrated data (BCD) 
and corrected BCD (CBCD) 
products, which were provided by the Spitzer Science
Center through their standard data processing pipeline version S18.18.0.
Data obtained from S18.18.0 need to be corrected in the 5.8 and
8.0 $\mu$m bands due to an error in calibration, so they have been
multiplied by $0.968$ and $0.973$, respectively, following
the prescriptions given in the web page\footnote{See
http://irsa.ipac.caltech.edu/data/SPITZER/docs/irac/}.
For each raw image, the pipeline produces
a flux-calibrated image (BCD),
where all well-understood instrumental signatures
(e. g., muxbleed) are removed (and including dark-subtraction and flat-fielding). 
The pipeline also attempts to systematically find all saturated point 
sources in the images and fit them using an appropriate PSF that is 
matched to the unsaturated wings of the source. A dedicated software module replaces 
the saturated point source with an unsaturated point source that has the 
correct flux density of the point source. We checked that this works quite 
well on the short-exposure frames but not on the long-exposure ones.
Another pipeline module performs artefact correction (stray light, saturation,
muxstripe, column pulldown and banding) on the BCD files, creating
``corrected'' BCD (CBCD) files. For details, see the on-line
instrument handbook\footnote{http://irsa.ipac.caltech.edu/data/SPITZER/docs/irac/ iracinstrumenthandbook/}.

The observations were performed in high dynamic range (HDR) mode, i. e., 
two images were taken at each position, a short exposure one ($0.4$ s) and a long
exposure one (10.4 s). This allows one to also obtain photometry for bright
sources avoiding saturation.  The observed area was covered by using
the IRAC mapping mode, with a grid of $10 \times 8$ positions in
``array'' coordinates, spaced by $260 \arcsec$ 
(i. e., slightly less than the detector size) in both directions
defined by the detector axes.  The selected dithering
mode was ``cycling'', with a scale factor ``medium'' and a number of
positions equal to 5, i. e., for each grid position, 5 pairs 
(the short- and  long-exposure integration) of 
dithered images were taken, randomly distributed up to a maximum distance
of 119 pixels and with a median distance of 53
pixels. Therefore, most of the points in the mosaic result from
the overlap of up to five images,
although small areas can be covered by more than 5 images (where
the dithered patterns of adjacent grid positions 
overlap) and few smaller areas may be covered by less than 5
images.  

Mosaicking and co-adding of all frames were carried out by using MOPEX
version 18.4.9,
yielding a pair of larger images (short-  and long-exposure)
per band with a pixel size $0\farcs 6 \times 0\farcs 6$ (about half
the native pixel size). We used a dual-outlier rejection strategy with
conservative settings and experimented different values of the
parameter RThreshold. The long-exposure CBCD frames at $5.8$ and $8.0$
$\mu$m displayed intense stripes associated with bright sources,
absent in the corresponding BCD frames. We found out that these are features 
caused by the artefact-mitigation pipeline. Consequently, we constructed
the long-exposure $5.8$ and $8.0$ $\mu$m images using the BCD files rather than the CBCD ones. 
On the other hand, as for the short-exposure mosaics, only a few CBCDs  at $8.0$ $\mu$m were
found heavily affected by the problem and replaced by the
corresponding BCDs. In all other cases we used the CBCDs.  

\section{Spitzer/IRAC photometry}
\label{sp:ir:ph}

We checked that any artefact had been removed efficiently from the 
final mosaicked images, except for the bandwidth effect in the long-exposure
$5.8$ and $8.0$ $\mu$m ones. This gives rise to ghosts associated with the brightest sources.
However, the corresponding artefacts were clearly recognisable and
removed from the final point-source photometry. Some optical banding, as well, due to internal 
radiation scattering in the arrays, is present in the long-exposure
$5.8$ and $8.0$ $\mu$m images yielding a small faint halo 
around the brightest areas. Being a diffuse structure, however, it is not expected to 
affect the final photometry significantly. 

Photometry was performed both on the
short-exposure images and on the long-exposure ones.
In each band, we set a saturation
limit to the long-exposure mosaic, conservatively derived by estimating the saturation limit in a single BCD following the
IRAC Instrument Handbook. The measured fluxes were converted into magnitudes by using
the zero magnitude flux densities given in Table~4.1 of the
IRAC Instrument Handbook (also given in Reach et al.\ \cite{reachetal}). 
The errors were calculated by using Eq.~(7) of Reach et al.\ (\cite{reachetal}),
i. e.,  neglecting electronic noise. We discarded a number of residual false detections 
by visual inspection of the images, while others were subsequently removed
by operating a colour-based selection during the analysis.

Short- and long-exposure photometry were
matched in each band, adding the bright sources from the former 
to the faint sources from the latter according to the following criterion.
We estimated a magnitude threshold 
for each band by requiring both that the average signal to noise ratio at the shorter exposure 
time was high enough, and that its corresponding flux
was well below the single BCD saturation
limit set at the longer exposure time.
The sources with flux above the threshold
were then taken from the short-exposure photometry, whereas 
those with flux below it were taken from the long-exposure photometry.
The chosen thresholds are [3.6]$=10.5$, [4.5]$=10$, [5.8]$=8$
and [8.0]$=8.5$. 

We cross-checked the photometry from the long- and short-exposure mosaics
in each band $I$ by computing the difference in magnitude of all sources
retrieved in both of them (with photometric
uncertainties less than $0.3$ mag). 
We plotted the histogram of the difference $\Delta m$
between long- and short-exposure measurements in bins
of $0.5$ $[I]$-magnitudes, sampling the sources over two different
image test areas. One area was 
chosen well inside the large ring-like structure (hence with little diffuse emission) and 
the other one inside the G353.2+0.9 projected area (with the most intense diffuse emission).
For brighter magnitude bins, the histograms appear to peak at values that are 
within $0.05$ mag from 0 (with a trend of the long-exposure measurements to be brighter)
following distributions resembling Gaussian curves. 
However, when moving towards fainter magnitudes, the distributions become less peaked, their 
width
increase, and a wing develops on the side where long-exposure measurements are brighter 
than short-exposure ones. The dispersion around the peak is always
narrower than the typical photometric error inside each magnitude bin,
This flatter distribution in the faint bins 
is more evident towards the test area with large diffuse emission,
although it does not seem to be correlated with the presence of diffuse emission.
This may be due either to the short-exposure photometry being nearer to its sensitivity limit, and/or
to a better sampling of the sky around point-sources for longer exposures
(i. e., better signal-to-noise). In addition, the distributions
for the $8.0$ $\mu$m band
do not exhibit the wing and are narrower than in the other bands. It appears that
we used a
DAOFIND finding threshold larger at $8.0$ $\mu$m  than in the other bands,
in units of one sigma of effective noise, leading to a smaller overlap
between short and long exposures. 

Whatever caused this trend in the distributions of magnitude difference,
the difference between long- and short-exposures
is of no concern at all at bright magnitudes (where the final
photometry is taken from the short-exposure mosaics), while we assume that
anyway the fainter fluxes are better characterised in the long-exposure mosaics
(yielding higher signal to noise ratios).

A few sources (much less than 1 \% of the total) exhibit large differences between short- 
and long-exposure magnitudes (larger than 2--3 times the width of the
$\Delta m$ distribution). In most cases, either they fall on some artefacts
in one of the two mosaics or they are residual cosmic rays passing the outlier
rejection stage of MOPEX. These sources were either removed or were assigned a 
photometric estimate from the image unaffected by the artefact. This further confirms that 
residual contamination from artefacts or cosmic rays is 
very low, well below $\sim 1$ \% of the total sample in each band.

\begin{figure*}
   \centering
   \includegraphics[width=10cm]{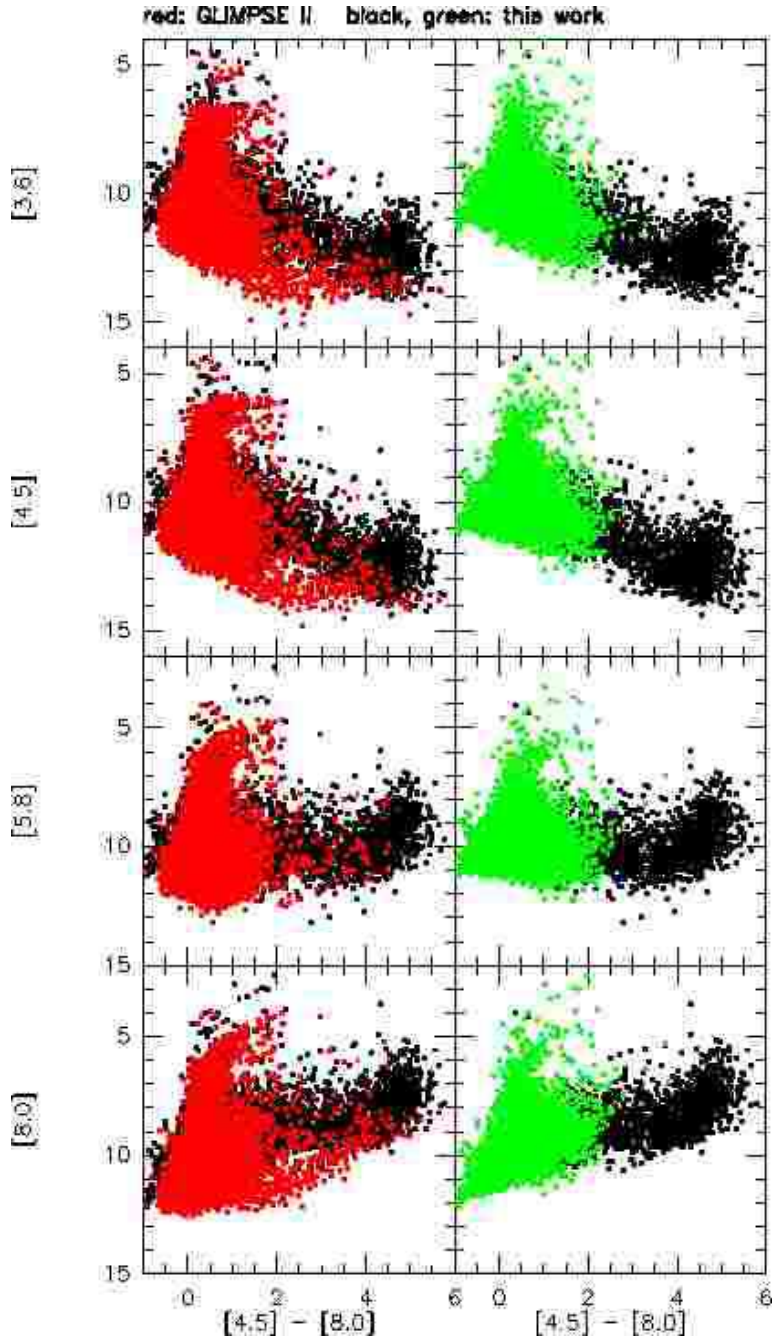}
   \caption{Magnitude-colour diagrams showing the magnitude
	in the various IRAC bands of the sources from the sample
        obtained in this work (black dots) as a function of [4.5]--[8.0]. 
        Panels on the same row display, from top to bottom, 
        the measurements for these sources in the
	$3.6$, $4.5$, $5.8$, and $8.0$ $\mu$m bands, respectively. 	
	In the left column panels, the sources from our sample
	are overlaid with those from GLIMPSE II (red dots). The green
	dots in the right column panels are those remaining from our photometry after
	removal of sources contaminated by PAH emission. Only objects detected in
	all four bands with photometric errors $< 0.3$ mag are plotted.
              \label{fig:cont1234}}
    \end{figure*}

Another consistency test that we performed consisted in comparing the photometry from this work with that
from the GLIMPSE survey. We retrieved all sources from the GLIMPSE II (epoch 1) archive
within $30 \arcmin$ from RA$=17$h $25$m and DEC$=-34\degr 18\arcmin$. 
The area covered by our data is smaller than that
and lies well inside the $30 \arcmin$ radius.
In the magnitude-colour diagrams of Fig.~\ref{fig:cont1234} showing magnitudes vs.\ [4.5]--[8.0], 
we have overlaid (left column) the GLIMPSE datapoints (red) and those from this work (black).
We have selected only objects detected in all four bands with 
photometric errors $< 0.3$ mag from the two samples.
It appears that the two photometry lists as a whole are consistent with each other, apart for
a large group of sources with 
[4.5]--[8.0]$>2$ much more numerous in our sample. These are sources contaminated by
PAH emission arising in the bright filaments that cross the region
(see Sect.~\ref{app:cont}). When these objects
are filtered out as explained in Sect.~\ref{app:cont}, 
the remaining ones (green dots, in the right column boxes) are distributed 
in the mag-colour diagram like the GLIMPSE datapoints. 

\begin{figure*}
   \centering
   \includegraphics[width=10cm]{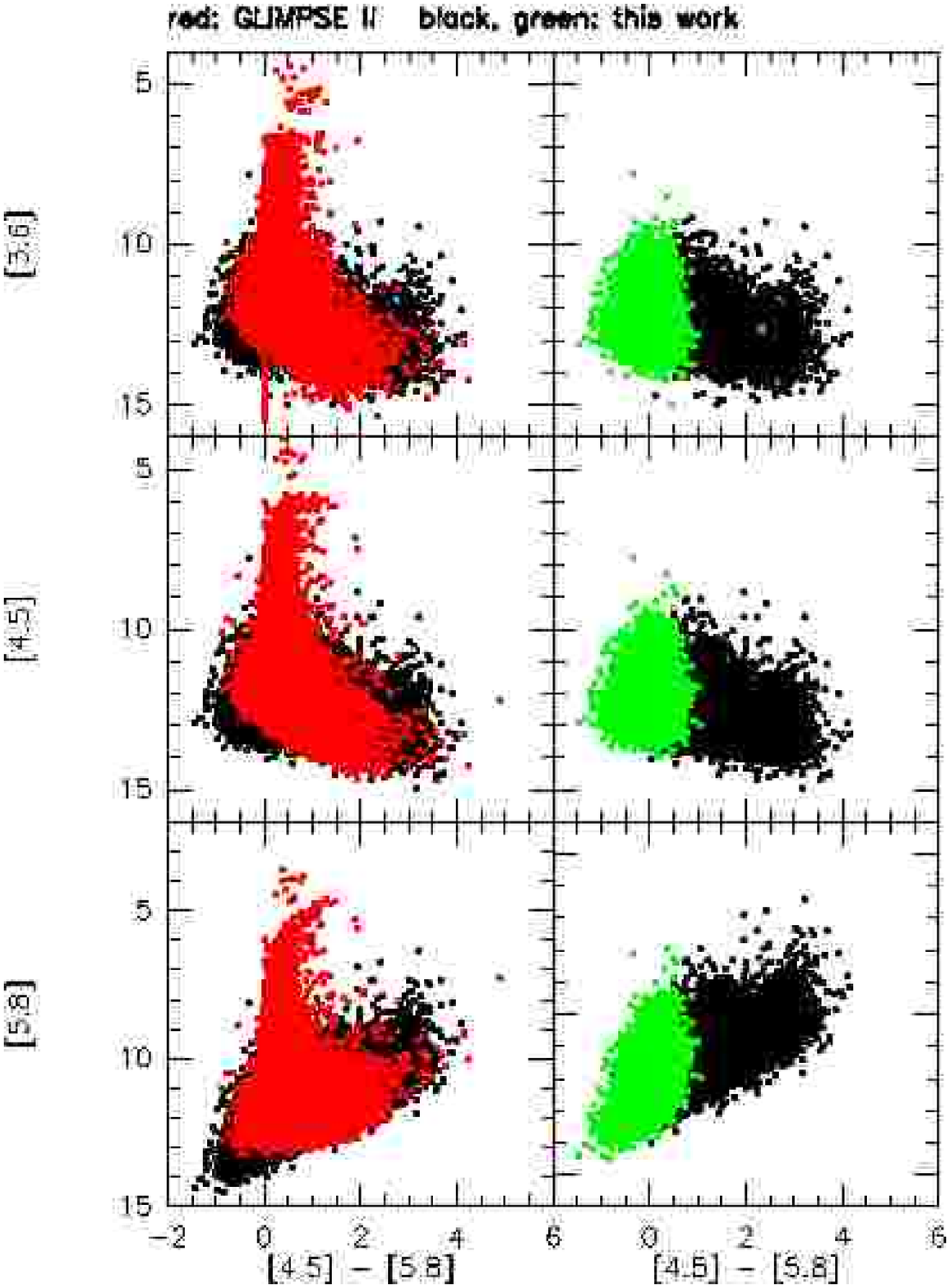}
   \caption{Magnitude-colour diagrams showing the magnitude
	of all sources from the sample obtained in this work (black dots) 
	{\em only} detected in the first three wavelength IRAC bands
	(i. e., $3.6$, $4.5$ and $5.8$ $\mu$m).
        Panels on the same row display, from top to bottom, 
        the measurements for these sources in the
	$3.6$, $4.5$, and $5.8$ $\mu$m bands, respectively, 	
	plotted
	as a function of [4.5]--[5.8]. In the left column panels,
	they are overlaid with all sources from GLIMPSE II (red dots)
        {\em at least} detected in the three shorter wavelength bands. The green
	dots in the right column panels are the sources remaining from our photometry after
	removal of those contaminated by PAH emission. Only objects 
	with photometric errors $< 0.3$ mag in the three shorter wavelength bands
	are plotted.
              \label{fig:cont123}}
    \end{figure*}

We repeated the same test by selecting all sources from our database with detections {\em only}
in the first three bands ($3.6$, $4.5$ and $5.8$ $\mu$m). 
In the magnitude-colour diagrams of Fig.~\ref{fig:cont123} 
their magnitudes are plotted vs.\ [4.5]--[5.8].
We have overlaid these (in black) with 
(boxes in the left column) the GLIMPSE datapoints with detections in {\em at least}
the first three bands (red).  We have selected only objects with
photometric errors $< 0.3$ mag in the three shorter wavelength bands from the two samples.
In this case the two photometric lists appear to be consistent, as well, apart from a slight
excess of red
sources in our sample with respect to the GLIMPSE sample. By removing all sources 
contaminated by PAH emission as explained in App.~\ref{app:cont} and plotting the
remaining ones (green dots, boxes in the right column), the excess of red sources disappears, 
confirming its origin. We note that the brightest sources in our sample with detection 
only in the first three bands have a
magnitude $\sim 10$ in all bands. Since stars have colours $\sim 0$ mag in all IRAC bands,
this indicates that the sources with detections only in the three shorter wavelength bands
are mostly stars too faint to be detected at $8.0$ $\mu$m, where the sensitivity is 
$[8.0] \sim 10-12$ (see bottom row of Fig.~\ref{fig:cont1234}).

\begin{figure*}
   \centering
   \includegraphics[width=10cm]{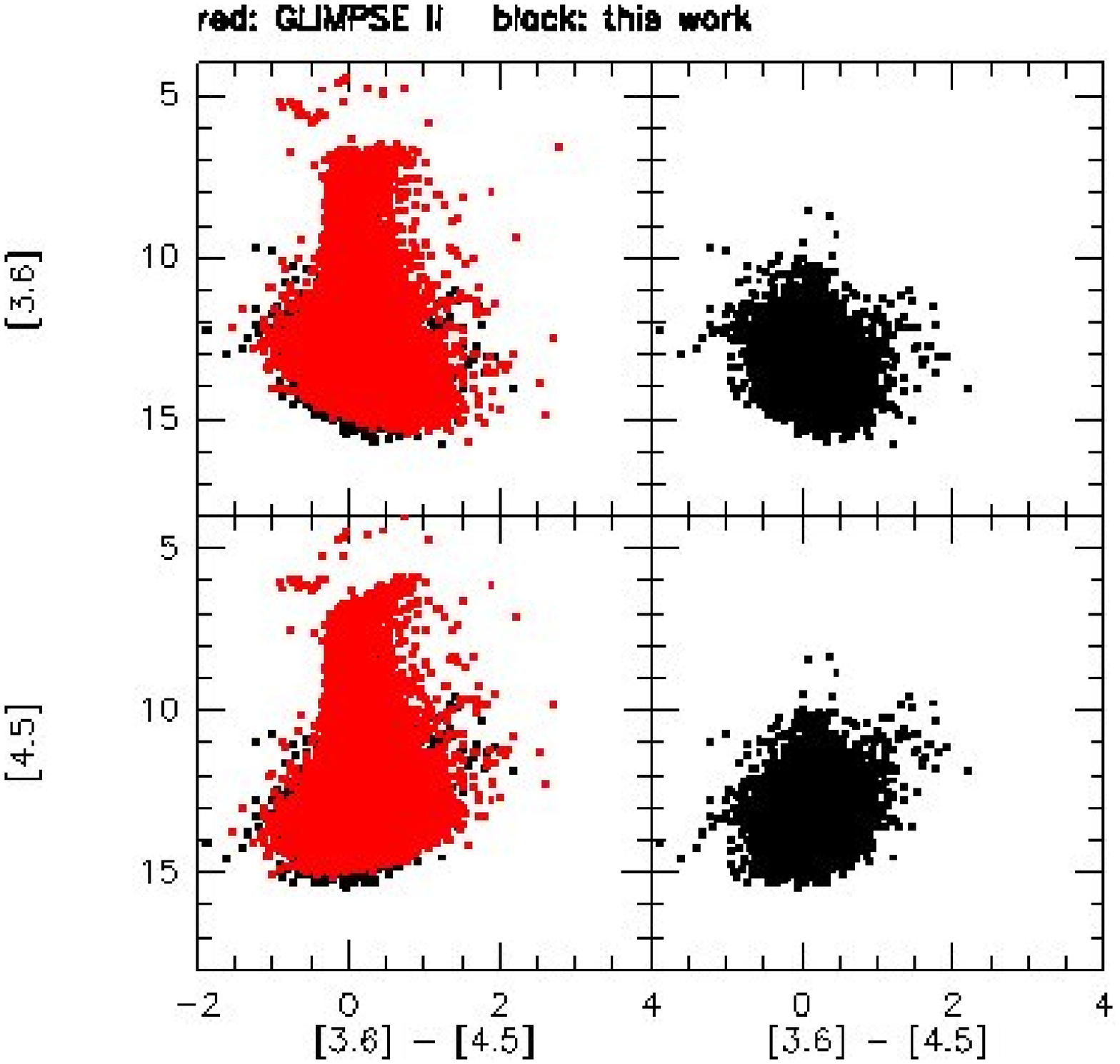}
   \caption{Magnitude-colour diagrams showing the magnitudes
	of all sources from the sample obtained in this work (black dots) 
	{\em only} detected in the two shorter wavelength IRAC bands
	(i. e., $3.6$ and $4.5$ $\mu$m).
        Panels on the same row display, from top to bottom, 
        the measurements for these sources in
	the $3.6$ and $4.5$ $\mu$m band, respectively, 	
	plotted
	as a function of [3.6]--[4.5]. On the left column panels,
	they are overlaid with all sources from GLIMPSE II (red dots)
        {\em at least} detected in the two shorter wavelength bands.  Only objects 
	with photometric errors $< 0.3$ mag in the two shorter wavelength bands
	are plotted.
              \label{fig:cont12}}
    \end{figure*}

Finally, in Fig.~\ref{fig:cont12} we plot magnitudes vs. [3.6]--[4.5] for all
sources from our database with detections {\em only} in the first two bands
and photometric errors $< 0.3$ mag (black dots). These are overlaid 
(boxes in the left column) with datapoints
from the GLIMPSE sample with detection in {\em at least} the first two bands
(red dots). The two photometric lists appear again to be consistent with each other. 
The brightest sources from our database
are a bit fainter than those in Fig.~\ref{fig:cont123}, suggesting that sources
detected only in the first two bands are again mostly stars too faint to
be detected in the two longer wavelength bands.

\begin{figure*}
   \centering
   \includegraphics[width=10cm]{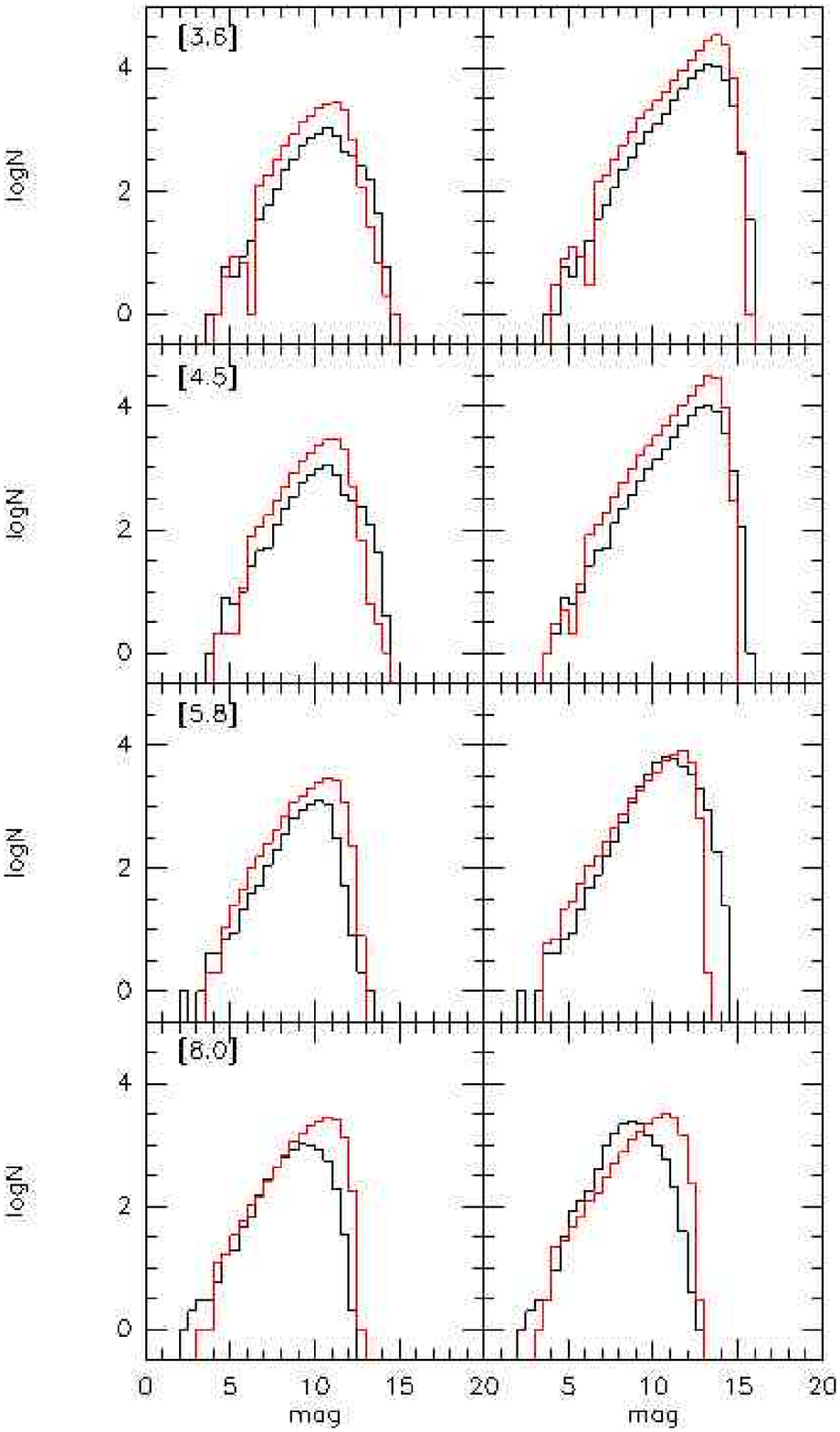}
   \caption{Number of sources per magnitude bins of $0.5$ mag, from
	our database (black line) and the GLIMPSE sample (red line).
	The boxes in each row refer to the band indicated on the
	upper left corner.
	For the diagrams on the left column we used the sources detected
 	in all four IRAC bands with photometric errors $< 0.3$ mag,
	whereas for those on the right column we used all sources
	detected in at least the corresponding band (with photometric errors $< 0.3$ mag
	in that band).
              \label{fig:histo}}
    \end{figure*}

One striking thing that appears clearly from Figs.~\ref{fig:cont1234}, ~\ref{fig:cont123}
and ~\ref{fig:cont12} is that the sensitivity limits of our photometry are about
the same as those of the GLIMPSE photometry. Nevertheless, our photometry does result from images
integrated at least for $10.4$ s and, on average, corresponding to a total exposure time
of 52 s. On the other hand, the GLIMPSE II (epoch 1) photometry is produced using two
images per position with exposure times $1.2$ s each. This property of the 
limiting sensitivities is demonstrated
in Fig~\ref{fig:histo}, comparing histograms of number of sources grouped in 0.5-mag magnitude
bins from our sample (black) and the GLIMPSE one (red). For the diagrams on the left column, we 
have used only the sources detected in all four bands with photometric errors $<0.3$ mag, whereas
for the diagrams on the right column we have used all sources detected in at least one 
wavelength band (with photometric errors $<0.3$ mag in that band). 
In any case, the distributions from the two samples are about equal, except
for the different number of sources retrieved (as said the selected GLIMPSE field 
is larger than the target one). Both exhibit an increase in the number of sources 
starting from low magnitudes, a turn off roughly at the same magnitude, and a steep decrease.
This clearly indicates that the completeness limit (and the sensitivity limit) in each band is roughly
the same for the two samples. 

We noticed a similar effect when comparing our short- and long-exposure photometry.
The latter is about one magnitude ($1.5$ magnitudes at $5.8$ $\mu$m) deeper than the former,
although the sensitivity values listed in the IRAC handbook would suggest a difference of at least
3--4 magnitudes. This means that our photometry is limited by a source of noise that
does not depend on the integration time. This can only be confusion due to the large number
of sources in the field, given that we are observing along a line of sight close to the
Galactic Centre. This is confirmed by Ram\'{i}rez et al. (\cite{ramirez}) who, in their IRAC survey of the
Galactic Centre, found that their photometry is limited by confusion, as well. Using images with average
total integration time of 6 s, they quote completeness limits which are even less than
those we can infer from Fig~\ref{fig:histo} for the GLIMPSE sample (with $2.4$ s total exposure time).
The histograms of number of sources vs.\ magnitude that they show, exhibit the same abrupt turn-off
at the faint end as ours. Furthermore, Fang et al.\ (\cite{fang})
achieve only slightly better sensitivities compared to GLIMPSE
using IRAC images of NGC~6357 of $10.4$ s exposure time. 
Therefore, we can conclude that our photometry is limited by source confusion.

\section{Completeness of Spitzer photometry}
\label{compl:app}

\begin{figure*}
   \centering
   \includegraphics[width=10cm]{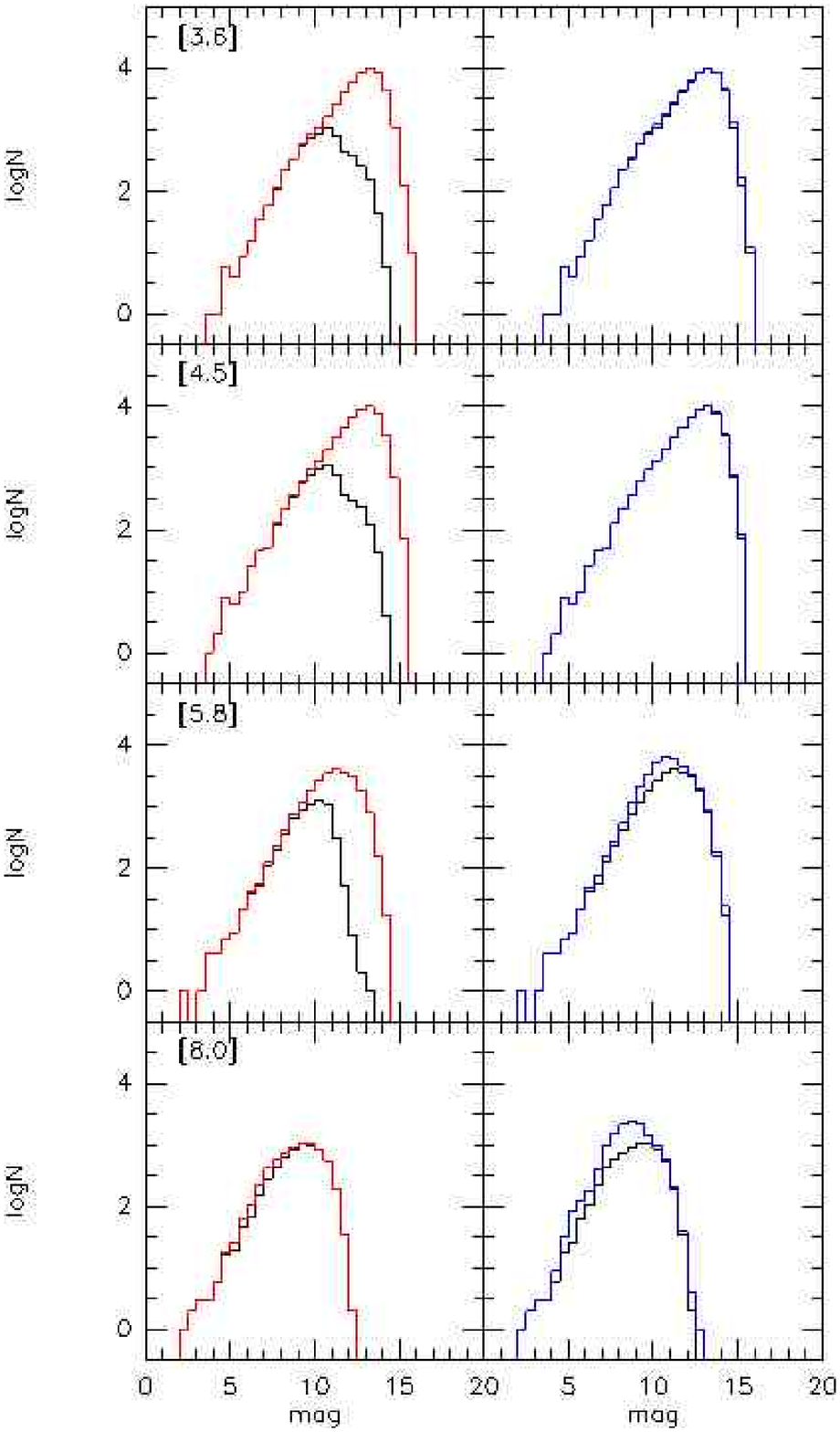}
   \caption{Number of sources retrieved in the $I$ band
        vs.\ [$I$] in bins of $0.5$ mag.
        The boxes in each row refer to the Spitzer/IRAC band $I$ indicated on the
        upper left  corner of the left-hand box. {\bf Upper two rows}:
        in the left column, we compare the $I$-band sources detected in all
        IRAC bands with photometric errors $<0.3$ mag (black line),
        and the sources detected in at least the first two IRAC bands
        with photometric errors $<0.3$ mag (red line);
        in the right column, we compare all sources detected
        in at least the first two IRAC bands,
        with photometric errors either $<0.3$ mag in both bands (black line),
        or $<0.3$ mag in at least the considered band $I$ (blue line).
        {\bf Lower two rows}:
        in the left column, we compare the $I$-band  sources detected in all
        IRAC bands, with photometric errors either $< 0.3$ mag in all bands (black line),
        or $<0.3$ mag in at least the considered band $I$ (red line);
        in the right column, we compare the $I$-band sources detected
        in all IRAC bands, with photometric errors $< 0.3$ mag
        in at least the considered band $I$ (black line),
        and the sources detected in at least the considered band $i$ with
        photometric errors $<0.3$ mag (blue line).
              \label{fig:histo_comp}}
    \end{figure*}

The photometric completeness in a band (i. e., the capability of retrieving most of the sources
up to a given magnitude in a field) mainly depends on the sensitivity of
the observations in that band. Because the sensitivity attained in the four IRAC bands
is different,
the completeness limit is fainter in the first two bands and
gets brighter by requiring simultaneous detection in more
than one band.  To study the effect, we constructed histograms of the 
magnitude distribution, in bins 0.5-mag wide, adopting several different selection criteria.
All histograms display an increase in the number of sources $N$ with increasing magnitude
up to a peak, and a subsequent  decrease due to loss of completeness.
A few examples are shown for all bands in Fig.~\ref{fig:histo_comp}.
In the following, by ``good'' detection we will mean one with a
photometric error $< 0.3$ mag. In addition, by ``completeness
degree'' we will mean the fraction of sources of a given
magnitude (irrespective of their nature) that are retrieved in the field.

To derive the completeness degree as a function of magnitude,
we first compared the histograms obtained in each IRAC band
from the following two samples of sources
(see the left column in Fig.~\ref{fig:histo_comp}). The first sample (S1) contains 
all the sources retrieved in a band $I$ with simultaneous
good detections in all 4 bands. The second sample (S2) contains 
all the sources retrieved in a band $I$ with good detections in $I$ and in at least
one adjacent band. Hence, different samples S2 can be obtained depending on
the choice of the adjacent band (e. g., we used $3.6$ and $4.5$ $\mu$m for
each of these two bands, $5.8$ and $4.5$ $\mu$m for the $5.8$ $\mu$m band,
and $5.8$ and $8.0$ $\mu$m for the $8.0$ $\mu$m band).
Hereafter, we will refer to the distribution of magnitude $[I]$ for samples 
S1 and S2 as S1[$I$] and S2[$I$], respectively.
At their  peak, the distribution of S1[$3.6$] and S1[$4.5$] are a factor 
$\sim 0.6-0.7$ of that of S2[$3.6$] and
S2[$4.5$], respectively. This ratio increases to $> 0.8$ one magnitude below
(i.\ e., brighter than) the peak.

The distributions of 
S1[$8.0$] and S2[$8.0$] (not shown) are qualitatively similar, although the ratio S1[$8.0$] to S2[$8.0$]
is $\sim 0.72$ one magnitude below the peak of S1[$8.0$].
However, such a comparison is less significant than for other bands,
given that most of the sources without a detection
in any of the first two bands are probably just blobs of diffuse emission
rather than stellar objects.

Based on the histograms of the magnitude distribution for sources with good detections 
in at least two
bands (i. e., S2[$I$]), we conservatively estimate that the
80 \% completeness limits of each sample S2 lie roughly one magnitude below its 
distribution
peak (set CL0). E. g., $[3.6]=12.25$, and $[4.5]=12.25$.
Therefore, based on the above assumption, we can derive 
$\sim 80$ \% completeness limits for sources with good detections in all IRAC bands
(set CL1): $[3.6]=10.25$, $[4.5]=9.75$, $[5.8]=9.25$, and
$[8.0]=8.75$. These are obtained simply by measuring the magnitude
at which the ratio of the distribution S1[$I$] to S2[$I$] is $\sim 0.8$, considering that this usually occurs
more than one magnitude below the peak of S2[$I$] (hence where S2[$I$] has been assumed 
to be almost complete).

The same exercise, but replacing the samples S1 with other samples 
obtained by requiring that
the selected sources are detected in at least the first three bands with photometric errors
$<0.3$ mag (S1a), yields
the following 80 \% completeness limits (set CL2):
$[3.6]=11.75$, $[4.5]=11.5$, and $[5.8]=10.25$.

Requiring that all the sources in a given band be also detected in other bands with
photometric errors $<0.3$ mag can be quite a demanding prescription,
causing otherwise good detections in that band
to be removed because the detection
in just another band has an error $>0.3$ mag. We checked that this is only a minor problem
by noting the following.
The histograms of the magnitude distribution
obtained by selecting $I$-band sources detected in all four bands,
but requiring good detections in all bands for one histogram (S1[$I$])
and at least in the $I$ band for the other (S4[$I$]),
both peak at the same magnitude, irrespectively of $I$. The ratio
S1[$I$] to S4[$I$] slightly
decreases with increasing magnitudes down to $> 80$ \% at the peak.
The same is found by selecting $I$-band sources detected in the first three bands,
but either requiring good detections in all three bands
(S1a[$I$]) or at least in the $I$ band (sample S6[$I$]).
In this case, the ratio of the distribution S1a[$I$] to S6[$I$] is $> 75$ \% at the peak.
For $I = 3.6$ and $I = 4.5$ $\mu$m bands, we compared the histograms of
$I$-band sources with good detections in at least the first two bands
(S2[$I$])
and $I$-band sources detected in at least the first two bands but with good detection 
in at least the $I$ band (sample S8[$I$]). These histograms
(see the two upper rows in the right column of Fig.~\ref{fig:histo_comp}
are very similar each other, with ratio of distribution S2[$I$] to S8[$I$] $> 95$ \%
at the peak.
By replacing the distribution S8[$I$] 
with those of sources with at least a good detection in one of the first two bands
(S8[$3.6$] $\cup$ S8[$4.5$]),
the ratio decreases to $> 86$ \% at the histogram peaks. Therefore, the effect
of requiring good detections in at least the first two bands should
imply a completeness limit not too different from
that in each single lower wavelength band.

The lower two rows of Fig.~\ref{fig:histo_comp} show
that $5.6$- and $8.0$-$\mu$m-band sources with at least a good detection in the considered
band, peak at slightly lower magnitudes than the corresponding sub-sample obtained
by requiring a good detection in all bands.  As said, we
believe that most of the sources in the two upper wavelength bands without simultaneous detections
in the first two bands are mostly blobs of diffuse emission 
or even artefacts. So the statistics of all detections (multiple as well as single) 
in those bands is very likely to be biased.

Finally, we expect quite different completeness limits in areas without and with diffuse emission.
To test this, we repeated all experiments described above but by selecting 1) only sources contained
in an area with at most faint diffuse emission, i. e., inside the large ring, and 2) only
sources inside the area containing G353.2+0.9. There is no difference in the histogram peak locations 
for the sets CL0, CL1, and CL2, at $3.6$ and $4.5$ $\mu$m, between the complete samples
discussed above and those of the sources inside the ring. On the other hand, the peaks of the histograms
for $5.8$ and $8.0$ $\mu$m sources inside the ring shift to larger magnitudes.
This is probably because the crowding of sources far from diffuse emission areas is much less
severe in the two longer wavelength IRAC bands. 
Apparently, in the first two bands the total (i. e., with sources from the whole field) samples are dominated
by sources not falling towards diffuse emission, hence by crowding effects.

As far as sources inside the area containing G353.2+0.9 are concerned, 
the peaks of the histograms shift to lower magnitudes
for sets CL0, CL1, and CL2, as expected, excluding CL1 at $3.6$ and $4.5$ $\mu$m
whose peaks are one magnitude above (i. e., fainter than) those of the total samples.
We suspect that this is due to the chance associations of faint sources
in the first two bands and blobs of diffuse emission in the two upper wavelength bands, 
artificially
increasing the detection efficiency when requiring good detections in all bands.
In any case, the source statistics is quite poor. Thus, we believe that only the
estimates from CL0 (i. e., from the first two bands) are reliable in areas of intense 
diffuse emission.

The derived completeness limits (at a $70-80$ \% level)
for different detection requirements and areas in the
images are listed in Table~\ref{compl:tab}.

\begin{table*}
\caption{Estimated completeness limits.
\label{compl:tab}}
\centering                          
\begin{tabular}{ c c c c c c c c c c}        
\hline\hline                 
Image area & \multicolumn{4}{c}{good detections in} &
\multicolumn{3}{c}{good detections in} & \multicolumn{2}{c}{good detections in} \\
   & \multicolumn{4}{c}{4 bands} & \multicolumn{3}{c}{the 3 shorter $\lambda$ bands} &
\multicolumn{2}{c}{the 2 shorter $\lambda$ bands} \\
 & [3.6] & [4.5] & [5.8] & [8.0] & [3.6] & [4.5] & [5.8] & [3.6] & [4.5] \\
\hline                        
Faint diffuse emission & $10.25$ & $9.75$ & $9.75$ & $9.75$
        & $11.75$ & $11.5$ & $11.25$ & $12.25$ & $12.25$ \\
Intense diffuse emission & -- & -- & $7.25$ & $6.75$
        & -- & -- & $7.25$ & $10.75$ & $10.75$\\
\hline                                   
\end{tabular}
\end{table*}

\section{Contamination of YSOs due to other red sources 
in the Spitzer photometry}
\label{app:cont}
\begin{figure*}
   \centering
   \includegraphics[width=10cm]{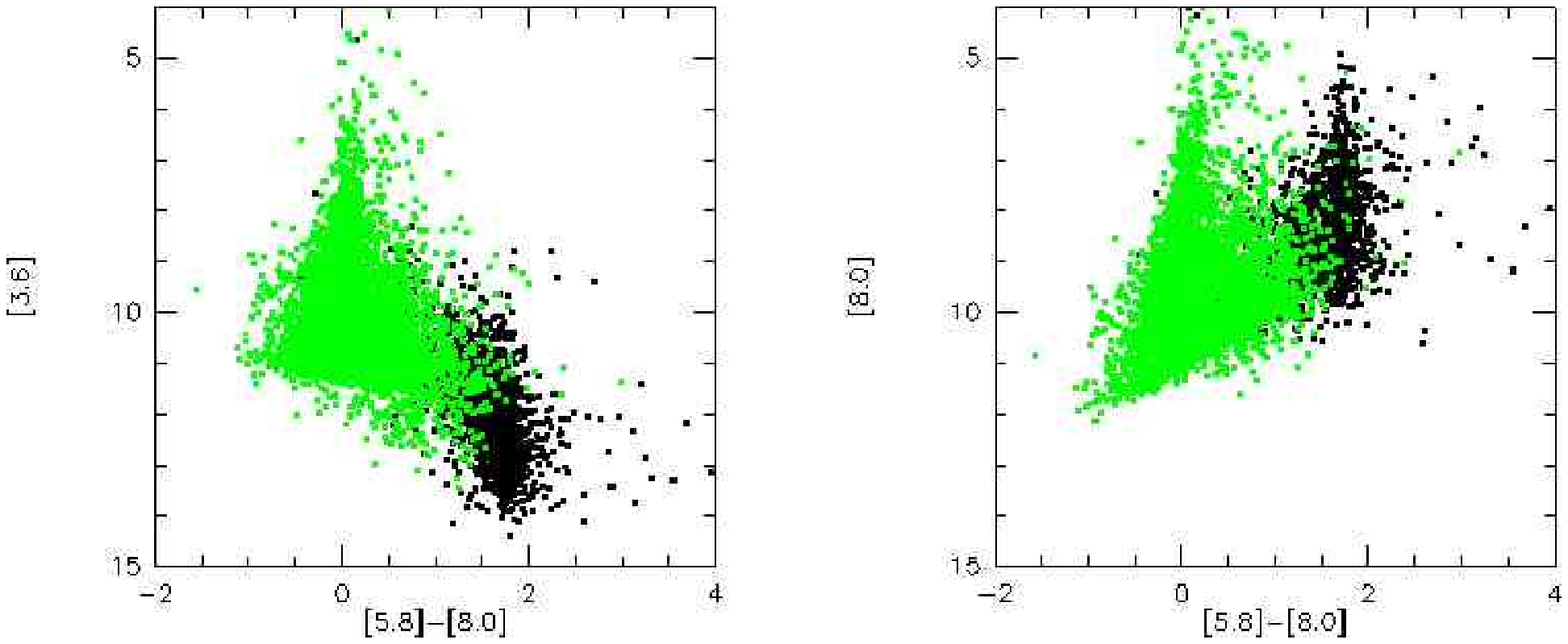}
   \caption{Magnitude vs.\ [5.8]--[8.0] in the $3.6$ $\mu$m (left)
	and $8.0$ $\mu$m (right) bands for all sources in our
	sample with detections in all four bands and photometric errors
	$< 0.3$ mag (black dots). Overlaid (green dots), all
	remaining sources after removal of the various contaminants
	as explained in the text.
              \label{fig:betcont}}
    \end{figure*}

Figure~\ref{fig:cont1234} clearly shows a prominent group of sources with [4.5]--[8.0] $>2$ 
in our sample, much more numerous than in the GLIMPSE sample.
This is better evidenced in Fig.~\ref{fig:betcont}, by plotting [3.6] and [8.0]
vs.\ [5.8]--[8.0] for sources detected in all four bands and having
photometric errors $<0.3$ mag. The [4.5] vs.\ [5.8]--[8.0] plot (not shown) is similar
to the [3.6] vs.\ [5.8]--[8.0] one, whereas the [5.8] vs.\ [5.8]--[8.0] plot (not shown) is similar
to the [8.0] vs.\ [5.8]--[8.0] one. A population of red sources centred at 
[5.8]--[8.0] $\sim 1.7-1.8$ is quite easily identified. These sources appear to be faint
at $3.6$ and $4.5$ $\mu$m ($< 10$ mag), but bright at $5.8$ and $8.0$ $\mu$m
($> 10$ mag). On the other hand, the group of sources centred at [5.8]--[8.0] $\sim 0$
are mostly stars.

\begin{figure*}
   \centering
   \includegraphics[width=10cm]{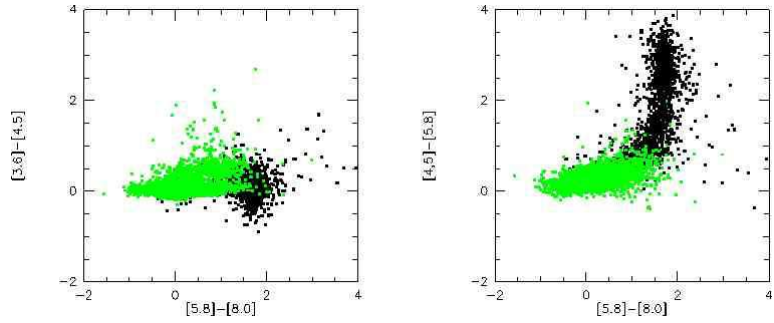}
   \caption{[3.6]--[4.5] vs.\ [5.8]--[8.0] (left) and
        [4.5]--[5.8] vs.\ [5.8]--[8.0] (right) 
	for all sources in our
	sample detected in all four bands with photometric errors
	$< 0.3$ mag (black dots). Overlaid (green dots), all
	remaining sources after removal of the various contaminants
	as explained in the text.
              \label{fig:colcont}}
    \end{figure*}

Plotting the position of the sources with [5.8]--[8.0] $> 1$ on 
each of the IRAC images, we found that almost all
appear to concentrate towards the filaments and areas of diffuse emission,
whereas sources with [5.8]--[8.0] $< 1$ are anti-correlated with those areas.
This suggests that most of the red sources are likely to be sources
contaminated by PAH emission rather than actual intrinsically red objects.
In fact, Zavagno et al.\ (\cite{zavagno}) found that analogous filaments in
RCW~79 have colours [3.6]--[4.5] $\sim 0.1$ and [5.8]--[8.0] $\sim 1.8$.
The same average colours can be derived from Fig.~\ref{fig:colcont} for the red sources
in our sample. These red sources probably arise due to random association of
faint sources in the two shorter wavelength bands and bright blobs of diffuse emission
in the two longer wavelength bands. All these blobs are bright, because
of both the lower sensitivity and saturation problems towards those areas. 
Figure~\ref{fig:colcont} (right panel) shows well this datapoint population,
whose colour [5.8]--[8.0] is remarkably constant ($\sim 1.7-1.8$ mag) whereas
[4.5]--[5.8] spreads between $\sim 1$ and $\sim 4$.
In other words, there is a sort of break between the emission at $3.6$ and $4.5$ $\mu$m 
([3.6]--[4.5] $\sim 0$) and that at $5.8$ and $8.0$ $\mu$m ([5.8]--[8.0] $\sim 1.7-1.8$ mag),
which points to a random association rather than intrinsically
red objects, as well. 

\begin{figure}
   \centering
   \includegraphics[width=6cm]{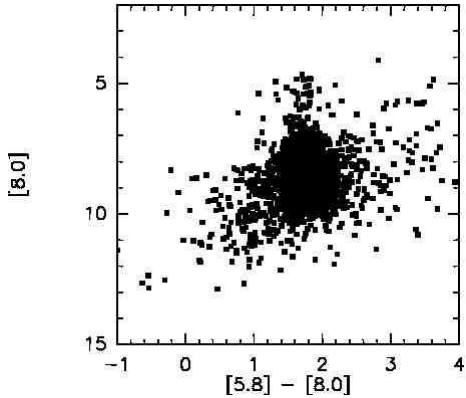}
   \caption{[8.0] vs.\ [5.8]--[8.0] 
	for all sources in our
	sample only detected at $5.8$ and $8.0$ $\mu$m.
              \label{fig:cont_34}}
    \end{figure}

A plot of [8.0] vs.\ [5.4]--[8.0] is 
shown in Fig.~\ref{fig:cont_34} for sources only detected at $5.8$ and $8.0$
$\mu$m; again, they are mostly concentrated at [5.4]--[8.0] $\sim 1.7 - 1.8$ and
are distributed like the red sources in Fig.~\ref{fig:cont1234}. Spatially, they concentrate
towards filaments and diffuse emission areas, as well. If they had counterparts in the two
lower wavelengths, they would correspond to even redder sources 
(so red to remain undetected at those wavelengths because lying below the
sensitivity limits) than those with detections in all
four bands. Consequently, it is more plausible that the objects exhibiting [5.4]--[8.0] $\sim 1.7 - 1.8$
are mostly blobs of diffuse emission.

To understand why this population of red sources is scarcely represented in
the GLIMPSE survey (see Fig.~\ref{fig:cont1234}), we plotted the position of 
all GLIMPSE sources detected in all four bands with photometric errors $<0.3$
mag. Unlike the red sources in our sample, few objects fall towards the filaments
and H\textsc{ii} regions. Therefore, the analogues of our red sources 
were filtered out of the GLIMPSE database because of their adopted selection criteria.

To remove the identified PAH-contaminants, we used the prescriptions given by
Gutermuth et al.\ (\cite{gutermuth}). We also removed all other possible 
contaminants following Gutermuth et al.\ (\cite{gutermuth}), i. e., AGNs,
star forming galaxies and unresolved shock emission. This is a conservative choice,
since we are observing through the Galactic plane and contamination
due to galaxies should be negligible. As can be seen in Figs.~\ref{fig:cont1234}
through \ref{fig:colcont} (green dots), the PAH contaminants appear to have been
removed quite satisfactorily.
On the other hand, for sources only detected in the first three bands,
we applied the criterion for PAH-contamination but not the others,
which require a knowledge of the flux at $8.0$ $\mu$m as well. 

   \begin{figure}
   \centering
   \includegraphics[width=6cm]{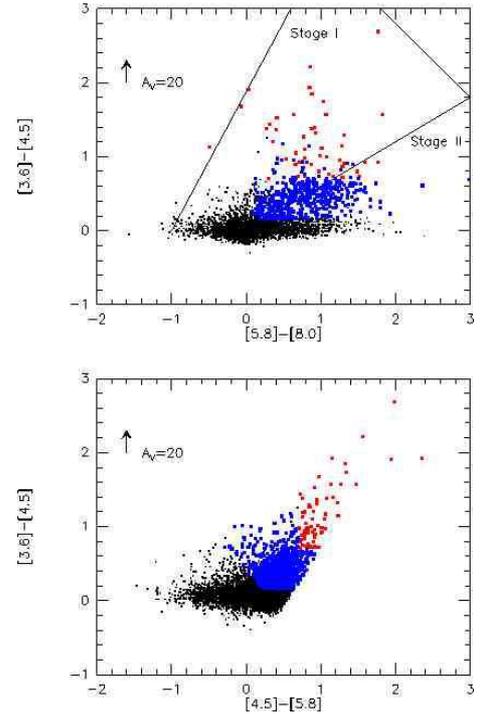}
      \caption{IRAC colour-colour diagrams. Top panel,
        [3.6]--[4.5]  vs.\ [5.8]--[8.0] for sources
           detected in all four bands and with photometric
           errors $<0.3$ mag. Bottom panel,
           [3.6]--[4.5]  vs.\ [4.5]--[5.8] for
           the same sources as above, plus those only detected in the three
           shorter wavelength bands (with photometric errors $< 0.3$ mag).
           The two samples have been cleaned of contaminants as explained
           in the text. Following the prescriptions of Gutermuth et al.\ 
           (\protect\cite{gutermuth}\protect)
           for their identification,
           Class I sources are in red and Class II sources in blue. 
           Stars are marked as black dots. A reddening
           corresponding to $A_{V} = 20$ mag (according to
           Indebetouw et al.\ \protect\cite{indebetouw}\protect) is marked by an arrow.
           In the top panel, the regions occupied by Stage I and
           Stage II sources (Robitaille et al.\ \protect\cite{robi:clas}\protect) are
           also outlined.
         \label{fig:col-col}}
   \end{figure}
%

The final efficiency of the cleaning process can be checked by plotting the colours of Class I
and Class II sources (see Fig.~\ref{fig:col-col}), identified after removal
of contaminants by using the colour criteria of Gutermuth et al.\ (\cite{gutermuth}). 
In the top panel of Fig.~\ref{fig:col-col}, the regions occupied by Stage I and
Stage II sources (following the scheme proposed by Robitaille et al.\ \cite{robi:clas},
who slightly modified the original classification scheme resulting in Class I and II sources)
are also outlined. 
By following the colour criteria of Allen et al.\ (\cite{allen:clas}), we would have
obtained similar results for Class I and Class II sources. The colour-colour
diagram showing [3.6]--[4.5] vs.\ [5.8]--[8.0] is very similar to that obtained
by Fang et al.\ (\cite{fang}), although a look at their Fig.~5a shows
the effects of the different classification criteria they adopted.

\section{Multi-wavelength classification of YSOs in the SofI field}
\label{compl:nir-irac-x}

The evolutionary stages of the cluster members can be derived by using either
NIR/MIR colours or their association with X-ray emission.
Nevertheless, different methods definitely suffer from different selection effects
and this has to be accounted for when comparing the obtained results. 

In this paper, we adopted several classification schemes based on: $JHK_{s}$, [3.6][4.5][5.8][8.0],
[3.6][4.5][5.8], $JH(HK_{s})$[4.5] colours, X-ray emission. All these indicators allow us to
identify the embedded protostar population, the classical T-Tauri population and
the weak-line T-Tauri population out of the whole cluster stellar population. However,
when computing number ratios we have to check for any difference in completeness 
that might arise because of the different sensitivities attained in each band.

\begin{figure*}
   \centering
   \includegraphics[width=10cm]{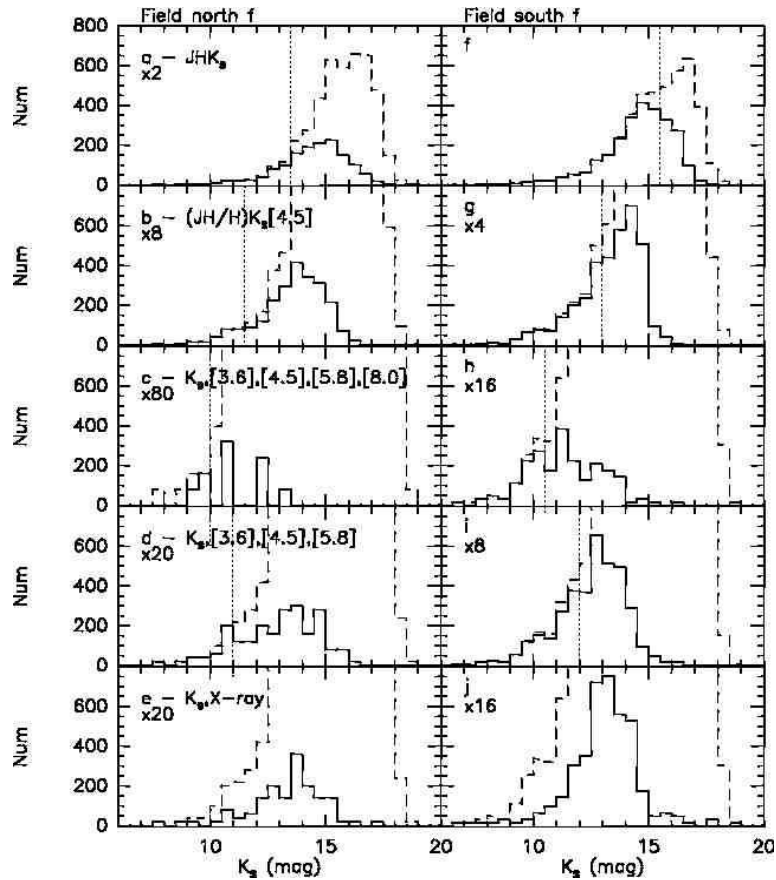}
   \caption{Number of $K_{s}$ sources per magnitude in bins of $0.5$ mag
        (dashed line)
        compared with (full line): the number of $K_{s}$ sources with also good detections
        both at $J$ and $H$ ({\bf a}, northern field; {\bf f} southern field);
	the number of $K_{s}$ sources with also good detections either at both $J$ and $H$
        or at least at $H$, and in the IRAC 4.5 $\mu$m band ({\bf b}, northern field; 
	{\bf g} southern field);
	the number of $K_{s}$ sources with also good detections in all four IRAC bands
          ({\bf c}, northern field; {\bf h} southern field);
	the number of $K_{s}$ sources with also good detections in the first three IRAC bands
          ({\bf d}, northern field; {\bf i} southern field);
         the number of $K_{s}$ sources that exhibits an X-ray emission ({\bf e}, northern field; 
	{\bf j} southern field).
	We only took into account sources with photometric uncertainty $< 0.3$ mag in all 
	the considered bands.
 In addition, the samples plotted in panels c, d, h, i, have been 
 	rid of sources with colours indicating PAH contamination as explained
        in App.~\ref{app:cont}.
	On the upper left corner of each box, a {\it xN} tag indicates whether the histograms 
	plotted have
	been multiplied by a constant $N$ to expand the scale.
	The vertical dotted lines mark the $K_{s}$ values at which $\sim 80$ \% of the $K_{s}$ 
	sources
	are still retrieved when requiring simultaneous valid detections in other bands.
              \label{fig:mw:comp}}
    \end{figure*}

To derive these differences, we reasonably assume that our $K_{s}$ image 
is the most sensitive of all available bands, either SofI or IRAC. On the one hand, many
$K_{s}$ sources are missed in $J$ and $H$ because of extinction. On the other hand, the 
observations in the IRAC 3.6 and 4.5 $\mu$m bands
are less extincted, but not as deep as $K_{s}$ and with a worse spatial resolution.
Therefore, $K_{s}$ can be considered as our reference band.
Consequently, we can compute the maximum $K_{s}$ for which at least 80 \% of the $K_{s}$ sources
are also detected i) in the $JH$ bands, ii) in the 4.5 $\mu$m bands AND either in the $JH$  
bands or in at least the $H$ band, iii) in all four IRAC bands and iv) in the first three 
IRAC bands. We will refer to the derived $K_{s}$ in each case as the 
''80 \% completeness limit'' throughout this section. 
We constructed the $K_{s}$ luminosity functions for all these classes (i--iv) of sources
and compared them with the $K_{s}$ luminosity function of the sources at least detected
in the $K_{s}$ band, both for the northern and the southern part of the SofI field. 
We only counted sources with photometric uncertainty $< 0.3$ mag in every 
band considered. The luminosity functions are compared in Fig.~\ref{fig:mw:comp}. For detections in three and four
IRAC bands, we discarded sources with IRAC colours typical of PAH contamination 
(see App.~\ref{app:cont}). We found however that this is important only for
sources with good detections in 4 IRAC bands located in the northern field, as expected because of the
contaminating diffuse emission from the H\textsc{ii} region. The 80 \% completeness limits are indicated with
vertical dotted lines in Fig.~\ref{fig:mw:comp}. 
They are also listed in Table~\ref{compl:nir-irac-x:tab}.

We note that the 80 \% completeness limit significantly varies from case to case, depending on
the colours used and the area of the field. For example, a lot of $K_{s}$ sources are missed particularly
in the northern SofI field when requiring valid detections in all four IRAC bands. This
is due to the problems suffered by the two upper wavelength IRAC bands in the regions of
diffuse emission outlined in the text. 
We also note that the $JHK_{s}$ colours mostly tend to filter out embedded stars, young
PMS stars and protostars, whereas IRAC colours mostly tend to filter out unreddened stars. 
Thus, we can hope to obtain a better 80 \% completeness limit by 
combining $K_{s}$ sources for which {\it at least} one of the colours from i) to iv) is available. 
However,
as shown in the 6th column of Table~\ref{compl:nir-irac-x:tab}, this only happens in the
northern SofI field.

The bottom row in Fig.~\ref{fig:mw:comp} shows the luminosity function of $K_{s}$ sources
that also exhibit an X-ray emission. Given that not all $K_{s}$ sources 
(particularly field stars, unlike T Tauri stars) have detectable
X-ray emission, this luminosity function cannot be directly compared with the one of
all sources with a valid $K_{s}$ detection. In the southern field, the luminosity function
exhibits a fast increase up to $K_{s} \sim 13$. This suggests that all $K_{s}$ sources
emitting X-rays are roughly complete down to that limit.
On the other hand, in the northern field the luminosity function is much less steep;
if this were mainly due to an increased difficulty in retrieving sources in the 
northern area of the $K_{s}$ image, lots of X-ray sources without a NIR
counterpart should be detected.  
Only 5 X-ray sources are actually detected without a NIR counterpart
there.
So, the flatter X-source $K_{s}$ luminosity function in the north 
may arise due to either a decrease in the probability
of detecting X-ray emission from embedded cluster members, or a greater $K_{s}$ magnitude
spread of cluster members caused by variable extinction. 
Therefore, it is difficult to estimate a 
$K_{s}$ completeness limit for X-ray emitting sources
in this case; it may lie in the range 11--13 mag. However, we note that 
for strongly variable extinction (as expected when moving from south to north)
this would not correspond to a particular mass limit.
 
\begin{table*}
\caption{$K_{s}$ values at which 80 \% of $K_{s}$ sources are still retrieved when also detected
                             in other bands.
\label{compl:nir-irac-x:tab}}     
\centering                          
\begin{tabular}{ c c c c c c }        
\hline\hline                 
Field &  $JHK_{s}$ & $JH$ or $H$,$K_{s}$[4.5] & [3.6][4.5][5.8][8.0] & [3.6][4.5][5.8] & all combined\\
\hline                        
north & $13.5$ & $11.5$ & $10.0$ & $11.0$  & $14.25$ \\
south & $15.5$ & $13.0$ & $10.5$ & $12.0$ & $15.5$ \\
\hline                                   
\end{tabular}
\end{table*}

\section{Individual NIR sources towards Pismis~24}
\label{ind:so}

\begin{figure}
   \centering
   \includegraphics[width=8cm]{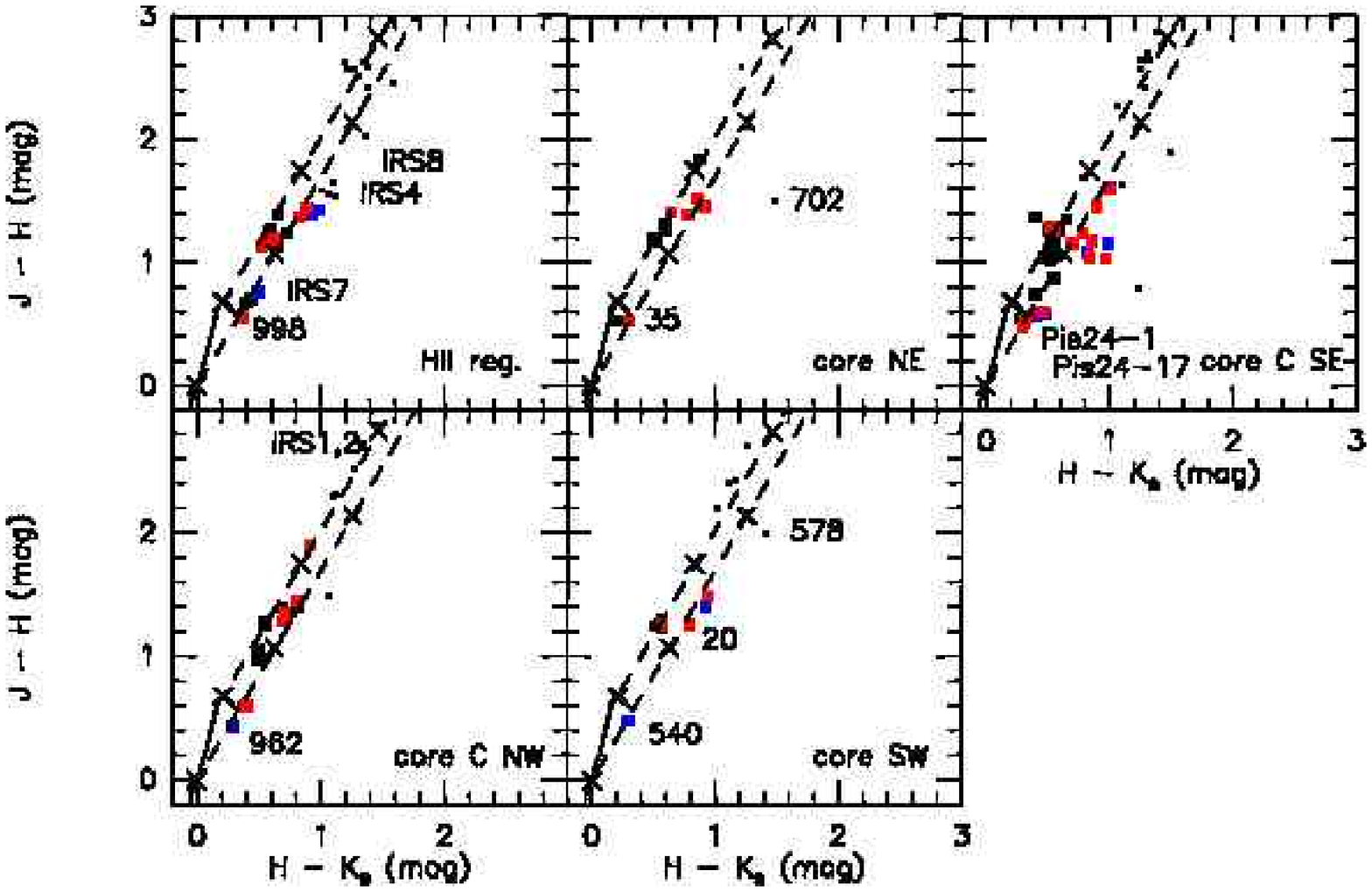}
      \caption{$J - H$ vs.\ $H-K_{s}$ for NIR sources towards
	the five areas (or sub-clusters) selected in the SofI field
 	of view. Sources with a $JHK_{s}$ colour excess 
	(and $H-K_{s}<1$) are marked by  blue
	squares, Sources with a $JH(HK_{s})[4.5]$ colour excess 
	(and $H-K_{s}<1$) are marked by red squares, other
	sources are marked by black squares (small squares for
         $H-K_{s} \ge 1$). The main sequence locus is also drawn in each diagram,
	along with reddening (dashed lines with crosses every
        $A_{V} = 10$ mag).
	The sources discussed in the text are labelled, as well. 
         \label{ccd:sub}}
   \end{figure}
%
%

\begin{figure}
   \centering
   \includegraphics[width=8cm]{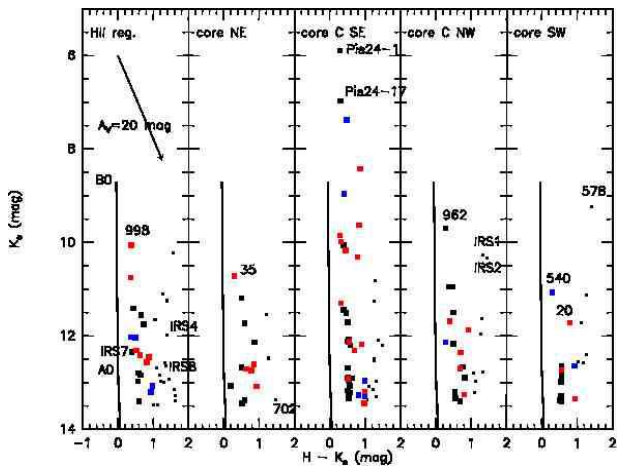}
      \caption{$K_{s}$ vs.\ $H-K_{s}$ for NIR sources towards
	the five areas (or sub-clusters) selected in the SofI field
 	of view. Sources with a $JHK_{s}$ colour excess 
	(and $H-K_{s}<1$) are marked by blue
	squares, sources with a $JH(HK_{s})[4.5]$ colour excess 
	(and $H-K_{s}<1$) are marked by red squares, other
	sources are marked by black squares (small squares for
         $H-K_{s} \ge 1$). The ZAMS is also drawn in each diagram,
	whereas an arrow in the first box from left marks 
        an extinction $A_{V} = 20$ mag. The sources discussed in the text
	are labelled, as well. 
         \label{cmd:sub}}
   \end{figure}

The nature of the most massive stars in each sub-cluster of Pis24 core
can be
better investigated by considering the corresponding $J - H$ vs.\ $H-K_{s}$ 
(Fig.~\ref{ccd:sub}) and $K_{s}$ vs.\ $H-K_{s}$ diagrams (Fig.~\ref{cmd:sub}).
Clearly, all the O-type stars lie towards the south-eastern part of sub-cluster core C,
whereas the other star concentrations are more reminiscent of the small embedded young clusters
where typically intermediate-mass stars form.
\newline
\noindent
{\bf Core SW}: 
it is the only sub-cluster, apart from core C SE, with a possible earlier-than-B0 star.
Our NIR source \# 578 is, in fact, bright enough and exhibits
a clear $JHK_{s}$ colour excess, suggesting it is a massive young star. 
\newline
\noindent
{\bf H\textsc{ii} region}: the most massive
star towards the H\textsc{ii} region, our NIR source \# 998, is 
a B2 star located on the western part
of the bar. The other bright ($K_{s} \sim 10.22$) star is located just west of
the elephant trunk but is probably a background giant. 
The star at the tip of the elephant trunk (IRS4 of Felli et al.\ \cite{felli:90}) 
is a B0--B4 star
with a $JHK_{s}$ colour excess. IRS7 and IRS8 of Felli et al.\ (\cite{felli:90}) also exhibit
a $JHK_{s}$ colour excess, but appear to be less massive than IRS4. We found that 
IRS1 and IRS2 were misplaced by Felli et al.\ (\cite{felli:90}): 
by comparing their Fig.~4 with our Fig.~\ref{pismisJHK}
it can easily be verified that these two sources lie $\sim 15\arcsec$ further south
than the bar (in their Fig.~7), i. e., towards the north-western part 
of sub-cluster core C and not in the bar.  In addition, Fig.~\ref{cmd:sub} and
Fig.~\ref{ccd:sub} clearly show that they are probably background giants.
\newline
\noindent
{\bf Core C (NW)}: the most massive star is a B1--B2 star, our NIR
source \# 962. It is located near the border between the two parts of sub-cluster core C,
and it could be associated with C SE rather than the C NW.
\newline
\noindent
{\bf Core NE}: it is the only core region hosting an IRAC-retrieved Class I source
(with $K_{s} < 13.5$), namely  our NIR source \# 702, which has  a characteristic colour 
excess in the $J - H$ vs.\ $H - K_{s}$ diagram (Fig.~\ref{ccd:sub}), as well. 

%
%

\end{appendix}



\begin{thebibliography}{}

  \bibitem[1976]{all76} Allen, C.\ W.\ 1976, Astrophysical Quantities
	(3rd ed.), (London: Athlone press)

  \bibitem[2004]{allen:clas} Allen, L.\ E., Calvet, N., D'Alessio, P.,
        et al.\ 2004, \apjs, 154, 363

  \bibitem[2011]{arthur} Arthur, S.\ J., Henney, W.\ J., Mellema, G., De Colle, F.,
	\& V\'{a}zquez-Semadeni, E.\ 2011, \mnras, 414, 1747


  \bibitem[2009]{baretal} Baraffe, I., Chabrier, G., \& Gallardo, J.\ 2009, 
	\apj, 702L, 27 

  \bibitem[2010]{becca10} Beccari, G., Spezzi, L., De Marchi, G., et al.\ 2010, 
	\apj, 720, 1108

  \bibitem[1979]{bessel} Bessel, M.\ S.\ 1979, \pasp, 91, 589

  \bibitem[1991]{bessel91} Bessel, M.\ S.\ 1991, \aj, 101, 662

  \bibitem[2011]{bisbas:11} Bisbas, T.\ G., W\"{u}nsch, R., Whitworth, A.\ P., Hubber, D.\ A.,
	\& Walch, S.\ 2011, \apj, 736, 142 

  \bibitem[2004]{bohi} Bohigas, J., Tapia, M., Roth, M., \& Ruiz, M.\ T.\ 2004, \aj, 127, 2826

  \bibitem[2011]{bonnell:11} Bonnell, I.\ A., Smith, R.\ J., Clark, P.\ C., \&
   Bate, M.\ R.\ 2011, \mnras, 410, 2339

  \bibitem[2011]{cappa11} Cappa, C.\ E., Barb\'{a}, R., Duronea, N.\ U., et al.\ 2011, \mnras, 415, 2844
 
  \bibitem[2003]{chabry} Chabrier, G.\ 2003, \pasp, 115, 763

  \bibitem[1983]{ch:kr} Chini, R., \& Kr\"{u}gel, E.\ 1983, \aap, 117, 289

  \bibitem[2007]{churchwell} Churchwell, E., Watson, D.\ F., Povich, M. S.,
 	et al.\ 2007, \apj, 670, 428


  \bibitem[2011]{dale:11} Dale, J.\ E., \& Bonnell, I.\ 2011, \mnras, 414, 321

  \bibitem[2012a]{dale:12a} Dale, J.\ E., Ercolano, B., \& Bonnell, I.\ 2012a, \mnras, 424, 377

  \bibitem[2012b]{dale:12b} Dale, J.\ E., Ercolano, B., \& Bonnell, I.\ 2012b, \mnras, 
                   427, 2852	

  \bibitem[2006]{damke} Damke, G., Barb\'{a}, R., Morrell, N.\ I., et al\ 
      2006, RevMexAA, 26, 181

  \bibitem[2005]{deha} Deharveng, L., Zavagno, A., \& Caplan, J.\ 
           2005, \aap, 433, 565 

  \bibitem[2011]{deha:zava} Deharveng, L., \& Zavagno, A.\
           2011, Observations of star formation triggered by H\textsc{ii} regions, in
           Computational Star Formation, Proceedings of the International Astronomical Union,
           ed. J.\ Alves, B.\ G.\ Elmegreen, J.\ M.\ Girart, \and V.\ Trimble,
           IAU Symposium, Vol.\ 270, 239 


 \bibitem[2011a]{dema11a} De Marchi, G., Panagia, N., Romaniello, M., et al.\ 2011a, 
           \apj, 740, 11

 \bibitem[2011b]{dema11b} De Marchi, G., Paresce, F., Panagia, N., et al.\ 2011b, 
           \apj, 739, 27

 \bibitem[2013]{dema13} De Marchi, G., Beccari, G., \& Panagia, N.\ 2013, 
           \apj, 775, 68

 \bibitem[2009]{deSilva} De Silva, G.\ M., Gibson, B.\ K., Lattanzio, J., \& Asplund, M.\
            2009, \aap, 500, L25

  \bibitem[2002]{dias} Dias, W.\ S., Alessi, B.\ S., Moitinho, A.,
     	\& L\'{e}pine, J.\ R.\ D.\ 2002, \aap, 389, 871

  \bibitem[2000]{dolphin} Dolphin, A.\ E.\ 2000, \pasp, 112, 1383 

  \bibitem[2008]{dotter} Dotter, A., Chaboyer, B., Jevremovi\'{c}, D., et al.\
	 2008, \apjs, 178, 89 


  \bibitem[2001]{ducati} Ducati, J.\ R., Bevilacqua, C.\ M., Rembold, S.\ B.,
         \&  Ribeiro, D.\ 2001, \apj, 558, 309

  \bibitem[2012]{fang} Fang, M., van Boekel, R., King, R.\ R., et al.\
 	2012, \aap, 539, 119 

  \bibitem[2004]{fazioetal} Fazio, G.\ G., Hora, J.\ L.,  
	Allen, L.\ E., et al.\ 2004,
	\apjs, 154, 10

  \bibitem[1999]{fm:99} Feigelson, E.\ D., \& Montmerle, T.\ 1999,
	\araa, 37, 363 

  \bibitem[1990]{felli:90} Felli, M., Persi, P., Roth, M., et al.\ 1990, \aap, 232, 477

  \bibitem[1999]{fitz} Fitzpatrick, E.\ L.\ 1999, PASP, 111, 63 

  \bibitem[2012]{flaccomio} Flaccomio, E., Micela, G., \& Sciortino, S.\ 2012, \aap, 548, A85

  \bibitem[2012]{genkru} Gendelev, L., \& Krumholz, M.\ R.\ 2012, \apj, 745, 158

 \bibitem[2014]{getman} Getman, K.\ V., Feigelson, E.\ D., Kuhn, M.\ A., et al.\  2014,
		\apj, 787, 108

  \bibitem[2012]{giannetti} Giannetti, A., Brand, J., Massi, F.,
	Tieftrunk, A., \& Beltr\'{a}n, M.\ T.\ 2012, \aap, 538, A41

  \bibitem[2012]{gouliermis} Gouliermis, D.\ A., Schmeja, S., Dolphin,
	A.\ E., et al.\ 2012, \apj, 748, 64

  \bibitem[2012]{gratton} Gratton, R.\ G., Carretta, E., \& Bragaglia, A.\ 2012,
       \aapr, 20, 50

  \bibitem[2010]{grit} Gritschneder, M., Burkert, A., Naab, T., \&
	Walch, S.\ 2010, \apj, 723, 971 

  \bibitem[2014]{groh} Groh, J. H., Meynet, G., Ekstr\"{o}m, S., \&
	Georgy, C.\ 2014, \aap, 564, A30 

  \bibitem[2007]{guarcello} Guarcello, M.\ G., Prisinzano, L., Micela, G.,
	et al.\ 2007, \aap, 462, 245

  \bibitem[2009]{gutermuth} Gutermuth, R.\ A., Megeath, S.\ T.,  
	Myers, P.\ C., et al.\ 2009,
	\apjs, 184, 18

  \bibitem[2011]{gvaramadze} Gvaramadze, V.\ V.,
   	Kniazev, A.\ Y., Kroupa, P., \& Oh, S.\ 2011, \aap, 535, A29  

  \bibitem[1981]{h&h} Habets, G.\ M.\ H.\ J., \& Heintze, J.\ R.\ W.\
	1981, \aaps, 46, 193 

  \bibitem[2001]{haisch} Haisch, K.\ E., Lada, E.\ A., \& Lada, C.\ J.\ 2001,
	\apj, 553, L153

  \bibitem[2001]{hartie} Hartmann, L.\ 2001, \aj, 121, 1030


  \bibitem[1997]{hillen} Hillenbrand, L.\ A.\ 1997, \aj, 113, 1733 

  \bibitem[1995]{holtzman} Holtzman, J. A., Burrows, C.\ J., Casertano, S.,
     et al.\ 1995, \pasp, 107, 1065


  \bibitem[2005]{indebetouw} Indebetouw, R., Mathis, J.\ S., Babler, B.\ L.,
	et al.\ 2005, \apj, 619, 931

  \bibitem[2011]{jeffries} Jeffries, R.\ D., Littlefair, S.\ P., Naylor, T., 
         \& Mayne, N.\ J.\ 2011, \mnras, 418, 1948

  \bibitem[2014]{jeffries2} Jeffries, R.\ D., Jackson, R.\ J., Cottaar, M., 
	  et al.\ 2014, \aap, 563, A94 

  \bibitem[1995]{ke:ha} Kenyon, S.\ J., \& Hartmann, L.\ 1995, \apjs, 101, 117


  \bibitem[1983]{Koorn}
 	Koornneef, J.\ 1983, \aap, 128, 84

   \bibitem[1993]{kroupa}
        Kroupa, P., Tout, C.\ A., \& Gilmore, G.\ 1993, \mnras, 262, 545

 \bibitem[2014]{kuhn}
	Kuhn, M.\ A., Feigelson, E.\ D., Getman, K.\ V., et al.\ 2014, \apj, 
				787, 107

  \bibitem[2003]{L&L} Lada, C.\ J., \& Lada, E.\ A., 2003, \araa, 41, 57.

\bibitem[2001]{basel} Lejeune, T., \& Schaerer, D.\ 2001, \aap, 366, 538

 \bibitem[2014]{lima} Lima, E.\ F., Bica, E., Bonatto, C., \& Saito, R.\ K.\
                            2014, \aap, 568, A16

  \bibitem[1984]{lortet} Lortet, M.-C., Testor, G., \& Niemela, V.\ 
          1984, \aap, 140, 24

  \bibitem[2007]{maw} Ma\'{i}z Apell\'{a}niz, J., Walborn, N.\ R.,
	Morrell, N.\ I., Niemela, V.\ S., \& Nelan, E.\ P.\
	2007, \apj, 660, 1480

  \bibitem[2009]{mushy} Maschberger, T., \& Kroupa, P.\ 2009, \mnras, 395, 931  

  \bibitem[2001]{massey01} Massey, P., DeGioia-Eastwood, K., \&
       Waterhouse, E.\ 2001, \aj, 121, 1050 
 
  \bibitem[1997]{massi97} Massi, F., Brand, J., \& Felli, M.\ 1997,
		\aap, 320, 972 

  \bibitem[2006]{massi06} Massi, F., Testi, L., \& Vanzi, L.\ 2006,
		\aap, 448, 1007 


  \bibitem[2006]{mellema} Mellema, G., Arthur, S.\ J., Henney, W.\ J.,
            Iliev, I.\ T., \& Shapiro, P.\ R.\ 2006, \apj, 647, 397

  \bibitem[1997]{meyer} Meyer, M.\ R., Calvet, N., \& Hillenbrand, L.\ A.\ 1997,
             \aj, 114, 288

  \bibitem[1973]{mo:vo} Moffat, A.\ F.\ J., \& Vogt, N.\ 1973, \aaps, 10, 135

  \bibitem[1998]{moor} Moorwood, A., Cuby, J.-G., \& Lidman, C.\ 
	1998, ESO Mess., 91, 9

  \bibitem[1978]{neckel} Neckel, T.\ 1978, \aap, 69, 51

  \bibitem[2013]{padoan} Padoan, P., Federrath, C., Chabrier, G., et al.\ 2013,
      The Star Formation Rate of Molecular Clouds. In Protostar and Planet VI,
      ed.\ H.\ Beuther, R.\ S.\ Klessen, C.\ P.\ Dullemond, \& Th.\ Henning,
      University of Arizona Press (arXiv:1312.5365)

  \bibitem[1999]{ps99}
 	Palla, F., \& Stahler, S.\ W.\ 1999, \apj, 525, 772

  \bibitem[2000]{ps00}
 	Palla, F., \& Stahler, S.\ W.\ 2000, \apj, 540, 255

  \bibitem[2012]{Par:Mey}
	Parker, R.\ J., \& Meyer, M.\ R.\ 2012, \mnras, 427, 637

  \bibitem[1998]{perss} Persson, S.\ E., Murphy, D.\ C., Krzeminski, W.,
	Roth, M., \& Rieke, M.\ J.\ 1998, \aj, 116, 2475

  \bibitem[2002]{p&z} Preibisch, Th., \& Zinnecker, H.\ 2002, \aj, 123, 1613

  \bibitem[2012]{preiby} Preibisch, Th.\ 2012, RAA, 12, 1
  
  \bibitem[2014]{erre} R core team 2014, R: A Language and Environment
           for Statistical Computing, R Foundation for Statistical Computing,
           Vienna, Austria (http://www.R-project.org)

  \bibitem[2008]{ramirez} Ram\'{i}rez, S.\ V., Arendt, R.\ G., Sellgren, K.,
	et al.\ 2008, \apjs, 175, 147 

  \bibitem[2005]{reachetal} Reach, W.\ T., Megeath, S.\ T., Cohen, M., et al.\
	2005, PASP, 117, 978 

  \bibitem[2014]{reid14} Reid, M.\ J., Menten, K.\ M., Brunthaler, A., et al.\
	2014, \apj, 783, 130

  \bibitem[1985]{r&l}
    	Rieke, G.\ H., \& Lebofsky, M.\ J.\ 1985, \apj, 288, 618

  \bibitem[2006]{robi:clas} Robitaille, T.\ P., Whitney, B.\ A., Indebetouw, R., 
	Wood, K., \& Denzmore, P.\ 2006, \apjs, 167, 256

  \bibitem[2008]{robi:agb} Robitaille, T.\ P., Meade, M.\ R., 
	Babler, B.\ L., et al.\ 2008, \aj, 136, 2413


  \bibitem[2012]{russ:12}
	Russeil, D., Zavagno, A., Adami, C., et al.\ 2012, \aap, 538, A142

  \bibitem[1998]{scalo} Scalo, J.\ 1998, The IMF Revisited: A Case for Variations. In
 The Stellar Initial Mass Function,
 Proc.\ of the 38th Herstmonceux Conf., ed.\ G.\ Gilmore, \& D.\ Howell,
 ASP Conf.\ Ser., Vol.\ 142, 201

  \bibitem[2000]{siess} Siess, L., Dufour, E., \& Forestini, M.\ 2000, \aap, 358, 593
 
  \bibitem[1978]{spitzer} Spitzer, L.\ 1968, Diffuse Matter in Space (New York:Interscience)

  \bibitem[2004]{stolte} Stolte, A., Brandner, W., Brandl, B., Zinnecker, H.,
	\& Grebel, E.\ K.\ 2004, \aj, 128, 765

  \bibitem[2008]{StraLau} Straizys, V., \& Laugalys, V., 2008, BaltA, 17, 253 


  \bibitem[2014]{tan} Tan, J.\ C., Beltr{\'a}n, M.\ T., Caselli, P., et al.\ 2014,
      Massive Star Formation. In Protostars and Planets VI,
      ed.\ H.\ Beuther, R.\ S.\ Klessen, C.\ P.\ Dullemond, \& Th.\ Henning,
      University of Arizona Press (arXiv:1402.0919)

  \bibitem[2011]{pisa}
	Tognelli, E., Prada Moroni, P.\ G., \& Degl'Innocenti, S.\ 2011, \aap, 533,A109

 

  \bibitem[2011]{walch:11} Walch, S., Whitworth, A., Bisbas, T.,
        W\"{u}nsch, R., \& Hubber, D.\ A.\ 2013, \mnras, 435, 917  

  \bibitem[1995]{w&j} Wand, M.\ P., \& Jones, M.\ C., 1995, Kernel Smoothing,
     Chapman and Hall, London 

  \bibitem[2007]{wang07} Wang, J., Townsley, L.\ K., Feigelson, E.\ D., et al.\
                                      2007, \apjs, 168, 100 
 
  \bibitem[2010]{WV10} Weidner, C., \& Vink, J.\ S.\ 2010, \aap, 524, A98

  \bibitem[2010]{west:10}
        Westmoquette, M.\ S., Slavin, J.\ D., Smith, L.\ J., \&
	Gallagher, J.\ S.\ III 2010, \mnras, 402, 152

  \bibitem[2003]{whitney} Whitney, B.\ A., Wood, K., Bjorkman, J.\ E., \& Cohen, M.\ 
        2003, \apj, 598, 1079

  \bibitem[2004]{whitney2} Whitney, B.\ A., Indebetouw, R., Bjorkman, J.\ E., \& Wood, K.\
        2004, \apj, 617, 1177

  \bibitem[2007]{wins07} Winston, E., Megeath, S.\ T., Wolk, S.\ J., et al.\
					2007, \apj, 669, 493

  \bibitem[2006]{zavagno} Zavagno, A., Deharveng, L., Comer\'{o}n, F., et al.\
	2006, \aap, 446, 171

  \bibitem[2007]{zin:yo} Zinnecker H., \& Yorke, H.\ W.\ 2007, \araa, 45, 481

\end{thebibliography}
\end{document}